\documentclass[traditabstract, longauth]{aa}
\usepackage{graphicx}
\usepackage[dvips]{color} 
\usepackage{epsfig}
\usepackage{amssymb,amsmath}
\usepackage{longtable}
\usepackage[breaklinks, colorlinks, citecolor=blue]{hyperref}
\usepackage{breakurl}
\usepackage{natbib}
\usepackage{multirow}
\usepackage{url}
\usepackage{txfonts}
\DeclareGraphicsExtensions{.eps,.ps,.jpg,.png}
\bibpunct{(}{)}{;}{a}{}{,}

\def\setsymbol#1#2{\expandafter\def\csname #1\endcsname{#2}}
\def\getsymbol#1{\csname #1\endcsname}

\def\Planck{{\it Planck\/}}

\def\HeJT{$^4$He-JT}

\newbox\tablebox    \newdimen\tablewidth
\def\leaderfil{\leaders\hbox to 5pt{\hss.\hss}\hfil}
\def\endPlancktable{\tablewidth=\columnwidth 
    $$\hss\copy\tablebox\hss$$
    \vskip-\lastskip\vskip -2pt}
\def\endPlancktablewide{\tablewidth=\textwidth 
    $$\hss\copy\tablebox\hss$$
    \vskip-\lastskip\vskip -2pt}
\def\tablenote#1 #2\par{\begingroup \parindent=0.8em
    \abovedisplayshortskip=0pt\belowdisplayshortskip=0pt
    \noindent
    $$\hss\vbox{\hsize\tablewidth \hangindent=\parindent \hangafter=1 \noindent
    \hbox to \parindent{\sup{\rm #1}\hss}\strut#2\strut\par}\hss$$
    \endgroup}
\def\doubleline{\vskip 3pt\hrule \vskip 1.5pt \hrule \vskip 5pt}

\def\L2{\ifmmode L_2\else $L_2$\fi}

\def\DeltaT{\ifmmode \Delta T\else $\Delta T$\fi}
\def\deltat{\ifmmode \Delta t\else $\Delta t$\fi}
\def\fknee{\ifmmode f_{\rm knee}\else $f_{\rm knee}$\fi}
\def\Fmax{\ifmmode F_{\rm max}\else $F_{\rm max}$\fi}
\def\solar{\ifmmode{\rm M}_{\mathord\odot}\else${\rm M}_{\mathord\odot}$\fi}

\def\inv{\ifmmode^{-1}\else$^{-1}$\fi}
\def\mo{\ifmmode^{-1}\else$^{-1}$\fi}
\def\sup#1{\ifmmode ^{\rm #1}\else $^{\rm #1}$\fi}
\def\expo#1{\ifmmode \times 10^{#1}\else $\times 10^{#1}$\fi}
\def\,{\thinspace}
\def\lsim{\mathrel{\raise .4ex\hbox{\rlap{$<$}\lower 1.2ex\hbox{$\sim$}}}}
\def\gsim{\mathrel{\raise .4ex\hbox{\rlap{$>$}\lower 1.2ex\hbox{$\sim$}}}}

\def\simprop{\mathrel{\raise .4ex\hbox{\rlap{$\propto$}\lower 1.2ex\hbox{$\sim$}}}}
\def\deg{\ifmmode^\circ\else$^\circ$\fi}
\def\pdeg{\ifmmode $\setbox0=\hbox{$^{\circ}$}\rlap{\hskip.11\wd0 .}$^{\circ}
          \else \setbox0=\hbox{$^{\circ}$}\rlap{\hskip.11\wd0 .}$^{\circ}$\fi}
\def\arcs{\ifmmode {^{\scriptstyle\prime\prime}}
          \else $^{\scriptstyle\prime\prime}$\fi}
\def\arcm{\ifmmode {^{\scriptstyle\prime}}
          \else $^{\scriptstyle\prime}$\fi}
\newdimen\sa  \newdimen\sb
\def\parcs{\sa=.07em \sb=.03em
     \ifmmode \hbox{\rlap{.}}^{\scriptstyle\prime\kern -\sb\prime}\hbox{\kern -\sa}
     \else \rlap{.}$^{\scriptstyle\prime\kern -\sb\prime}$\kern -\sa\fi}
\def\parcm{\sa=.08em \sb=.03em
     \ifmmode \hbox{\rlap{.}\kern\sa}^{\scriptstyle\prime}\hbox{\kern-\sb}
     \else \rlap{.}\kern\sa$^{\scriptstyle\prime}$\kern-\sb\fi}
\def\ra[#1 #2 #3.#4]{#1\sup{h}#2\sup{m}#3\sup{s}\llap.#4}
\def\dec[#1 #2 #3.#4]{#1\deg#2\arcm#3\arcs\llap.#4}
\def\deco[#1 #2 #3]{#1\deg#2\arcm#3\arcs}
\def\rra[#1 #2]{#1\sup{h}#2\sup{m}}

\def\dots{\relax\ifmmode \ldots\else $\ldots$\fi}
\def\WHzsr{\ifmmode $W\,Hz\mo\,sr\mo$\else W\,Hz\mo\,sr\mo\fi}
\def\mHz{\ifmmode $\,mHz$\else \,mHz\fi}
\def\GHz{\ifmmode $\,GHz$\else \,GHz\fi}
\def\mKs{\ifmmode $\,mK\,s$^{1/2}\else \,mK\,s$^{1/2}$\fi}
\def\muKs{\ifmmode \,\mu$K\,s$^{1/2}\else \,$\mu$K\,s$^{1/2}$\fi}
\def\muKRJs{\ifmmode \,\mu$K$_{\rm RJ}$\,s$^{1/2}\else \,$\mu$K$_{\rm RJ}$\,s$^{1/2}$\fi}
\def\muKHz{\ifmmode \,\mu$K\,Hz$^{-1/2}\else \,$\mu$K\,Hz$^{-1/2}$\fi}
\def\MJysr{\ifmmode \,$MJy\,sr\mo$\else \,MJy\,sr\mo\fi}
\def\MJysrmK{\ifmmode \,$MJy\,sr\mo$\,mK$_{\rm CMB}\mo\else \,MJy\,sr\mo\,mK$_{\rm CMB}\mo$\fi}
\def\microns{\ifmmode \,\mu$m$\else \,$\mu$m\fi}

\def\muK{\ifmmode \,\mu$K$\else \,$\mu$\hbox{K}\fi}
\def\microK{\ifmmode \,\mu$K$\else \,$\mu$\hbox{K}\fi}
\def\muW{\ifmmode \,\mu$W$\else \,$\mu$\hbox{W}\fi}
\def\kms{\ifmmode $\,km\,s$^{-1}\else \,km\,s$^{-1}$\fi}
\def\kmsMpc{\ifmmode $\,\kms\,Mpc\mo$\else \,\kms\,Mpc\mo\fi}
\setsymbol{LFI:center:frequency:70GHz:units}{70.3\,GHz}
\setsymbol{LFI:center:frequency:44GHz:units}{44.1\,GHz}
\setsymbol{LFI:center:frequency:30GHz:units}{28.5\,GHz}

\setsymbol{LFI:center:frequency:70GHz}{70.3}
\setsymbol{LFI:center:frequency:44GHz}{44.1}
\setsymbol{LFI:center:frequency:30GHz}{28.5}

\setsymbol{LFI:center:frequency:LFI18:Rad:M:units}{71.7\GHz}
\setsymbol{LFI:center:frequency:LFI19:Rad:M:units}{67.5\GHz}
\setsymbol{LFI:center:frequency:LFI20:Rad:M:units}{69.2\GHz}
\setsymbol{LFI:center:frequency:LFI21:Rad:M:units}{70.4\GHz}
\setsymbol{LFI:center:frequency:LFI22:Rad:M:units}{71.5\GHz}
\setsymbol{LFI:center:frequency:LFI23:Rad:M:units}{70.8\GHz}
\setsymbol{LFI:center:frequency:LFI24:Rad:M:units}{44.4\GHz}
\setsymbol{LFI:center:frequency:LFI25:Rad:M:units}{44.0\GHz}
\setsymbol{LFI:center:frequency:LFI26:Rad:M:units}{43.9\GHz}
\setsymbol{LFI:center:frequency:LFI27:Rad:M:units}{28.3\GHz}
\setsymbol{LFI:center:frequency:LFI28:Rad:M:units}{28.8\GHz}
\setsymbol{LFI:center:frequency:LFI18:Rad:S:units}{70.1\GHz}
\setsymbol{LFI:center:frequency:LFI19:Rad:S:units}{69.6\GHz}
\setsymbol{LFI:center:frequency:LFI20:Rad:S:units}{69.5\GHz}
\setsymbol{LFI:center:frequency:LFI21:Rad:S:units}{69.5\GHz}
\setsymbol{LFI:center:frequency:LFI22:Rad:S:units}{72.8\GHz}
\setsymbol{LFI:center:frequency:LFI23:Rad:S:units}{71.3\GHz}
\setsymbol{LFI:center:frequency:LFI24:Rad:S:units}{44.1\GHz}
\setsymbol{LFI:center:frequency:LFI25:Rad:S:units}{44.1\GHz}
\setsymbol{LFI:center:frequency:LFI26:Rad:S:units}{44.1\GHz}
\setsymbol{LFI:center:frequency:LFI27:Rad:S:units}{28.5\GHz}
\setsymbol{LFI:center:frequency:LFI28:Rad:S:units}{28.2\GHz}

\setsymbol{LFI:center:frequency:LFI18:Rad:M}{71.7}
\setsymbol{LFI:center:frequency:LFI19:Rad:M}{67.5}
\setsymbol{LFI:center:frequency:LFI20:Rad:M}{69.2}
\setsymbol{LFI:center:frequency:LFI21:Rad:M}{70.4}
\setsymbol{LFI:center:frequency:LFI22:Rad:M}{71.5}
\setsymbol{LFI:center:frequency:LFI23:Rad:M}{70.8}
\setsymbol{LFI:center:frequency:LFI24:Rad:M}{44.4}
\setsymbol{LFI:center:frequency:LFI25:Rad:M}{44.0}
\setsymbol{LFI:center:frequency:LFI26:Rad:M}{43.9}
\setsymbol{LFI:center:frequency:LFI27:Rad:M}{28.3}
\setsymbol{LFI:center:frequency:LFI28:Rad:M}{28.8}
\setsymbol{LFI:center:frequency:LFI18:Rad:S}{70.1}
\setsymbol{LFI:center:frequency:LFI19:Rad:S}{69.6}
\setsymbol{LFI:center:frequency:LFI20:Rad:S}{69.5}
\setsymbol{LFI:center:frequency:LFI21:Rad:S}{69.5}
\setsymbol{LFI:center:frequency:LFI22:Rad:S}{72.8}
\setsymbol{LFI:center:frequency:LFI23:Rad:S}{71.3}
\setsymbol{LFI:center:frequency:LFI24:Rad:S}{44.1}
\setsymbol{LFI:center:frequency:LFI25:Rad:S}{44.1}
\setsymbol{LFI:center:frequency:LFI26:Rad:S}{44.1}
\setsymbol{LFI:center:frequency:LFI27:Rad:S}{28.5}
\setsymbol{LFI:center:frequency:LFI28:Rad:S}{28.2}

\setsymbol{LFI:white:noise:sensitivity:70GHz:units}{134.7\muKs}
\setsymbol{LFI:white:noise:sensitivity:44GHz:units}{164.7\muKs}
\setsymbol{LFI:white:noise:sensitivity:30GHz:units}{143.4\muKs}

\setsymbol{LFI:white:noise:sensitivity:70GHz}{134.7}
\setsymbol{LFI:white:noise:sensitivity:44GHz}{164.7}
\setsymbol{LFI:white:noise:sensitivity:30GHz}{143.4}

\setsymbol{LFI:white:noise:sensitivity:LFI18:Rad:M:units}{512.0\muKs}
\setsymbol{LFI:white:noise:sensitivity:LFI19:Rad:M:units}{581.4\muKs}
\setsymbol{LFI:white:noise:sensitivity:LFI20:Rad:M:units}{590.8\muKs}
\setsymbol{LFI:white:noise:sensitivity:LFI21:Rad:M:units}{455.2\muKs}
\setsymbol{LFI:white:noise:sensitivity:LFI22:Rad:M:units}{492.0\muKs}
\setsymbol{LFI:white:noise:sensitivity:LFI23:Rad:M:units}{507.7\muKs}
\setsymbol{LFI:white:noise:sensitivity:LFI24:Rad:M:units}{462.2\muKs}
\setsymbol{LFI:white:noise:sensitivity:LFI25:Rad:M:units}{413.6\muKs}
\setsymbol{LFI:white:noise:sensitivity:LFI26:Rad:M:units}{478.6\muKs}
\setsymbol{LFI:white:noise:sensitivity:LFI27:Rad:M:units}{277.7\muKs}
\setsymbol{LFI:white:noise:sensitivity:LFI28:Rad:M:units}{312.3\muKs}
\setsymbol{LFI:white:noise:sensitivity:LFI18:Rad:S:units}{465.7\muKs}
\setsymbol{LFI:white:noise:sensitivity:LFI19:Rad:S:units}{555.6\muKs}
\setsymbol{LFI:white:noise:sensitivity:LFI20:Rad:S:units}{623.2\muKs}
\setsymbol{LFI:white:noise:sensitivity:LFI21:Rad:S:units}{564.1\muKs}
\setsymbol{LFI:white:noise:sensitivity:LFI22:Rad:S:units}{534.4\muKs}
\setsymbol{LFI:white:noise:sensitivity:LFI23:Rad:S:units}{542.4\muKs}
\setsymbol{LFI:white:noise:sensitivity:LFI24:Rad:S:units}{399.2\muKs}
\setsymbol{LFI:white:noise:sensitivity:LFI25:Rad:S:units}{392.6\muKs}
\setsymbol{LFI:white:noise:sensitivity:LFI26:Rad:S:units}{418.6\muKs}
\setsymbol{LFI:white:noise:sensitivity:LFI27:Rad:S:units}{302.9\muKs}
\setsymbol{LFI:white:noise:sensitivity:LFI28:Rad:S:units}{285.3\muKs}

\setsymbol{LFI:white:noise:sensitivity:LFI18:Rad:M}{512.0}
\setsymbol{LFI:white:noise:sensitivity:LFI19:Rad:M}{581.4}
\setsymbol{LFI:white:noise:sensitivity:LFI20:Rad:M}{590.8}
\setsymbol{LFI:white:noise:sensitivity:LFI21:Rad:M}{455.2}
\setsymbol{LFI:white:noise:sensitivity:LFI22:Rad:M}{492.0}
\setsymbol{LFI:white:noise:sensitivity:LFI23:Rad:M}{507.7}
\setsymbol{LFI:white:noise:sensitivity:LFI24:Rad:M}{462.2}
\setsymbol{LFI:white:noise:sensitivity:LFI25:Rad:M}{413.6}
\setsymbol{LFI:white:noise:sensitivity:LFI26:Rad:M}{478.6}
\setsymbol{LFI:white:noise:sensitivity:LFI27:Rad:M}{277.7}
\setsymbol{LFI:white:noise:sensitivity:LFI28:Rad:M}{312.3}
\setsymbol{LFI:white:noise:sensitivity:LFI18:Rad:S}{465.7}
\setsymbol{LFI:white:noise:sensitivity:LFI19:Rad:S}{555.6}
\setsymbol{LFI:white:noise:sensitivity:LFI20:Rad:S}{623.2}
\setsymbol{LFI:white:noise:sensitivity:LFI21:Rad:S}{564.1}
\setsymbol{LFI:white:noise:sensitivity:LFI22:Rad:S}{534.4}
\setsymbol{LFI:white:noise:sensitivity:LFI23:Rad:S}{542.4}
\setsymbol{LFI:white:noise:sensitivity:LFI24:Rad:S}{399.2}
\setsymbol{LFI:white:noise:sensitivity:LFI25:Rad:S}{392.6}
\setsymbol{LFI:white:noise:sensitivity:LFI26:Rad:S}{418.6}
\setsymbol{LFI:white:noise:sensitivity:LFI27:Rad:S}{302.9}
\setsymbol{LFI:white:noise:sensitivity:LFI28:Rad:S}{285.3}

\setsymbol{LFI:knee:frequency:70GHz:units}{29.5\mHz}
\setsymbol{LFI:knee:frequency:44GHz:units}{56.2\mHz}
\setsymbol{LFI:knee:frequency:30GHz:units}{113.7\mHz}

\setsymbol{LFI:knee:frequency:70GHz}{29.5}
\setsymbol{LFI:knee:frequency:44GHz}{56.2}
\setsymbol{LFI:knee:frequency:30GHz}{113.7}

\setsymbol{LFI:knee:frequency:LFI18:Rad:M:units}{16.3\mHz}
\setsymbol{LFI:knee:frequency:LFI19:Rad:M:units}{15.1\mHz}
\setsymbol{LFI:knee:frequency:LFI20:Rad:M:units}{18.7\mHz}
\setsymbol{LFI:knee:frequency:LFI21:Rad:M:units}{37.2\mHz}
\setsymbol{LFI:knee:frequency:LFI22:Rad:M:units}{12.7\mHz}
\setsymbol{LFI:knee:frequency:LFI23:Rad:M:units}{34.6\mHz}
\setsymbol{LFI:knee:frequency:LFI24:Rad:M:units}{46.2\mHz}
\setsymbol{LFI:knee:frequency:LFI25:Rad:M:units}{24.9\mHz}
\setsymbol{LFI:knee:frequency:LFI26:Rad:M:units}{67.6\mHz}
\setsymbol{LFI:knee:frequency:LFI27:Rad:M:units}{187.4\mHz}
\setsymbol{LFI:knee:frequency:LFI28:Rad:M:units}{122.2\mHz}
\setsymbol{LFI:knee:frequency:LFI18:Rad:S:units}{17.7\mHz}
\setsymbol{LFI:knee:frequency:LFI19:Rad:S:units}{22.0\mHz}
\setsymbol{LFI:knee:frequency:LFI20:Rad:S:units}{8.7\mHz}
\setsymbol{LFI:knee:frequency:LFI21:Rad:S:units}{25.9\mHz}
\setsymbol{LFI:knee:frequency:LFI22:Rad:S:units}{15.8\mHz}
\setsymbol{LFI:knee:frequency:LFI23:Rad:S:units}{129.8\mHz}
\setsymbol{LFI:knee:frequency:LFI24:Rad:S:units}{100.9\mHz}
\setsymbol{LFI:knee:frequency:LFI25:Rad:S:units}{38.9\mHz}
\setsymbol{LFI:knee:frequency:LFI26:Rad:S:units}{58.9\mHz}
\setsymbol{LFI:knee:frequency:LFI27:Rad:S:units}{104.4\mHz}
\setsymbol{LFI:knee:frequency:LFI28:Rad:S:units}{40.7\mHz}

\setsymbol{LFI:knee:frequency:LFI18:Rad:M}{16.3}
\setsymbol{LFI:knee:frequency:LFI19:Rad:M}{15.1}
\setsymbol{LFI:knee:frequency:LFI20:Rad:M}{18.7}
\setsymbol{LFI:knee:frequency:LFI21:Rad:M}{37.2}
\setsymbol{LFI:knee:frequency:LFI22:Rad:M}{12.7}
\setsymbol{LFI:knee:frequency:LFI23:Rad:M}{34.6}
\setsymbol{LFI:knee:frequency:LFI24:Rad:M}{46.2}
\setsymbol{LFI:knee:frequency:LFI25:Rad:M}{24.9}
\setsymbol{LFI:knee:frequency:LFI26:Rad:M}{67.6}
\setsymbol{LFI:knee:frequency:LFI27:Rad:M}{187.4}
\setsymbol{LFI:knee:frequency:LFI28:Rad:M}{122.2}
\setsymbol{LFI:knee:frequency:LFI18:Rad:S}{17.7}
\setsymbol{LFI:knee:frequency:LFI19:Rad:S}{22.0}
\setsymbol{LFI:knee:frequency:LFI20:Rad:S}{8.7}
\setsymbol{LFI:knee:frequency:LFI21:Rad:S}{25.9}
\setsymbol{LFI:knee:frequency:LFI22:Rad:S}{15.8}
\setsymbol{LFI:knee:frequency:LFI23:Rad:S}{129.8}
\setsymbol{LFI:knee:frequency:LFI24:Rad:S}{100.9}
\setsymbol{LFI:knee:frequency:LFI25:Rad:S}{38.9}
\setsymbol{LFI:knee:frequency:LFI26:Rad:S}{58.9}
\setsymbol{LFI:knee:frequency:LFI27:Rad:S}{104.4}
\setsymbol{LFI:knee:frequency:LFI28:Rad:S}{40.7}

\setsymbol{LFI:slope:70GHz:units}{$-1.03$\mHz}
\setsymbol{LFI:slope:44GHz:units}{$-0.89$\mHz}
\setsymbol{LFI:slope:30GHz:units}{$-0.87$\mHz}

\setsymbol{LFI:slope:70GHz}{$-1.03$}
\setsymbol{LFI:slope:44GHz}{$-0.89$}
\setsymbol{LFI:slope:30GHz}{$-0.87$}

\setsymbol{LFI:slope:LFI18:Rad:M:units}{$-1.04$\mHz}
\setsymbol{LFI:slope:LFI19:Rad:M:units}{$-1.09$\mHz}
\setsymbol{LFI:slope:LFI20:Rad:M:units}{$-0.69$\mHz}
\setsymbol{LFI:slope:LFI21:Rad:M:units}{$-1.56$\mHz}
\setsymbol{LFI:slope:LFI22:Rad:M:units}{$-1.01$\mHz}
\setsymbol{LFI:slope:LFI23:Rad:M:units}{$-0.96$\mHz}
\setsymbol{LFI:slope:LFI24:Rad:M:units}{$-0.83$\mHz}
\setsymbol{LFI:slope:LFI25:Rad:M:units}{$-0.91$\mHz}
\setsymbol{LFI:slope:LFI26:Rad:M:units}{$-0.95$\mHz}
\setsymbol{LFI:slope:LFI27:Rad:M:units}{$-0.87$\mHz}
\setsymbol{LFI:slope:LFI28:Rad:M:units}{$-0.88$\mHz}
\setsymbol{LFI:slope:LFI18:Rad:S:units}{$-1.15$\mHz}
\setsymbol{LFI:slope:LFI19:Rad:S:units}{$-1.00$\mHz}
\setsymbol{LFI:slope:LFI20:Rad:S:units}{$-0.95$\mHz}
\setsymbol{LFI:slope:LFI21:Rad:S:units}{$-0.92$\mHz}
\setsymbol{LFI:slope:LFI22:Rad:S:units}{$-1.01$\mHz}
\setsymbol{LFI:slope:LFI23:Rad:S:units}{$-0.95$\mHz}
\setsymbol{LFI:slope:LFI24:Rad:S:units}{$-0.73$\mHz}
\setsymbol{LFI:slope:LFI25:Rad:S:units}{$-1.16$\mHz}
\setsymbol{LFI:slope:LFI26:Rad:S:units}{$-0.79$\mHz}
\setsymbol{LFI:slope:LFI27:Rad:S:units}{$-0.82$\mHz}
\setsymbol{LFI:slope:LFI28:Rad:S:units}{$-0.91$\mHz}

\setsymbol{LFI:slope:LFI18:Rad:M}{$-1.04$}
\setsymbol{LFI:slope:LFI19:Rad:M}{$-1.09$}
\setsymbol{LFI:slope:LFI20:Rad:M}{$-0.69$}
\setsymbol{LFI:slope:LFI21:Rad:M}{$-1.56$}
\setsymbol{LFI:slope:LFI22:Rad:M}{$-1.01$}
\setsymbol{LFI:slope:LFI23:Rad:M}{$-0.96$}
\setsymbol{LFI:slope:LFI24:Rad:M}{$-0.83$}
\setsymbol{LFI:slope:LFI25:Rad:M}{$-0.91$}
\setsymbol{LFI:slope:LFI26:Rad:M}{$-0.95$}
\setsymbol{LFI:slope:LFI27:Rad:M}{$-0.87$}
\setsymbol{LFI:slope:LFI28:Rad:M}{$-0.88$}
\setsymbol{LFI:slope:LFI18:Rad:S}{$-1.15$}
\setsymbol{LFI:slope:LFI19:Rad:S}{$-1.00$}
\setsymbol{LFI:slope:LFI20:Rad:S}{$-0.95$}
\setsymbol{LFI:slope:LFI21:Rad:S}{$-0.92$}
\setsymbol{LFI:slope:LFI22:Rad:S}{$-1.01$}
\setsymbol{LFI:slope:LFI23:Rad:S}{$-0.95$}
\setsymbol{LFI:slope:LFI24:Rad:S}{$-0.73$}
\setsymbol{LFI:slope:LFI25:Rad:S}{$-1.16$}
\setsymbol{LFI:slope:LFI26:Rad:S}{$-0.79$}
\setsymbol{LFI:slope:LFI27:Rad:S}{$-0.82$}
\setsymbol{LFI:slope:LFI28:Rad:S}{$-0.91$}

\setsymbol{LFI:FWHM:70GHz:units}{13\parcm01}
\setsymbol{LFI:FWHM:44GHz:units}{27\parcm92}
\setsymbol{LFI:FWHM:30GHz:units}{32\parcm65}

\setsymbol{LFI:FWHM:70GHz}{13.01}
\setsymbol{LFI:FWHM:44GHz}{27.92}
\setsymbol{LFI:FWHM:30GHz}{32.65}

\setsymbol{LFI:FWHM:LFI18:units}{13\parcm39}
\setsymbol{LFI:FWHM:LFI19:units}{13\parcm01}
\setsymbol{LFI:FWHM:LFI20:units}{12\parcm75}
\setsymbol{LFI:FWHM:LFI21:units}{12\parcm74}
\setsymbol{LFI:FWHM:LFI22:units}{12\parcm87}
\setsymbol{LFI:FWHM:LFI23:units}{13\parcm27}
\setsymbol{LFI:FWHM:LFI24:units}{22\parcm98}
\setsymbol{LFI:FWHM:LFI25:units}{30\parcm46}
\setsymbol{LFI:FWHM:LFI26:units}{30\parcm31}
\setsymbol{LFI:FWHM:LFI27:units}{32\parcm65}
\setsymbol{LFI:FWHM:LFI28:units}{32\parcm66}

\setsymbol{LFI:FWHM:LFI18}{13.39}
\setsymbol{LFI:FWHM:LFI19}{13.01}
\setsymbol{LFI:FWHM:LFI20}{12.75}
\setsymbol{LFI:FWHM:LFI21}{12.74}
\setsymbol{LFI:FWHM:LFI22}{12.87}
\setsymbol{LFI:FWHM:LFI23}{13.27}
\setsymbol{LFI:FWHM:LFI24}{22.98}
\setsymbol{LFI:FWHM:LFI25}{30.46}
\setsymbol{LFI:FWHM:LFI26}{30.31}
\setsymbol{LFI:FWHM:LFI27}{32.65}
\setsymbol{LFI:FWHM:LFI28}{32.66}

\setsymbol{LFI:FWHM:uncertainty:LFI18:units}{0.170\arcm}
\setsymbol{LFI:FWHM:uncertainty:LFI19:units}{0.174\arcm}
\setsymbol{LFI:FWHM:uncertainty:LFI20:units}{0.170\arcm}
\setsymbol{LFI:FWHM:uncertainty:LFI21:units}{0.156\arcm}
\setsymbol{LFI:FWHM:uncertainty:LFI22:units}{0.164\arcm}
\setsymbol{LFI:FWHM:uncertainty:LFI23:units}{0.171\arcm}
\setsymbol{LFI:FWHM:uncertainty:LFI24:units}{0.652\arcm}
\setsymbol{LFI:FWHM:uncertainty:LFI25:units}{1.075\arcm}
\setsymbol{LFI:FWHM:uncertainty:LFI26:units}{1.131\arcm}
\setsymbol{LFI:FWHM:uncertainty:LFI27:units}{1.266\arcm}
\setsymbol{LFI:FWHM:uncertainty:LFI28:units}{1.287\arcm}

\setsymbol{LFI:FWHM:uncertainty:LFI18}{0.170}
\setsymbol{LFI:FWHM:uncertainty:LFI19}{0.174}
\setsymbol{LFI:FWHM:uncertainty:LFI20}{0.170}
\setsymbol{LFI:FWHM:uncertainty:LFI21}{0.156}
\setsymbol{LFI:FWHM:uncertainty:LFI22}{0.164}
\setsymbol{LFI:FWHM:uncertainty:LFI23}{0.171}
\setsymbol{LFI:FWHM:uncertainty:LFI24}{0.652}
\setsymbol{LFI:FWHM:uncertainty:LFI25}{1.075}
\setsymbol{LFI:FWHM:uncertainty:LFI26}{1.131}
\setsymbol{LFI:FWHM:uncertainty:LFI27}{1.266}
\setsymbol{LFI:FWHM:uncertainty:LFI28}{1.287}

\setsymbol{HFI:center:frequency:100GHz:units}{100\,GHz}
\setsymbol{HFI:center:frequency:143GHz:units}{143\,GHz}
\setsymbol{HFI:center:frequency:217GHz:units}{217\,GHz}
\setsymbol{HFI:center:frequency:353GHz:units}{353\,GHz}
\setsymbol{HFI:center:frequency:545GHz:units}{545\,GHz}
\setsymbol{HFI:center:frequency:857GHz:units}{857\,GHz}

\setsymbol{HFI:center:frequency:100GHz}{100}
\setsymbol{HFI:center:frequency:143GHz}{143}
\setsymbol{HFI:center:frequency:217GHz}{217}
\setsymbol{HFI:center:frequency:353GHz}{353}
\setsymbol{HFI:center:frequency:545GHz}{545}
\setsymbol{HFI:center:frequency:857GHz}{857}

\setsymbol{HFI:Ndetectors:100GHz}{8}
\setsymbol{HFI:Ndetectors:143GHz}{11}
\setsymbol{HFI:Ndetectors:217GHz}{12}
\setsymbol{HFI:Ndetectors:353GHz}{12}
\setsymbol{HFI:Ndetectors:545GHz}{3}
\setsymbol{HFI:Ndetectors:857GHz}{4}

\setsymbol{HFI:FWHM:Maps:100GHz:units}{9\parcm88}
\setsymbol{HFI:FWHM:Maps:143GHz:units}{7\parcm18}
\setsymbol{HFI:FWHM:Maps:217GHz:units}{4\parcm87}
\setsymbol{HFI:FWHM:Maps:353GHz:units}{4\parcm65}
\setsymbol{HFI:FWHM:Maps:545GHz:units}{4\parcm72}
\setsymbol{HFI:FWHM:Maps:857GHz:units}{4\parcm39}
\setsymbol{HFI:FWHM:Maps:100GHz}{9.88}
\setsymbol{HFI:FWHM:Maps:143GHz}{7.18}
\setsymbol{HFI:FWHM:Maps:217GHz}{4.87}
\setsymbol{HFI:FWHM:Maps:353GHz}{4.65}
\setsymbol{HFI:FWHM:Maps:545GHz}{4.72}
\setsymbol{HFI:FWHM:Maps:857GHz}{4.39}

\setsymbol{HFI:beam:ellipticity:Maps:100GHz}{1.15}
\setsymbol{HFI:beam:ellipticity:Maps:143GHz}{1.01}
\setsymbol{HFI:beam:ellipticity:Maps:217GHz}{1.06}
\setsymbol{HFI:beam:ellipticity:Maps:353GHz}{1.05}
\setsymbol{HFI:beam:ellipticity:Maps:545GHz}{1.14}
\setsymbol{HFI:beam:ellipticity:Maps:857GHz}{1.19}

\setsymbol{HFI:FWHM:Mars:100GHz:units}{9\parcm37}
\setsymbol{HFI:FWHM:Mars:143GHz:units}{7\parcm04}
\setsymbol{HFI:FWHM:Mars:217GHz:units}{4\parcm68}
\setsymbol{HFI:FWHM:Mars:353GHz:units}{4\parcm43}
\setsymbol{HFI:FWHM:Mars:545GHz:units}{3\parcm80}
\setsymbol{HFI:FWHM:Mars:857GHz:units}{3\parcm67}

\setsymbol{HFI:FWHM:Mars:100GHz}{9.37}
\setsymbol{HFI:FWHM:Mars:143GHz}{7.04}
\setsymbol{HFI:FWHM:Mars:217GHz}{4.68}
\setsymbol{HFI:FWHM:Mars:353GHz}{4.43}
\setsymbol{HFI:FWHM:Mars:545GHz}{3.80}
\setsymbol{HFI:FWHM:Mars:857GHz}{3.67}

\setsymbol{HFI:beam:ellipticity:Mars:100GHz}{1.18}
\setsymbol{HFI:beam:ellipticity:Mars:143GHz}{1.03}
\setsymbol{HFI:beam:ellipticity:Mars:217GHz}{1.14}
\setsymbol{HFI:beam:ellipticity:Mars:353GHz}{1.09}
\setsymbol{HFI:beam:ellipticity:Mars:545GHz}{1.25}
\setsymbol{HFI:beam:ellipticity:Mars:857GHz}{1.03}

\setsymbol{HFI:CMB:relative:calibration:100GHz}{$\lsim 1\%$}
\setsymbol{HFI:CMB:relative:calibration:143GHz}{$\lsim 1\%$}
\setsymbol{HFI:CMB:relative:calibration:217GHz}{$\lsim 1\%$}
\setsymbol{HFI:CMB:relative:calibration:353GHz}{$\lsim 1\%$}
\setsymbol{HFI:CMB:relative:calibration:545GHz}{}
\setsymbol{HFI:CMB:relative:calibration:857GHz}{}

\setsymbol{HFI:CMB:absolute:calibration:100GHz}{$\lsim 2\%$}
\setsymbol{HFI:CMB:absolute:calibration:143GHz}{$\lsim 2\%$}
\setsymbol{HFI:CMB:absolute:calibration:217GHz}{$\lsim 2\%$}
\setsymbol{HFI:CMB:absolute:calibration:353GHz}{$\lsim 2\%$}
\setsymbol{HFI:CMB:absolute:calibration:545GHz}{}
\setsymbol{HFI:CMB:absolute:calibration:857GHz}{}

\setsymbol{HFI:FIRAS:gain:calibration:accuracy:statistical:100GHz}{}
\setsymbol{HFI:FIRAS:gain:calibration:accuracy:statistical:143GHz}{}
\setsymbol{HFI:FIRAS:gain:calibration:accuracy:statistical:217GHz}{}
\setsymbol{HFI:FIRAS:gain:calibration:accuracy:statistical:353GHz}{2.5\%}
\setsymbol{HFI:FIRAS:gain:calibration:accuracy:statistical:545GHz}{1\%}
\setsymbol{HFI:FIRAS:gain:calibration:accuracy:statistical:857GHz}{0.5\%}

\setsymbol{HFI:FIRAS:gain:calibration:accuracy:systematic:100GHz}{}
\setsymbol{HFI:FIRAS:gain:calibration:accuracy:systematic:143GHz}{}
\setsymbol{HFI:FIRAS:gain:calibration:accuracy:systematic:217GHz}{}
\setsymbol{HFI:FIRAS:gain:calibration:accuracy:systematic:353GHz}{}
\setsymbol{HFI:FIRAS:gain:calibration:accuracy:systematic:545GHz}{7\%}
\setsymbol{HFI:FIRAS:gain:calibration:accuracy:systematic:857GHz}{7\%}

\setsymbol{HFI:FIRAS:zero:point:accuracy:100GHz:units}{0.8\MJysr}
\setsymbol{HFI:FIRAS:zero:point:accuracy:143GHz:units}{}
\setsymbol{HFI:FIRAS:zero:point:accuracy:217GHz:units}{}
\setsymbol{HFI:FIRAS:zero:point:accuracy:353GHz:units}{1.4\MJysr}
\setsymbol{HFI:FIRAS:zero:point:accuracy:545GHz:units}{2.2\MJysr}
\setsymbol{HFI:FIRAS:zero:point:accuracy:857GHz:units}{1.7\MJysr}

\setsymbol{HFI:FIRAS:zero:point:accuracy:100GHz}{0.8}
\setsymbol{HFI:FIRAS:zero:point:accuracy:143GHz}{}
\setsymbol{HFI:FIRAS:zero:point:accuracy:217GHz}{}
\setsymbol{HFI:FIRAS:zero:point:accuracy:353GHz}{1.4}
\setsymbol{HFI:FIRAS:zero:point:accuracy:545GHz}{2.2}
\setsymbol{HFI:FIRAS:zero:point:accuracy:857GHz}{1.7}

\setsymbol{HFI:unit:conversion:100GHz:units}{0.2415\MJysrmK}
\setsymbol{HFI:unit:conversion:143GHz:units}{0.3694\MJysrmK}
\setsymbol{HFI:unit:conversion:217GHz:units}{0.4811\MJysrmK}
\setsymbol{HFI:unit:conversion:353GHz:units}{0.2883\MJysrmK}
\setsymbol{HFI:unit:conversion:545GHz:units}{0.05826\MJysrmK}
\setsymbol{HFI:unit:conversion:857GHz:units}{0.002238\MJysrmK}

\setsymbol{HFI:unit:conversion:100GHz}{0.2415}
\setsymbol{HFI:unit:conversion:143GHz}{0.3694}
\setsymbol{HFI:unit:conversion:217GHz}{0.4811}
\setsymbol{HFI:unit:conversion:353GHz}{0.2883}
\setsymbol{HFI:unit:conversion:545GHz}{0.05826}
\setsymbol{HFI:unit:conversion:857GHz}{0.002238}

\setsymbol{HFI:colour:correction:alpha=-2:V1.01:100GHz}{0.9893}
\setsymbol{HFI:colour:correction:alpha=-2:V1.01:143GHz}{0.9759}
\setsymbol{HFI:colour:correction:alpha=-2:V1.01:217GHz}{1.0007}
\setsymbol{HFI:colour:correction:alpha=-2:V1.01:353GHz}{1.0028}
\setsymbol{HFI:colour:correction:alpha=-2:V1.01:545GHz}{1.0019}
\setsymbol{HFI:colour:correction:alpha=-2:V1.01:857GHz}{0.9889}

\setsymbol{HFI:colour:correction:alpha=0:V1.01:100GHz}{1.0008}
\setsymbol{HFI:colour:correction:alpha=0:V1.01:143GHz}{1.0148}
\setsymbol{HFI:colour:correction:alpha=0:V1.01:217GHz}{0.9909}
\setsymbol{HFI:colour:correction:alpha=0:V1.01:353GHz}{0.9888}
\setsymbol{HFI:colour:correction:alpha=0:V1.01:545GHz}{0.9878}
\setsymbol{HFI:colour:correction:alpha=0:V1.01:857GHz}{1.0014}

\begin{document}

\title{\textit{Planck} early results. III. First assessment of the Low Frequency Instrument in-flight performance}

%
\author{\small
A. Mennella\inst{22, 37}
\and
M. Bersanelli\inst{22, 39}
\and
R. C. Butler\inst{38}
\and
A. Curto\inst{47}
\and
F. Cuttaia\inst{38}
\and
R. J. Davis\inst{50}
\and
J. Dick\inst{57}
\and
M. Frailis\inst{37}
\and
S. Galeotta\inst{37}
\and
A. Gregorio\inst{23}
% ...
}
\institute{\small
Aalto University Mets\"{a}hovi Radio Observatory, Mets\"{a}hovintie 114, FIN-02540 Kylm\"{a}l\"{a}, Finland\\
\and
Agenzia Spaziale Italiana Science Data Center, c/o ESRIN, via Galileo Galilei, Frascati, Italy\\
\and
Agenzia Spaziale Italiana, Viale Liegi 26, Roma, Italy\\
\and
Astroparticule et Cosmologie, CNRS (UMR7164), Universit\'{e} Denis Diderot Paris 7, B\^{a}timent Condorcet, 10 rue A. Domon et L\'{e}onie Duquet, Paris, France\\
\and
Australia Telescope National Facility, CSIRO, P.O. Box 76, Epping, NSW 1710, Australia\\
\and
CNRS, IRAP, 9 Av. colonel Roche, BP 44346, F-31028 Toulouse cedex 4, France\\
\and
California Institute of Technology, Pasadena, California, U.S.A.\\
\and
Centre of Mathematics for Applications, University of Oslo, Blindern, Oslo, Norway\\
\and
DA-Design Oy, Keskuskatu 29, Jokioinen, Finland\\
\and
DTU Space, National Space Institute, Juliane Mariesvej 30, Copenhagen, Denmark\\
\and
Departamento de F\'{\i}sica, Universidad de Oviedo, Avda. Calvo Sotelo s/n, Oviedo, Spain\\
\and
Departamento de Ingenier\'{i}a de Comunicaciones, Universidad de Cantabria, Plaza de la Ciencia, 39005 Santander, Spain\\
\and
Departamento de Matem\'{a}ticas, Estad\'{\i}stica y Computaci\'{o}n, Universidad de Cantabria, Avda. de los Castros s/n, Santander, Spain\\
\and
Department of Physics \& Astronomy, University of British Columbia, 6224 Agricultural Road, Vancouver, British Columbia, Canada\\
\and
Department of Physics and Astronomy, University of Southern California, Los Angeles, California, U.S.A.\\
\and
Department of Physics, Gustaf H\"{a}llstr\"{o}min katu 2a, University of Helsinki, Helsinki, Finland\\
\and
Department of Physics, University of California, Berkeley, California, U.S.A.\\
\and
Department of Physics, University of California, One Shields Avenue, Davis, California, U.S.A.\\
\and
Department of Physics, University of California, Santa Barbara, California, U.S.A.\\
\and
Dipartimento di Fisica G. Galilei, Universit\`{a} degli Studi di Padova, via Marzolo 8, 35131 Padova, Italy\\
\and
Dipartimento di Fisica, Universit\`{a} La Sapienza, P. le A. Moro 2, Roma, Italy\\
\and
Dipartimento di Fisica, Universit\`{a} degli Studi di Milano, Via Celoria, 16, Milano, Italy\\
\and
Dipartimento di Fisica, Universit\`{a} degli Studi di Trieste, via A. Valerio 2, Trieste, Italy\\
\and
Dipartimento di Fisica, Universit\`{a} di Ferrara, Via Saragat 1, 44122 Ferrara, Italy\\
\and
Dipartimento di Fisica, Universit\`{a} di Roma Tor Vergata, Via della Ricerca Scientifica, 1, Roma, Italy\\
\and
Dipartimento di Matematica, Universit\`{a} di Roma Tor Vergata, Via della Ricerca Scientifica, 1, Roma, Italy\\
\and
Dpto. Astrof\'{i}sica, Universidad de La Laguna (ULL), E-38206 La Laguna, Tenerife, Spain\\
\and
European Space Agency, ESAC, Planck Science Office, Camino bajo del Castillo, s/n, Urbanizaci\'{o}n Villafranca del Castillo, Villanueva de la Ca\ {n}ada, Madrid, Spain\\
\and
European Space Agency, ESOC, Robert-Bosch-Str. 5, Darmstadt, Germany\\
\and
European Space Agency, ESTEC, Keplerlaan 1, 2201 AZ Noordwijk, The Netherlands\\
\and
Haverford College Astronomy Department, 370 Lancaster Avenue, Haverford, Pennsylvania, U.S.A.\\
\and
Helsinki Institute of Physics, Gustaf H\"{a}llstr\"{o}min katu 2, University of Helsinki, Helsinki, Finland\\
\and
INAF - Osservatorio Astrofisico di Arcetri, Largo Enrico Fermi 5, Firenze, Italy\\
\and
INAF - Osservatorio Astrofisico di Catania, Via S. Sofia 78, Catania, Italy\\
\and
INAF - Osservatorio Astronomico di Padova, Vicolo dell'Osservatorio 5, Padova, Italy\\
\and
INAF - Osservatorio Astronomico di Roma, via di Frascati 33, Monte Porzio Catone, Italy\\
\and
INAF - Osservatorio Astronomico di Trieste, Via G.B. Tiepolo 11, Trieste, Italy\\
\and
INAF/IASF Bologna, Via Gobetti 101, Bologna, Italy\\
\and
INAF/IASF Milano, Via E. Bassini 15, Milano, Italy\\
\and
ISDC Data Centre for Astrophysics, University of Geneva, ch. d'Ecogia 16, Versoix, Switzerland\\
\and
Infrared Processing and Analysis Center, California Institute of Technology, Pasadena, CA 91125, U.S.A.\\
\and
Institut d'Astrophysique Spatiale, CNRS (UMR8617) Universit\'{e} Paris-Sud 11, B\^{a}timent 121, Orsay, France\\
\and
Institut d'Astrophysique de Paris, CNRS UMR7095, Universit\'{e} Pierre \& Marie Curie, 98 bis boulevard Arago, Paris, France\\
\and
Institute of Astronomy, University of Cambridge, Madingley Road, Cambridge CB3 0HA, U.K.\\
\and
Institute of Theoretical Astrophysics, University of Oslo, Blindern, Oslo, Norway\\
\and
Instituto de Astrof\'{\i}sica de Canarias, C/V\'{\i}a L\'{a}ctea s/n, La Laguna, Tenerife, Spain\\
\and
Instituto de F\'{\i}sica de Cantabria (CSIC-Universidad de Cantabria), Avda. de los Castros s/n, Santander, Spain\\
\and
Istituto di Fisica del Plasma, CNR-ENEA-EURATOM Association, Via R. Cozzi 53, Milano, Italy\\
\and
Jet Propulsion Laboratory, California Institute of Technology, 4800 Oak Grove Drive, Pasadena, California, U.S.A.\\
\and
Jodrell Bank Centre for Astrophysics, Alan Turing Building, School of Physics and Astronomy, The University of Manchester, Oxford Road, Manchester, M13 9PL, U.K.\\
\and
LERMA, CNRS, Observatoire de Paris, 61 Avenue de l'Observatoire, Paris, France\\
\and
Lawrence Berkeley National Laboratory, Berkeley, California, U.S.A.\\
\and
Max-Planck-Institut f\"{u}r Astrophysik, Karl-Schwarzschild-Str. 1, 85741 Garching, Germany\\
\and
MilliLab, VTT Technical Research Centre of Finland, Tietotie 3, Espoo, Finland\\
\and
National Radio Astronomy Observatory, 520 Edgemont Road, Charlottesville VA 22903-2475, U.S.A.\\
\and
Nokia Corporation, It\"{a}merenkatu 11-13, Helsinki, Finland\\
\and
SISSA, Astrophysics Sector, via Bonomea 265, 34136, Trieste, Italy\\
% ...
}

\date{Received 9 January 2011 / Accepted 9 May 2011}

\abstract{The scientific performance of the Planck Low Frequency Instrument (LFI) after one year of in-orbit operation is presented. We describe the main optical parameters and discuss photometric calibration, white noise sensitivity, and noise properties. A preliminary evaluation of the impact of the main systematic effects is presented. For each of the performance parameters, we outline the methods used to obtain them from the flight data and provide a comparison with pre-launch ground assessments, which are essentially confirmed in flight.}

 
\titlerunning{LFI in flight performance}
\authorrunning{}
\maketitle

%

\section{Introduction}
\label{sec_introduction}

    The \Planck\footnote{\Planck\ (http://www.esa.int/\Planck ) is a project of the European Space Agency (ESA) with instruments provided by 
two scientific consortia funded by ESA member states (in particular the lead 
countries France and Italy), with contributions from NASA (USA) and telescope reflectors provided by 
a collaboration between ESA and a scientific consortium led and funded by Denmark.}
mission was designed and developed to produce a full-sky survey of the cosmic microwave 
background (CMB) with unprecedented accuracy, both in temperature and polarisation. 
The need to control and remove astrophysical foregrounds with exquisite precision led to the 
requirement of intensive spectral coverage over a broad range. \Planck\ features nine frequency 
bands, roughly logarithmically spaced, in the range 27-900\,GHz. As an additional product, 
the nine \Planck\ sky maps will provide the community with a variety of new astrophysical 
data in this rich and largely unexplored frequency domain. 

Two complementary cryogenic instruments employing different technologies, the Low Frequency 
Instrument (LFI) and the High frequency Instrument (HFI), share the \Planck\ focal plane 
\citep{bersanelli2010,lamarre2010}.  The LFI is an array of coherent microwave receivers in Ka, Q, and V frequency bands. The instrument is based on state-of-the-art indium phosphide (InP) cryogenic high electron mobility transistor (HEMT) amplifiers, implemented in a differential system using blackbody loads as reference signals. The LFI front-end is cooled to $\sim 20$\,K for optimal sensitivity, while the reference loads, connected with the HFI front-end unit, are cooled to $\sim 4.5$\,K. 

Following the successful launch of the \Planck\ satellite (Kourou, 14 May 2009), the LFI 
instrument has been working flawlessly and has been operated according to the mission plan 
\citep{planck2011-1.1}. Starting on 4 June 2009, the LFI entered the 
calibration and performance verification (CPV) phase, dedicated to instrument testing and 
radiometer tuning, exploiting the different thermal configuration during cool-down \citep{gregorio2011}. 
On 13 August 2009, the LFI instrument, together with HFI and the 
entire satellite system, was set in nominal observing mode and has been continuously surveying the sky since then. 
Details of the Planck mission and scanning strategy are given
in \citet{planck2011-1.1}.

In this paper we provide a description of the in-flight behaviour of \Planck-LFI and an 
evaluation of its performance based on the first year of science operations. As planned since 
the early design phase, the LFI calibration strategy is based on a combination of ground and in-orbit measurements. Where possible, key instrument parameters have been measured both in pre-flight and in-orbit tests, and thus can be compared for self-consistency.  Ground calibration results have been
obtained in three main test campaigns, corresponding to three levels of instrument integration, i.e., at single radiometer level \citep{villa2010}, at instrument level \citep{mennella2010}, and during the \Planck\ system-level campaign carried out at CSL\footnote{Centre Spatial de Li\'ege}.  In the latter campaign (August 2008), the instruments were operated in conditions very similar to flight.

\Planck\ is based on a novel concept of active cooling, which supports for the first time space-borne 0.1\,K bolometers and 20\,K HEMT amplifiers. Much of the complexity of the \Planck\ payload comes from its cryo-chain and from the very demanding stability requirements set by the two instruments \citep{planck2011-1.3}.  In particular, monitoring the performance and stability of 
the LFI InP amplifiers and phase switches provides ground-breaking technological insight 
on these devices in view of possible future projects. The performance analysis 
given in this paper refers to the first 12 months of data, and represents a first assessment of the instrument behaviour and systematics adequate to support the \Planck\ early release compact source catalog (ERCSC; \citealt{planck2011-1.10}) and the early science papers (see \Planck\ papers in this volume). 
In particular, no analysis of polarisation performance is given here. A complete characterisation
of the LFI instrument behaviour, including a full discussion of systematic
effects and trend analysis, will be provided 
with the January 2013 data release.


\section{Instrument}
\label{sec_instrument_design}

   \subsection{Instrument configuration}
\label{sec_instrument_configuration}
        
\begin{figure*}
        \begin{center}
            \includegraphics[width=18cm]{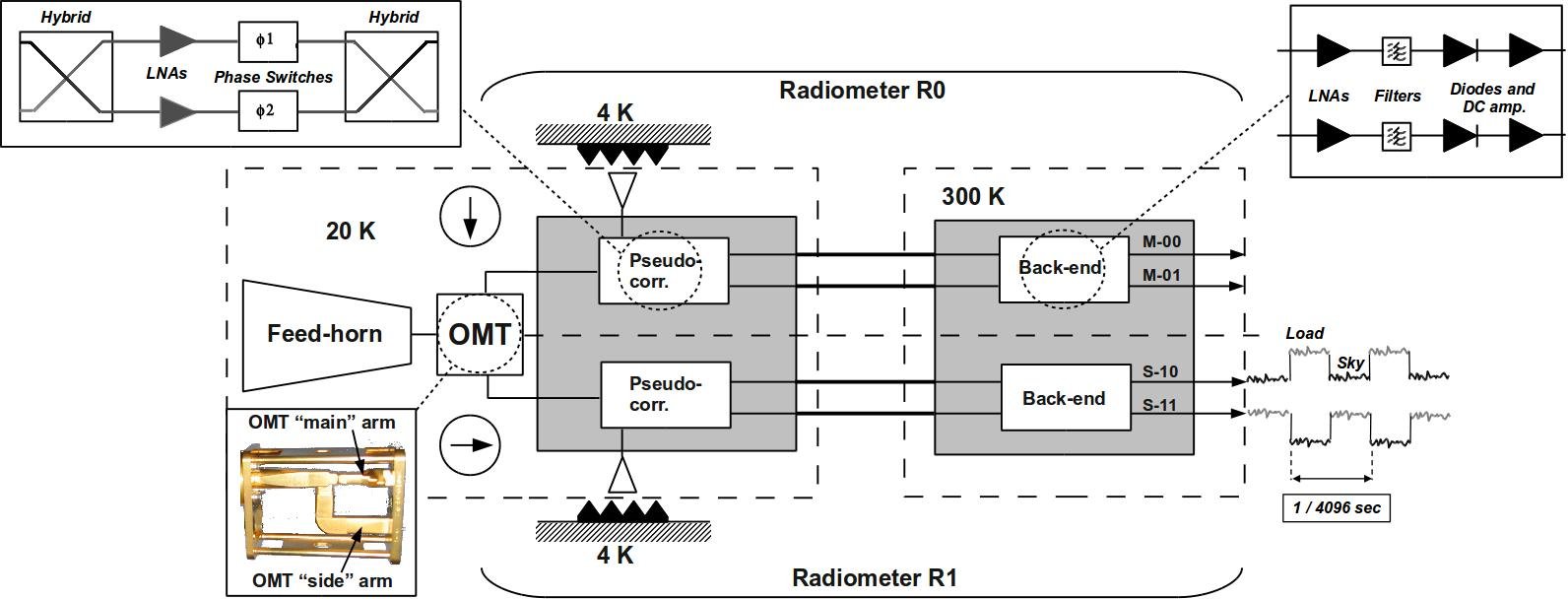}
        \end{center}
        \caption{Schematic of a complete RCA from feed-horn to analog voltage output, with insets showing the OMT, the details of the 20\,K pseudo-correlator, and the details of the back-end radio-frequency amplification, low-pass filtering, detection, and DC amplification.}
        \label{fig_rca_schematic}
    \end{figure*}

The instrument (Fig.~\ref{fig_lfi_instrument}) consists of a $\sim 20$ K focal plane unit hosting the corrugated feed horns, orthomode transducers (OMTs), and receiver front-end modules (FEMs). A set of 44~composite waveguides \citep{d'arcangelo2009a} interfaced with the three V-groove radiators \citep{planck2011-1.3} connects the front-end modules to the warm ($\sim 300$\,K) back-end unit (BEU), which contains further radio frequency amplification, detector diodes, and electronics for data acquisition and bias supply. Each LFI radiometer chain assembly (RCA) consists of two radiometers feeding two diode detectors (Fig.~\ref{fig_rca_schematic}), for a total of 44~detectors. The 11~RCAs are labelled by numbers from 18 to 28 as outlined in the right panel of Fig.~\ref{fig_lfi_instrument}.

    Figure~\ref{fig_rca_schematic} shows a schematic of a complete \hbox{RCA}. The feed horn is connected to an OMT, which splits the incoming radiation into two perpendicular linear polarisation components that propagate through two independent pseudo-correlation differential radiometers. These radiometers are labelled as \texttt{M} or \texttt{S} depending on the arm of the OMT they are connected to (``Main'' or ``Side'', see lower-left inset of Fig.~\ref{fig_rca_schematic}).

    In each radiometer, the sky signal coming from the OMT output is continuously compared with a stable 4\,K blackbody reference load mounted on the external shield of the HFI 4\,K box \citep{valenziano2009}. After being summed by a first hybrid coupler, the two signals are amplified by $\sim 30$\,\hbox{dB}. A phase shift alternating at 4096\,Hz between 0\deg\ and 180\deg\ is applied in one of the two amplification chains.  A second hybrid coupler separates back the sky and reference load components, which are further amplified and detected in the warm \hbox{BEU}.  The output voltage ranges from $-2.5$\,V to $+2.5$\,V.

    The output diodes are labelled with binary codes \texttt{00}, \texttt{01} (for radiometer \texttt{M}) and \texttt{10}, \texttt{11} (for radiometer \texttt{S}).  The four outputs of each radiometric chain are referred to as \texttt{M-00}, \texttt{M-01}, \texttt{S-10}, \texttt{S-11} (Fig.~\ref{fig_rca_schematic}).

    After detection, an analog circuit in the data acquisition electronics (DAE) removes a programmable offset to obtain a nearly null DC output voltage,    and a programmable gain is applied to match the signal level to the ADC input range.

    \begin{figure}[h!]
        \begin{center}
            \includegraphics[width=9cm]{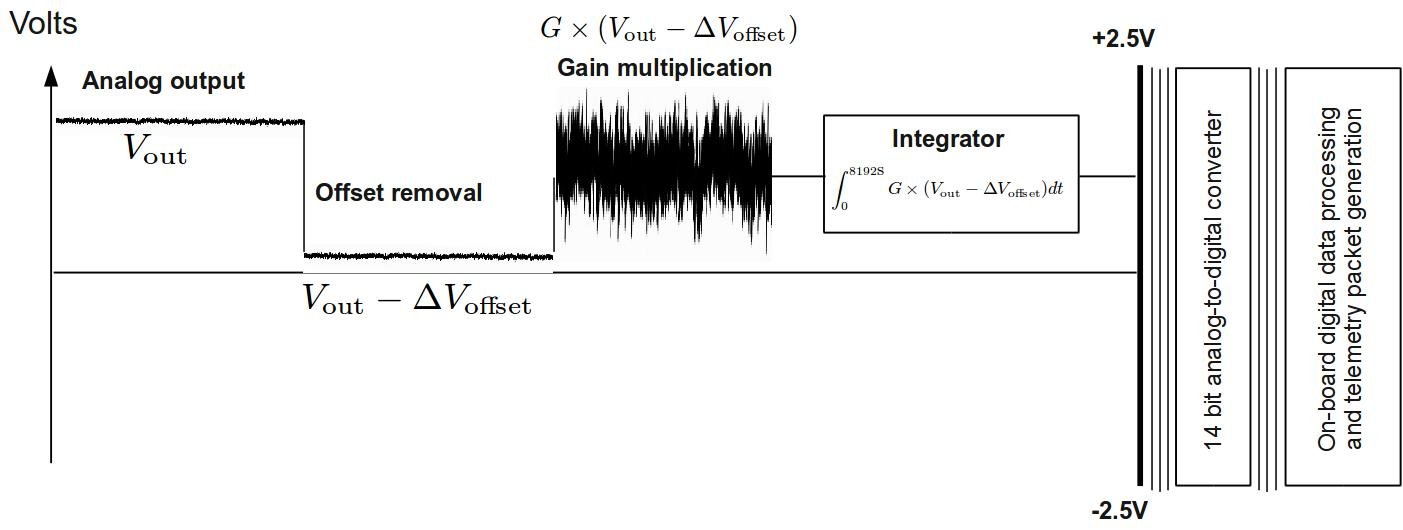}
        \end{center}
        \caption{Signal processing after detection with details about the analog offset and gain stages. The analog signal is digitised by a 14 bit analog-to-digital converter (ADC) and then processed.  Processing includes data rebinning, lossy requantisation, lossless compression, and telemetry packetization.}
        \label{fig_dae_gain_offset}
    \end{figure}

    After the ADC, data are downsampled, requantised, compressed according to a scheme described in \citet{herreros2009} and \citet{maris2009}, and assembled into telemetry packets. On the ground, telemetry packets are converted to volts, using the applied offset and gain factors, and split into sky and reference load time-ordered data (TOD).  


\subsection{Signal model}
\label{sec_signal_model}

    In the ideal case of a perflectly balanced radiometer, the differential power output for each of the four diodes can be written as follows \citep{seiffert2002,mennella2003,bersanelli2010}:

    \begin{equation}
        P_{{\rm out},0}^{\rm diode} = a\, G_{\rm tot}\,k\,\beta \left[ T_{\rm sky} + T_{\rm noise} - r\left( T_{\rm ref} + T_{\rm noise}\right) \right],
        \label{eq_p0}
    \end{equation}
    where $G_{\rm tot}$ is the total gain, $k$ is the Boltzmann constant, $\beta$ is the  bandwidth, and $a$ is the diode constant. $T_{\rm sky}$ and $T_{\rm ref}$ are the sky and reference load antenna temperatures at the inputs of the first hybrid, and $T_{\rm noise}$ is the receiver noise temperature. In this section we always refer to average quantities.  We omit, for simplicity, angled brackets, i.e., $T_{\rm sky}\equiv\langle T_{\rm sky}\rangle$, $T_{\rm ref}\equiv\langle T_{\rm ref}\rangle$, etc.
    
    The gain modulation factor, $r$, is defined by
    \begin{equation}
        r = \frac{T_{\rm sky} + T_{\rm noise}}{T_{\rm ref} + T_{\rm noise}},
        \label{eq_r}
    \end{equation}
    and is used to balance (in software) the temperature offset between the sky and reference load signals and minimise the residual 1/$f$ noise in the differential datastream. In nominal operating conditions, the gain modulation factor is in the range $0.84 < r < 1.00$, depending on channel.  This parameter is calculated each pointing period\footnote{A pointing period is the amount of time during which the \Planck\ telescope scans the same ring in the sky.} from the average uncalibrated total power data using the relationship:
    \begin{equation}
        r = V_{\rm sky} / V_{\rm ref}.
        \label{eq_r_v}
    \end{equation}
    
    Details about the gain modulation factor calculation and its implementation in the scientific pipeline can be found in \citet{mennella2003} and \citet{planck2011-1.6}.

    The white noise spectral density at the output of each diode is essentially independent of the reference-load absolute temperature and is given by
    \begin{equation}
        \Delta T_0^{\rm diode} = \frac{2\,(T_{\rm sky}+T_{\rm noise})}{\sqrt{\beta}}.
        \label{eq_deltat_diode_ideal}
    \end{equation}
If the front-end components are not perfectly balanced, then the separation of the sky and reference load signals after the second hybrid is not perfect and the outputs are mixed.  First-order deviations in white noise sensitivity from the ideal behaviour are caused mainly by noise temperature and phase-switch amplitude mismatches. Following the notation used in \citet{seiffert2002}, we define $\epsilon_{T_{\rm n}}$, the imbalance in front end noise temperature,  and $\epsilon_{A_1}$ and $\epsilon_{A_2}$, the imbalance in signal attenuation in the two states of the phase switch. Equation~\ref{eq_deltat_diode_ideal} for the two diodes of a slightly imbalanced radiometer then becomes
    
    \begin{equation}
        \left(\Delta T^{\rm diode}\right)^2 \approx  \left(\Delta T_0^{\rm diode}\right)^2
        \left( 1 \pm \frac{\epsilon_{A_1}-\epsilon_{A_2}}{2} + \alpha\epsilon_{T_{\rm n}}\right),
        \label{eq_deltat_diode_nonideal}
    \end{equation}
    which is identical for the two diodes apart from the sign of the term $(\epsilon_{A_1}-\epsilon_{A_2})/2$, representing the phase switch amplitude imbalance. This indicates that the isolation loss caused by this imbalance generates an anticorrelation between the white noise levels of the single-diode data streams.
    
    For this reason, the LFI scientific data streams are obtained by averaging the voltage outputs from the two diodes in each radiometer:
    \begin{equation}
        V_{\rm out}^{\rm rad} = w_1 V_{\rm out}^{\rm diode\, 1} + w_2 V_{\rm out}^{\rm diode\, 2},
        \label{eq_vout_radiometer}
    \end{equation}
    where $w_1$ and $w_2$ are inverse-variance weights calculated from the data as discussed in \citet{planck2011-1.6}.
    This way, the diode-diode anti-correlation is cancelled, and the radiometer white noise becomes
    
    \begin{equation}
        \Delta T^{\rm rad} \approx  \frac{\Delta T_0^{\rm diode}}{\sqrt{2}}
        \left( 1  + \alpha\epsilon_{T_{\rm n}}\right)^{1/2}.
        \label{eq_deltat_rad_nonideal}
    \end{equation}

    In Eqs.~(\ref{eq_deltat_diode_nonideal}) and (\ref{eq_deltat_rad_nonideal}), $\epsilon\ll 1$, while $\alpha$ is a term $\lesssim 1$ given by:
    \begin{equation}
        \alpha= \frac{T_{\rm noise}
        \left(2\,T_{\rm noise}+T_{\rm sky}+T_{\rm ref}\right)} {2\,\left(T_{\rm sky}+T_{\rm noise}\right)\left(T_{\rm ref}+T_{\rm noise}\right)}.
        \label{eq_alpha}
    \end{equation}
    
\begin{figure}[h!]
        \begin{center}
            \includegraphics[width=9cm]{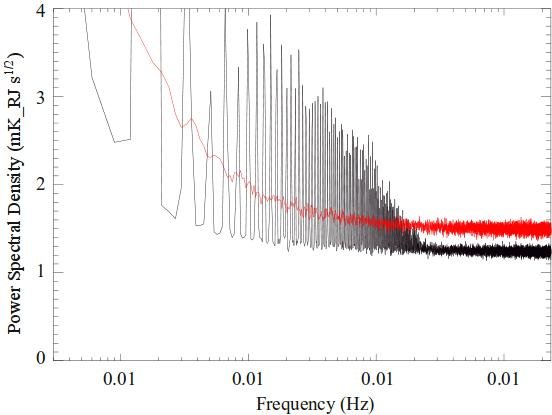}
        \end{center}

        \caption{Amplitude spectral density for weighted sum (black) and weighted difference (red) of \texttt{LFI28M} radiometer diodes.  A white noise component that is anti-correlated between the two timestreams accounts for the difference in white noise level seen in the plot. Further analysis steps use the weighed sum data.}
        \label{fig_psd_correlation}
    \end{figure}

    Figure~\ref{fig_psd_correlation} illustrates this anti-correlated noise for a representative LFI channel (\texttt{LFI28M}). In black we plot the amplitude spectral density (ASD) of the weighted sum of the diode pair corresponding to the radiometer. The sky signal is visible as a series of spikes at the satellite rotation frequency and harmonics. The 1/$f$ and white noise levels are clearly visible, too. In red we show the ASD of the weighted difference of the same two diodes, a case in which (as expected) the sky signal is removed. In the difference signal the white noise is nearly 20\% higher because of    the anti-correlated noise in the output of two diodes. In fact, if the diodes were uncorrelated then the sum and difference of their ASD's should show identical white noise levels.  

    While the effects of this correlated component and the proper propagation through the pipeline to maps could be calculated and corrected at the map level, we have found it more natural and effective to combine the diodes in the time domain, performing calibration and further processing on the combined time stream.


\subsection{In-band response and centre frequency}
\label{sec_band_response}

    The in-band receiver response has been thoroughly modelled and measured for all the LFI detectors during ground tests. The complete set of bandpass curves has been published in \citet{zonca2009}. From each curve we have derived the effective centre frequency according to
    \begin{equation}
        \nu_0 =  
        \frac{\int_{\nu_{\rm min}}^{\nu_{\rm max}} \nu\, g(\nu)\, d\nu} 
        {\int_{\nu_{\rm min}}^{\nu_{\rm max}} g(\nu)\, d\nu},
        \label{eq_centre_frequency}
    \end{equation}
    where $\Delta\nu = \nu_{\rm max} - \nu_{\rm min}$ is the receiver bandwidth and $g(\nu)$ is the bandpass response. Table~\ref{tab_centre_frequency_radiometer} gives the centre frequencies of the 22 LFI radiometers. For each radiometer, $g(\nu)$ is calculated by weight-averaging the bandpass response of the two individual diodes with the same weights used to average detector timestreams. For simplicity and for historical reasons, we will continue to refer to the three channels as the 30, 44, and 70\,GHz channels.

\begin{table}                    
\begingroup
\newdimen\tblskip \tblskip=5pt
\caption{LFI centre frequencies.}  
\label{tab_centre_frequency_radiometer}
\nointerlineskip
\vskip -3mm
\footnotesize
\setbox\tablebox=\vbox{
   \newdimen\digitwidth 
   \setbox0=\hbox{\rm 0} 
   \digitwidth=\wd0 
   \catcode`*=\active 
   \def*{\kern\digitwidth}
   \newdimen\signwidth 
   \setbox0=\hbox{+} 
   \signwidth=\wd0 
   \catcode`!=\active 
   \def!{\kern\signwidth}
\halign{\hbox to 1.3in{#\leaderfil}\tabskip=2em&
        \hfil#\hfil&
        \hfil#\hfil\tabskip=0pt\cr                             
\noalign{\doubleline}
\omit&\multispan2\hfil $\nu_0$\hfil\cr
\noalign{\vskip -4pt}
\omit&\multispan2\hrulefill\cr
\omit&Radiometer M&Radiometer S\cr
\omit\hfil RCA\hfil&[GHz]&[GHz]\cr
\noalign{\vskip 3pt\hrule\vskip 5pt}
{\bf V band; ``70\,GHz''}\cr
\noalign{\vskip 4pt}
\hglue 2em LFI18   & \getsymbol{LFI:center:frequency:LFI18:Rad:M}&\getsymbol{LFI:center:frequency:LFI18:Rad:S}\cr
\hglue 2em LFI19   & \getsymbol{LFI:center:frequency:LFI19:Rad:M}&\getsymbol{LFI:center:frequency:LFI19:Rad:S}\cr
\hglue 2em LFI20   & \getsymbol{LFI:center:frequency:LFI20:Rad:M}&\getsymbol{LFI:center:frequency:LFI20:Rad:S}\cr
\hglue 2em LFI21   & \getsymbol{LFI:center:frequency:LFI21:Rad:M}&\getsymbol{LFI:center:frequency:LFI21:Rad:S}\cr
\hglue 2em LFI22   & \getsymbol{LFI:center:frequency:LFI22:Rad:M}&\getsymbol{LFI:center:frequency:LFI22:Rad:S}\cr
\hglue 2em LFI23   & \getsymbol{LFI:center:frequency:LFI23:Rad:M}&\getsymbol{LFI:center:frequency:LFI23:Rad:S}\cr
\noalign{\vskip 2pt}
\hglue 3em {\bf Average}&\multispan2\hfil \getsymbol{LFI:center:frequency:70GHz}\hfil\cr
\noalign{\vskip 5pt}
{\bf Ka band; ``44\,GHz''}\cr
\noalign{\vskip 4pt}
\hglue 2em LFI24   & \getsymbol{LFI:center:frequency:LFI24:Rad:M}&\getsymbol{LFI:center:frequency:LFI24:Rad:S}\cr
\hglue 2em LFI25   & \getsymbol{LFI:center:frequency:LFI25:Rad:M}&\getsymbol{LFI:center:frequency:LFI25:Rad:S}\cr
\hglue 2em LFI26   & \getsymbol{LFI:center:frequency:LFI26:Rad:M}&\getsymbol{LFI:center:frequency:LFI26:Rad:S}\cr
\noalign{\vskip 2pt}
\hglue 3em {\bf Average}&\multispan2\hfil \getsymbol{LFI:center:frequency:44GHz}\hfil\cr
\noalign{\vskip 5pt}
{\bf K band; ``30\,GHz''}\cr
\noalign{\vskip 4pt}
\hglue 2em LFI27   & \getsymbol{LFI:center:frequency:LFI27:Rad:M}&\getsymbol{LFI:center:frequency:LFI27:Rad:S}\cr
\hglue 2em LFI28   & \getsymbol{LFI:center:frequency:LFI28:Rad:M}&\getsymbol{LFI:center:frequency:LFI28:Rad:S}\cr
 \noalign{\vskip 2pt}
\hglue 3em {\bf Average}&\multispan2\hfil \getsymbol{LFI:center:frequency:30GHz}\hfil\cr
\noalign{\vskip 5pt\hrule\vskip 3pt}}}
\endPlancktable
\endgroup
\end{table}
    
    Colour corrections, $C(\alpha)$, needed to derive the brightness temperature of a source with a power-law spectral index $\alpha$, are given in Table~\ref{tab_colour_corrections}. The values are averaged for the 11 RCAs and for the three frequency channels. Details about the definition of colour corrections are provided in \citet{planck2011-1.6}.

\begin{table*}                    
\begingroup
\newdimen\tblskip \tblskip=5pt
\caption{Colour corrections for the 11 LFI RCAs individually and averaged by frequency}
\label{tab_colour_corrections}
\nointerlineskip
\vskip -3mm
\footnotesize
\setbox\tablebox=\vbox{
   \newdimen\digitwidth 
   \setbox0=\hbox{\rm 0} 
   \digitwidth=\wd0 
   \catcode`*=\active 
   \def*{\kern\digitwidth}
   \newdimen\signwidth 
   \setbox0=\hbox{+} 
   \signwidth=\wd0 
   \catcode`!=\active 
   \def!{\kern\signwidth}
\halign{\hbox to 1.3in{#\leaderfil}\tabskip=2em&
        \hfil#\hfil&
        \hfil#\hfil&
        \hfil#\hfil&
        \hfil#\hfil&
        \hfil#\hfil&
        \hfil#\hfil&
        \hfil#\hfil\tabskip=0pt\cr                             
\noalign{\doubleline}
\omit&\multispan7\hfil S{\sc pectral} I{\sc ndex} $\alpha$\hfil\cr
\noalign{\vskip -4pt}
\omit&\multispan7\hrulefill\cr
\omit\hfil RCA\hfil&$-2.00$&$-1.00$&  0.00&  1.00&  2.00&  3.00&  4.00\cr
\noalign{\vskip 3pt\hrule\vskip 5pt}
                \texttt{LFI18}& 1.054&  1.028& 1.011& 1.003& 1.003 & 1.010& 1.026\cr
                \texttt{LFI19}& 1.170&  1.113& 1.066& 1.026& 0.994 & 0.969& 0.949\cr
                \texttt{LFI20}& 1.122&  1.079& 1.044& 1.017& 0.997 & 0.983& 0.975\cr
                \texttt{LFI21}& 1.087&  1.053& 1.028& 1.010& 1.000 & 0.996& 0.998\cr
                \texttt{LFI22}& 0.973&  0.971& 0.976& 0.988& 1.007 & 1.033& 1.066\cr
                \texttt{LFI23}& 1.015&  1.004& 0.999& 0.998& 1.003 & 1.012& 1.026\cr
\noalign{\vskip 3pt}
{\bf 70\,GHz average}&\bf 1.070&\bf 1.041&\bf 1.021&\bf 1.007&\bf 1.001&\bf 1.001&\bf 1.007\cr
\noalign{\vskip 8pt}
                \texttt{LFI24}& 1.028&  1.015& 1.007& 1.002& 1.000& 1.003& 1.009\cr
                \texttt{LFI25}& 1.039&  1.024& 1.013& 1.005& 1.000& 0.999& 1.000\cr
                \texttt{LFI26}& 1.050&  1.032& 1.017& 1.007& 1.000& 0.997& 0.997\cr
\noalign{\vskip 3pt}
{\bf 44\,GHz average}&\bf 1.039&\bf 1.024&\bf 1.012&\bf 1.004&\bf 1.000&\bf 0.999&\bf 1.002\cr
\noalign{\vskip 8pt}
                \texttt{LFI27}& 1.078&  1.049& 1.026& 1.010& 1.000& 0.996& 0.998\cr
                \texttt{LFI28}& 1.079&  1.049& 1.026& 1.009& 1.000& 0.997& 1.002\cr
\noalign{\vskip 3pt}
{\bf 30\,GHz average}&\bf 1.079&\bf 1.049&\bf 1.026&\bf 1.010&\bf 1.000&\bf 0.997&\bf 1.000\cr
\noalign{\vskip 5pt\hrule\vskip 3pt}}}
\endPlancktable
\endgroup
\end{table*}


\section{LFI operations}
\label{sec_operations}

   \subsection{Cooldown and tuning}
\label{sec_cooldown_tuning}

\begin{figure}[htb]
        \begin{center}
            \includegraphics[width=9cm]{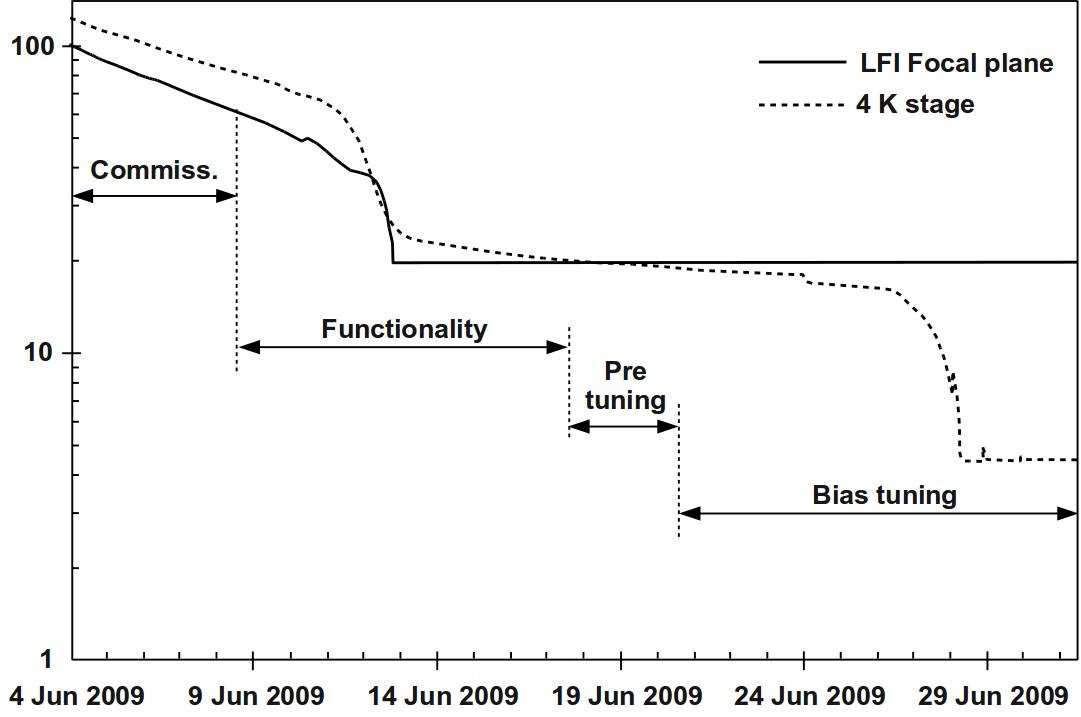}
            \caption{Cooldown curve of LFI focal plane and 4K stage. Functionality and bias tuning tests were carried out during this phase.}
            \label{fig_cooldown}
        \end{center}
    \end{figure}

    LFI operations began during the calibration, performance, and verification (CPV) phase of the mission. Functionality and tuning tests \citep{gregorio2011} were carried out, taking advantage of the varying temperature of the 4K-stage during cooldown (see Fig.~\protect\ref{fig_cooldown}).
    
    For optimal scientific performance, the bias voltages of the front-end low noise amplifiers (LNAs) and phase switch currents must be carefully tuned. Although radiometer tuning was performed during ground tests \citep{cuttaia2009}, the procedure was repeated in flight to account for possible changes in the electrical and thermal environment.

    Phase switch bias currents ($I_1$, $I_2$) of the 30\,GHz and 44\,GHz RCAs were tuned to optimise amplitude balance. This was done by exploring a two-dimensional bias surface around the optimal points found during ground tests. The results (repeated also at the end of the tuning campaign) confirmed the optimal points found before launch. The phase switches of the 70\,GHz RCAs, instead, were set to the maximum biases (1\,mA) and were not tuned.  Ground tests showed that the rise time was sensitive to bias currents, and decreased as the current increased from 0.5\,mA to 1\,mA \citep{gregorio2011}.

    LNA biases were tuned by exploring a large volume in the bias space of each amplifier.  The common drain voltage, $V_{\rm d}$, the gate voltage of the first stage, $V_{\rm g1}$, and the gate voltage common to the remaining stages, $V_{\rm g2}$, were sampled according to a ``hyper matrix'' (HYM) tuning strategy, in which several bias quadruplets $[(V_{\rm g1},V_{\rm g2})_{\rm LNA1},(V_{\rm g1},V_{\rm g2})_{\rm LNA2}]$ were varied for each radiometer. This strategy increased considerably the sampled parameter space with respect to ground tuning tests \citep{cuttaia2009}, and allowed us to fully characterise the radiometer performance in terms of noise temperature, isolation, and drain current balance.
    
    To define the broad bias regions to be deeply sampled by the HYM tuning procedure,  a pre-tuning phase was run with the instrument at 20\,K and the reference loads at 20.4\,\hbox{K}. The large imbalance between the sky and reference load signals provided us with enough voltage difference to estimate noise temperature.
    
    During the HYM tuning phase, smaller bias volumes around the optimal pre-tuning points were sampled at four different temperatures (see Fig.~\ref{fig_cooldown}) between 19.1\,K and 4\,K during the cooldown of the \HeJT\ cooler, allowing us also to characterize the response linearity \citep{mennella2009}. Drain voltages were also tuned for a limited subset of $V_{\rm g1},V_{\rm g2}$ quadruplets, making the overall bias space six-dimensional.
    
\begin{figure*}
        \begin{center}
            \includegraphics[width=9cm]{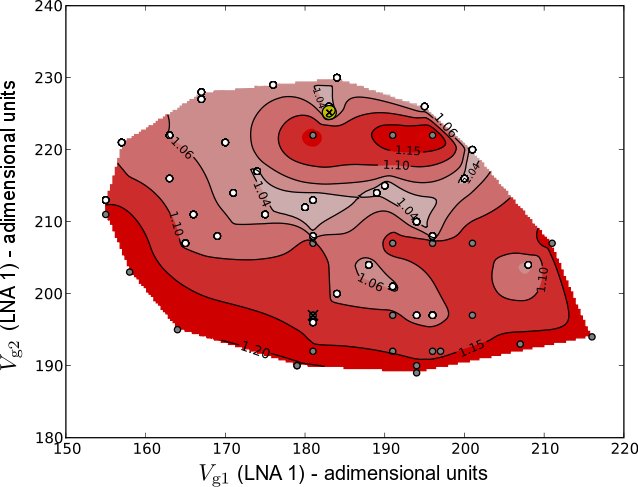}
            \includegraphics[width=9cm]{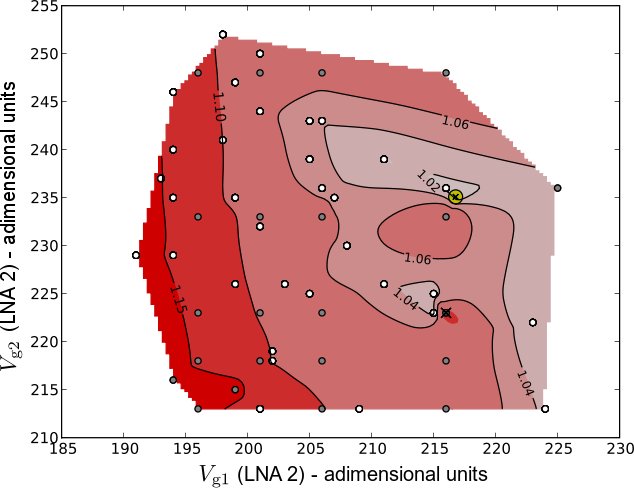}
            \caption{Example of condensed noise temperature maps for the two LNAs of radiometer \texttt{LFI21S}. Contour values represent the ratio of the noise temperature in a given bias configuration to the minimum noise temperature found in all tested configurations. The yellow crossed point is the chosen bias point.}
        \label{fig_TUN_Tnoise}
        \end{center}
    \end{figure*}

    Figure~\ref{fig_TUN_Tnoise} shows an example of ``condensed noise temperature'' maps.  These are contour plots in the $V_{\rm g1},V_{\rm g2}$ space for the LNAs of a given radiometer. Each point in the plot is the average of the best 20\% noise temperature values determined by the quadruplets sharing that particular $V_{\rm g1},V_{\rm g2}$ pair. The same approach was used to map isolation and drain currents. 

    Noise temperature and isolation maps were calculated by fitting data using both a linear and a non-linear response model. Minor effects caused by non-linear response were observed, as expected, only for 30 and 44\,GHz channels, confirming on-ground test results. As emphasized in \citet{mennella2009}, signal compression in 30 and 44\,GHz receivers is relevant only if the input range is of the order of $\gtrsim 1$\,K. Therefore it can be completely neglected in normal operations and in flight calibration, where the sky dynamic range is $\lesssim 10$\,mK. The optimal bias values found during tuning were close to those found during satellite-level  tests on the ground, with maximum deviations of about 10\%. Table~\ref{tab_summary_bias} summarizes the improvements in noise temperature and isolation between the ground and flight tests.

\begin{table}[h!]
\begingroup
\newdimen\tblskip \tblskip=5pt
\caption{Reduction in $T_{\rm noise}$ and increase in isolation (in dB) in flight compared to ground tests.}
\label{tab_summary_bias}
\nointerlineskip
\vskip -3mm
\footnotesize
\setbox\tablebox=\vbox{
   \newdimen\digitwidth 
   \setbox0=\hbox{\rm 0} 
   \digitwidth=\wd0 
   \catcode`*=\active 
   \def*{\kern\digitwidth}
   \newdimen\signwidth 
   \setbox0=\hbox{+} 
   \signwidth=\wd0 
   \catcode`!=\active 
   \def!{\kern\signwidth}
\halign{\hbox to 1.7in{#\leaderfil}\tabskip=1.0em&
        \hfil#\hfil&
        \hfil#\hfil&
        #\hfil\tabskip=0pt\cr                         
\noalign{\doubleline}
\omit&Min&Median&Max\cr
\omit&[\%]&[\%]&[\%]\cr
\noalign{\vskip 4pt\hrule\vskip 5pt}
$\delta\, T_{\rm noise} / T_{\rm noise}$& 0 & 4.2 & 12.5\cr
$\delta\, {\rm Isolation} / {\rm Isolation} [dB]$& 1 & 6 & 115\cr
\noalign{\vskip 5pt\hrule\vskip 3pt}}}
\endPlancktable
\endgroup
\end{table}

    The only large change was in radiometer \texttt{LFI21S}, for which isolation improved from $-7$\,dB measured on ground to $-16$\,dB measured in flight. There was a corresponding improvement in white noise sensitivity (see Sect.\,\ref{sec_sensitivity}).
    
\begin{figure}
\begin{center}
\includegraphics[width=9cm]{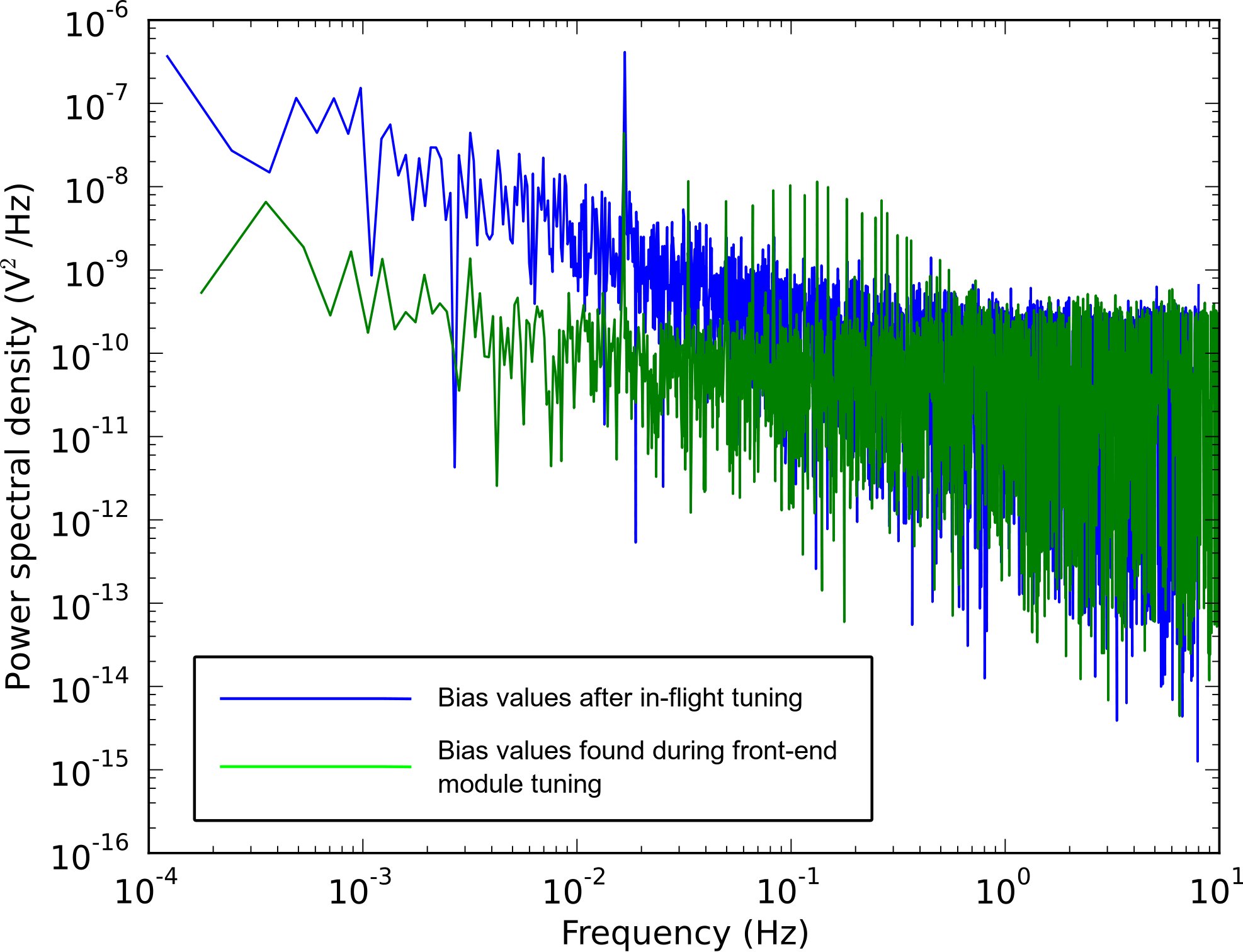}
\caption{\label{fig_LFI25_1_f_drift} Power spectral density of data from \texttt{LFI25M-00} detector with two different bias sets (Blue: optimal biases determined after flight tuning tests. Green: optimal biases determined during front-end module tuning before integration).}                            
\end{center}
\end{figure}

    After tuning, an unexpectedly high level of $1/f$ noise fluctuations was observed for the 44\,GHz RCAs, using either the new flight bias settings or the old ground ones. Dedicated tests showed that this instability was correlated with the phase switch configuration of 70\,GHz \texttt{LFI23}, and disappeared when the radiometers were biased with the optimal voltages found during the optimisation of the individual front-end modules before instrument integration (see Fig.~\ref{fig_LFI25_1_f_drift}).  This interaction between RCAs belonging to different frequency channels was unexpected and deeply investigated during \hbox{CPV}. The root cause was never fully established, but the most likely explanation was a parasitic oscillation triggered by unexpected cross-talk in the warm electronics. Details about investigations performed in flight to understand and solve this problem are reported in \citet{gregorio2011}. The final bias setting \citep{Davis2009}, characterised by a slightly higher power consumption ($\sim 40$\,mW) but similar noise and isolation performance, eliminated the  problem.
    
    Table~\ref{tab_flight_bias} summarises the bias settings chosen for the front-end amplifiers at the end of CPV. They have never been changed since the start of nominal operations.

\begin{table}                    
\begingroup
\newdimen\tblskip \tblskip=5pt
\caption{Origin of final bias settings.}
\label{tab_flight_bias}
\nointerlineskip
\vskip -3mm
\footnotesize
\setbox\tablebox=\vbox{
   \newdimen\digitwidth 
   \setbox0=\hbox{\rm 0} 
   \digitwidth=\wd0 
   \catcode`*=\active 
   \def*{\kern\digitwidth}
   \newdimen\signwidth 
   \setbox0=\hbox{+} 
   \signwidth=\wd0 
   \catcode`!=\active 
   \def!{\kern\signwidth}
\halign{\hbox to 1.3in{#\leaderfil}\tabskip=3em&
        #\hfil&
        #\hfil\tabskip=0pt\cr                         
\noalign{\doubleline}
\omit&\multispan2\hfil R{\sc adiometer}\hfil\cr
\noalign{\vskip -4pt}
\omit&\multispan2\hrulefill\cr
\omit\hfil RCA\hfil&\omit\hfil M\hfil&\omit\hfil S\hfil\cr
\noalign{\vskip 3pt\hrule\vskip 5pt}
                \texttt{LFI18}& Flight$^{\rm a}$&  Flight\cr
                \texttt{LFI19}& Ground$^{\rm b}$&  Ground\cr
                \texttt{LFI20}& Ground&  Ground\cr
                \texttt{LFI21}& Ground&  Flight\cr
                \texttt{LFI22}& Ground&  Ground\cr
                \texttt{LFI23}& Ground&  Ground\cr
\noalign{\vskip 8pt}
                \texttt{LFI24}& FEM$^{\rm c}$&  FEM\cr
                \texttt{LFI25}& FEM&  FEM\cr
                \texttt{LFI26}& FEM&  FEM\cr
\noalign{\vskip 8pt}
                \texttt{LFI27}& Ground&  Ground\cr
                \texttt{LFI28}& Ground&  Ground\cr
\noalign{\vskip 5pt\hrule\vskip 3pt}}}
\endPlancktable
\tablenote a Determined during in-flight tests.\par
\tablenote b Determined during satellite-level ground tests.\par
\tablenote c Determined during module-level ground tests.\par
\endgroup
\end{table}


The last tuning step of the CPV phase configures the signal processing unit (SPU) data compressor for each of the 44 LFI channels \citep{maris2009} so that the data fit into the allocated telemetry bandwidth without significant loss of sensitivity (defined below) from quantisation errors produced by the lossy compressor. Quantisation errors appear as an additional white noise component that can be characterised as the ratio of the quantisation rms $\varepsilon_q$ over the intrinsic rms $\sigma$ of the signal before compression.  The ratio $\varepsilon_q/\sigma$ can be monitored directly in-flight with the so-called ``calibration channel'', a telemetry mode that provides 15\,min of uncompressed data for each diode every day \citep{bersanelli2010}.

\begin{figure*}
\centering
\includegraphics[width=18cm]{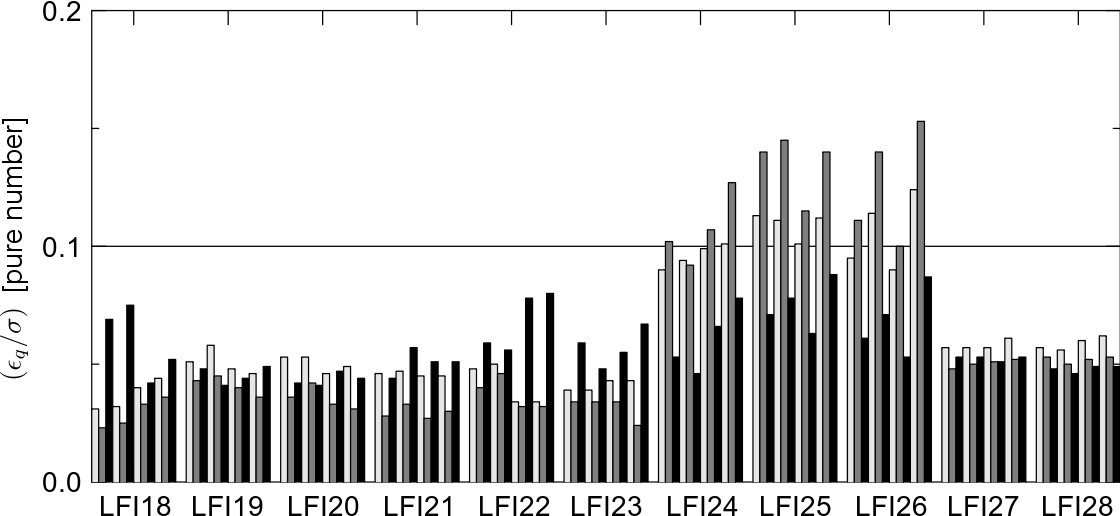}
\caption{\label{fig_quantBarchart} Average values of $\varepsilon_q/\sigma$ (quantisation error normalized to RMS) measured during the CPV tests for the 44 LFI detectors grouped by receiver.  For each channel, three bars show $\varepsilon_q/\sigma$ for the: (i) sky signal (white); (ii) reference load signal (gray); and (iii) differential signal (black). The $\varepsilon_q/\sigma < 0.1$ requirement in the difference data \citep{maris2009} is satisfied by every channel.}
\end{figure*}

    To limit the loss of sensitivity to less than 1\%, we require $\varepsilon_q/\sigma<0.1$ in the differenced data \citep{maris2009}. The results of the SPU calibration are summarized in Fig.~\ref{fig_quantBarchart} and Table~\ref{tab_quantErrorNumbers}. The $(\varepsilon_q/\sigma)_\mathrm{dif} < 0.1$ requirement in the differenced data is satisfied by every channel.

    \begin{table*}                    
    \begingroup
    \newdimen\tblskip \tblskip=5pt
    \caption{Normalized quantisation error $\varepsilon_q/\sigma$ for the $N$ channels at each frequency.  Values are plotted in Fig.~\ref{fig_quantBarchart}.}
    \label{tab_quantErrorNumbers}
    \nointerlineskip
    \vskip -3mm
    \footnotesize
    \setbox\tablebox=\vbox{
    \newdimen\digitwidth 
    \setbox0=\hbox{\rm 0} 
    \digitwidth=\wd0 
    \catcode`*=\active 
    \def*{\kern\digitwidth}
    \newdimen\signwidth 
    \setbox0=\hbox{+} 
    \signwidth=\wd0 
    \catcode`!=\active 
    \def!{\kern\signwidth}
    \halign{\hbox to 1.1in{#\leaderfil}\tabskip=2em&
            \hfil#\hfil&
            \hfil#\hfil\tabskip=1em&
            \hfil#\hfil&
            \hfil#\hfil\tabskip=2em&
            \hfil#\hfil\tabskip=1em&
            \hfil#\hfil&
            \hfil#\hfil\tabskip=2em&
            \hfil#\hfil\tabskip=1em&
            \hfil#\hfil&
            \hfil#\hfil\tabskip=0pt\cr                         
    \noalign{\doubleline}
    \omit&&\multispan9\hfil $\varepsilon_q/\sigma$ [\%]\hfil\cr
    \noalign{\vskip -4pt}
    \omit&&\multispan9\hrulefill\cr
    \omit&&\multispan3\hfil Sky\hfil&\multispan3\hfil Reference\hfil&\multispan3\hfil Difference\hfil\cr
    \noalign{\vskip -4pt}
    \omit&&\multispan3\hrulefill&\multispan3\hrulefill&\multispan3\hrulefill\cr
    \omit\hfil F{\sc requency}\hfil&\omit\hfil $N$\hfil&Median&Min&Max&Median&Min&Max&Median&Min&Max\cr
    \noalign{\vskip 3pt\hrule\vskip 5pt}
    70 & 24&  *4.5& 3.1& *5.8& *3.3& 2.3& *4.6& 5.1& 4.1& 8.0\cr
    44 & 12&  10.1& 9.0& 12.4& 12.1& 9.2& 15.3& 6.9& 4.6& 8.8\cr
    30 & *8&  *5.7& 5.6& *6.2& *5.1& 4.8& *5.3& 5.0& 4.6& 5.3\cr
    \noalign{\vskip 5pt\hrule\vskip 3pt}}}
    \endPlancktablewide
    \endgroup
    \end{table*}

    The stability of the SPU compressor is monitored daily by an automatic procedure that checks for variations of the telemetry and quantisation error and for near-saturation conditions. $\varepsilon_q$ has changed little during the first year of operations. Typical variations are much less than 1\%, with maximum variations around 1.7\%.


\subsection{Instrument response during cooldown}
\label{sec_instrument_response_during_cooling}

\begin{figure}[h]
        \begin{center}
            \includegraphics[width=9cm]{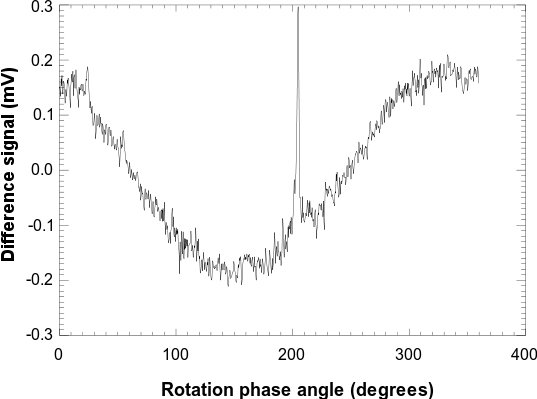}
            \caption{Phase-binned, uncalibrated, differential data from \texttt{LFI28M-00}, acquired on 14 June 2009 near the beginning of the performance verification phase.  Data for 100\,revolutions of the telescope have been averaged over 960~``phase bins'' of the rotation angle.  The CMB dipole signal, which was used for preliminary photometric calibration, and a spike measured while crossing the galactic plane, are clearly visible.}
            \label{fig_cpv_phasebinned}
        \end{center}
    \end{figure}

    LFI observations began during the cooldown, even before tuning.  Figure~\ref{fig_cpv_phasebinned} shows some of the first differential, uncalibrated output from \texttt{LFI28M-00}. 

\begin{figure*}
        \begin{center}
            \includegraphics[width=18cm]{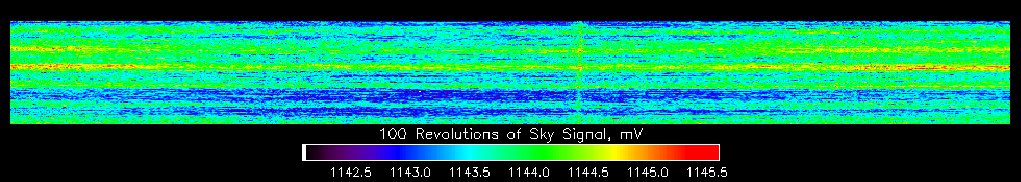}
            \includegraphics[width=18cm]{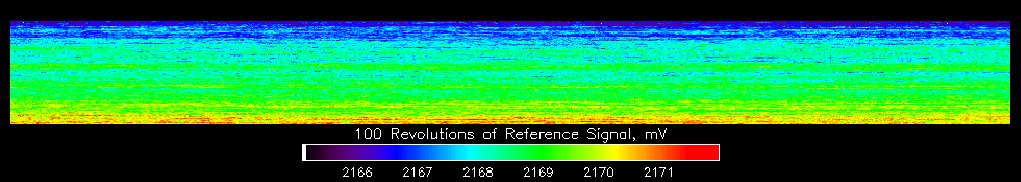}
            \includegraphics[width=18cm]{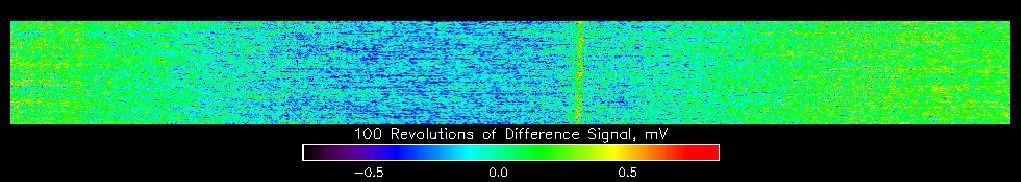}
            \caption{Phase binned uncalibrated pseudo-maps for \texttt{LFI28M-00}. Top: sky signal. Middle: reference load signal. Bottom: difference signal.  The horizontal axis is spin axis phase angle, the vertical axis is revolution number, and the color scale gives signal amplitude.  The mean value has been subtracted from each revolution in the difference map.}
            \label{fig_cpv_ps}
        \end{center}
    \end{figure*}

    Figure~\ref{fig_cpv_ps} shows a ``pseudo-map.''   The horizontal axis is spin axis phase angle, the vertical axis is revolution number, and the color scale gives signal amplitude.  The map shows the sky, reference, and difference data for 100~telescope revolutions on 14 June 2009. These are the same data that appear in the phase binned map in Fig.~\ref{fig_cpv_phasebinned}.   In the sky signal, one can see the galaxy spike, a hint of the CMB dipole, and fluctuations slower than the pointing period given by $1/f$ noise fluctuations in the total power voltage output.  In the reference load signal, one can see a significant gradient due to the rapid cooldown of the 4\,K stage during this period, as well as the same $1/f$ fluctuations seen in the sky signal. In the difference signal, one can see that correlated fluctuations in the sky and reference data are cancelled.  The mean value has been subtracted from each revolution in the difference map.  

\subsection{Operations after verification phase}
\label{sec_operations_after_cpv}

    Since the beginning of nominal LFI operations, no instrument parameters have been changed, with one exception. Starting on 11 August 2009, occasional uncommanded jumps in output signal in a single channel were seen, sometimes causing saturation in the ADC and temporary loss of science data.  These jumps were traced to single bit-flip changes in the DAE gain-setting circuit, presumably a result of single-event upsets (SEUs) from cosmic rays.  Initially, science data were lost in saturated channels until the gain could be reset during the next downlink period. Starting in October 2009, an automatic procedure resets all gain values every $\sim40$\,min.

\begin{figure}[htb]
        \begin{center}
            \includegraphics[width=9cm]{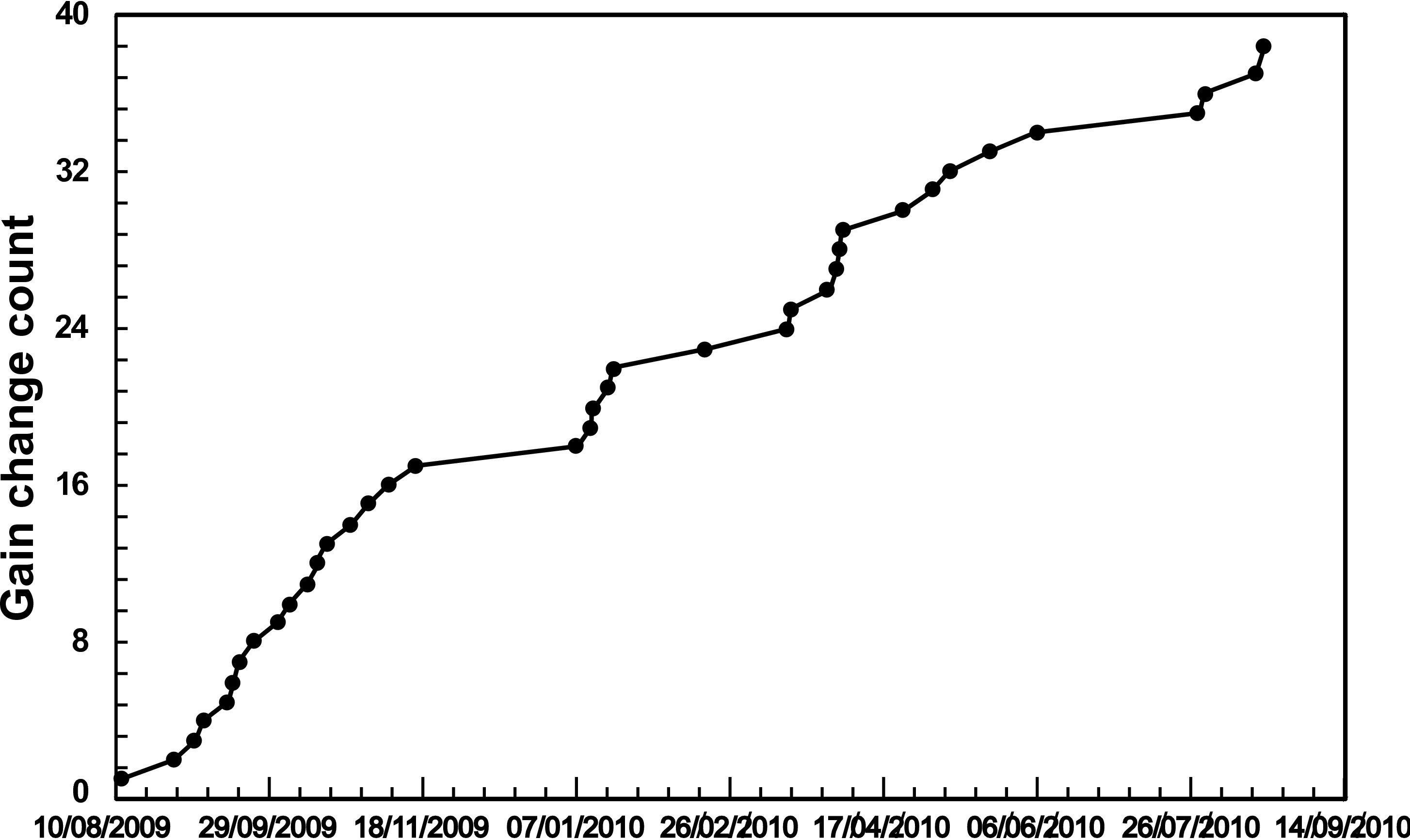}
            \caption{Cumulative number of gain change events during the first year of operations.}
            \label{fig_gainchange}
        \end{center}
    \end{figure}

    Figure~\protect\ref{fig_gainchange} shows the cumulative number of gain change events as a function of time during the first year of operation. A total of 38 were seen through 30 September 2010, corresponding to an average rate of about one event every 11 days.  Out of these 38 events, 13 saturated the ADC, leading to lost observation time from the corresponding detector, five did not saturate the ADC but the compression algorithm was not able to deal with the anomalous signal statistics, and 20 could be recovered simply by applying the correct gain after the event.

    In most cases the gain increased, but occasionally it decreased.   The bit flip occurs in different channels, on different bits, and the resulting value after the flip is not always the same.  The most significant bit (controlling the second amplifier stage) never flips, suggesting that the component undergoing a possible single event upset (SEU) is the programmable gain amplifier in the first amplifier stage, an Analog Devices AD526SD/883B.  The internal technology of the gain amplifiers in the two stages is different, therefore a different sensitivity to high energy events would not be too surprising.  Nevertheless, investigation\footnote{Bibliographic research on Electronic Radiation Response Information Center (ERRIC), NASA GFSC Radhome; ESA radiation effect database} revealed no information on SEU effects on AD526 devices.  The event rate is too low to reveal a correlation between solar activity and cosmic ray fluxes as measured by the onboard space radiation environment monitor (SREM; \citet{planck2011-1.1}); however, a significant population of high energy cosmic rays, larger than expected before launch,  has been observed by the HFI instrument \citep{planck2011-1.5}.

\subsection{Missing and usable data}
\label{sec_flagged_data}

    Since the start of the mission, only a small percentage of the LFI science data have been lost or considered unusable for science.  Table~\ref{tab_data_flags_percentage} gives the percentage of time lost to missing data, anomalies, and maneuvers for the period from 12 August 2009 to 7 June 2010.  
    
    The main source of missing data is telemetry packets where the arithmetic compression performed by the SPU is incorrect, causing a decompression error.  There were 15 such packets in all LFI channels, with negligible scientific impact.  For instance, for the entire 70\,GHz frequency channel, there are just 101 lost seconds.

    Anomalies include the DAE gain jumps described in Sect.\,\ref{sec_operations_after_cpv}, and other instabilities (see Sect.~\ref{sec_stability}) that make the data unsuitable for science.  

   Spacecraft maneuvers, from routine repointings of the spin axis to stationkeeping maneuvers (see \cite{planck2011-1.1}) cause by far the largest fraction of discarded data so far; however, we expect these data to be fully recovered after additional analysis of the startracker and gyroscope data.  If this expectation is met, and the performance of \Planck\ remains as it has been, well over 99\% of all observing time will be usable.

    \begin{table}                    
    \begingroup
    \newdimen\tblskip \tblskip=5pt
    \caption{Percentages of usable and unusable data for the period 8 August 2009 to 7 June 2010.  See text for explanation of the categories.}
    \label{tab_data_flags_percentage}
    \nointerlineskip
    \vskip -3mm
    \footnotesize
    \setbox\tablebox=\vbox{
    \newdimen\digitwidth 
    \setbox0=\hbox{\rm 0} 
    \digitwidth=\wd0 
    \catcode`*=\active 
    \def*{\kern\digitwidth}
    \newdimen\signwidth 
    \setbox0=\hbox{+} 
    \signwidth=\wd0 
    \catcode`!=\active 
    \def!{\kern\signwidth}
    \halign{\hbox to 1.3in{#\leaderfil}\tabskip=1.5em&
            \hfil#\hfil&
            \hfil#\hfil&
            \hfil#\hfil\tabskip=0pt\cr 
    \noalign{\doubleline}
    \omit&30\,GHz&44\,GHz&70\,GHz\cr
    \omit\hfil Category\hfil&[\%]&[\%]&[\%]\cr
    \noalign{\vskip 3pt\hrule\vskip 5pt}
    Missing&$1.6\times10^{-4}$&$2.7\times10^{-4}$&$3.9\times10^{-4}$\cr
    Anomalies&*0.4&*0.7&*0.4\cr
    Maneuvers&*8.3&*8.3&*8.3\cr
    Used&     91.3&91.0&91.3\cr
    \noalign{\vskip 5pt\hrule\vskip 3pt}}}
    \endPlancktable
    \endgroup
    \end{table}


\section{Beams and angular resolution}
\label{sec_beams_angular_resolution}

   The most accurate measurements of the LFI main beams have been made with Jupiter, the most powerful 
unresolved (to \Planck) celestial source in the LFI frequency range.
Since the LFI feed horns point to different positions on the sky, they 
detect the signal at different times.  Table~\ref{tab_jupiterOD} gives the dates of Jupiter observations in the Fall of 2009.

\begin{table}                    
\begingroup
\newdimen\tblskip \tblskip=5pt
\caption{Dates of Jupiter observations in 2009}
\label{tab_jupiterOD}
\nointerlineskip
\vskip -3mm
\footnotesize
\setbox\tablebox=\vbox{
   \newdimen\digitwidth 
   \setbox0=\hbox{\rm 0} 
   \digitwidth=\wd0 
   \catcode`*=\active 
   \def*{\kern\digitwidth}
   \newdimen\signwidth 
   \setbox0=\hbox{+} 
   \signwidth=\wd0 
   \catcode`!=\active 
   \def!{\kern\signwidth}
\halign{\hbox to 0.84in{#\leaderfil}\tabskip=1.0em&
        \hfil#\hfil&
        \hfil#\hfil\tabskip=0pt\cr 
\noalign{\doubleline}
\omit\hfil RCA \hfil&Operational Day&Date\cr
\noalign{\vskip 3pt\hrule\vskip 5pt}
\omit\bf 70\,GHz\hfil\cr
\hglue 1em18, 23&168--169&28 October--29 October\cr
\hglue 1em19, 22&169--170&29 October--30 October\cr
\hglue 1em18, 23&169--171&29 October--31 October\cr
\noalign{\vskip 5pt}
\omit\bf 44\,GHz\hfil\cr
\hglue 1em24&    170--171&30 October--31 October\cr
\hglue 1em25, 26&163--165&23 October--25 October\cr
\noalign{\vskip 5pt}
\omit\bf 30\,GHz\cr
\hglue 1em27, 28&170--172&30 October--1 November\cr
\noalign{\vskip 5pt\hrule\vskip 3pt}}}
\endPlancktable
\endgroup
\end{table}

The first step in extraction of the beams was to remove $1/f$-type noise from the data using the 
{\tt Madam} destriping map-making code.  Planets were masked during this process. Details of the \texttt{Madam}
destriper and of the pipeline implemented to extract beams from planet measurements is reported in \citet{planck2011-1.6}.

To map the beam, each sample contained in the selected timelines was projected in the $(u,v)$-plane 
perpendicular to the nominal line-of-sight (LOS) of the telescope (and at $85\deg$ to the satellite spin axis).  
The $u$ and $v$ coordinates are defined in terms of the usual spherical coordinates $(\theta,\phi)$:
\begin{eqnarray}
    u &=& \sin\theta\cos\phi, \nonumber\\
    v &=& \sin\theta\sin\phi.
    \label{eq_uv}
\end{eqnarray}

To increase the signal-to-noise ratio, data were binned in an angular region of 
2\arcm\ for the 70\,GHz channels and 4\arcm\ for the 30 and 44\,GHz channels.  We recovered all beams 
down to $-20$\,dB from the peak.  An elliptical Gaussian was fit to each beam for both {\tt M} and {\tt S} radiometers.  
The FWHM given in Table \ref{tab_beam-parameters} is the square root of the product of the major axis and minor axis 
FWHMs of the individual beams, averaged between {\tt M} and {\tt S} radiometers.  The uncertainties 
in Table~\ref{tab_beam-parameters} are the standard deviation of the mean of the $1\sigma$ statistical 
uncertainties of the fit. Although a small difference between the {\tt M} and {\tt S} beams caused by 
optics and receiver non-idealities can be expected, for the purpose of point-source extraction relevant 
for this paper the beams have been considered identical, as this effect is well within the statistical uncertainty.

The ellipticity and orientation of the beams are fundamental parameters. Exhaustive
results and details on the determination of all LFI beam parameters will be presented in the future.  For 
the purpose of this paper, Table~\ref{tab_beam-parameters} gives the typical FWHM and ellipticity averaged over each frequency channel.

\begin{figure*}
        \begin{center}
            \includegraphics[width=18cm]{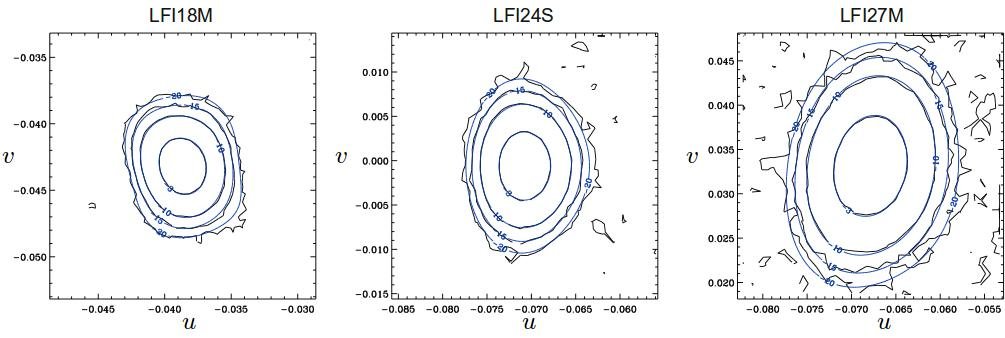}
            \caption{\label{fig_lfi_beams} Example (one for each frequency channel) of the LFI measured beams compared with simulations. 
            The simulated main beams have been computed in the co- and cross-polar basis according to the 
            Ludwig's third definition \citep{ludwig1973}, in spherical grids with 301 x 301 points defined with respect to the LOS
            frame. They are referred to the design telescope configuration.  
            In each plot the contours are the levels at $-3$, $-10$, $-15$, and $-20$\,dB 
            from the corresponding power peak. The simulations have been carried out in the 
            transmitting mode using GRASP9 software. Physical optics and physical theory
            of diffraction has been used on both reflectors.}
        \end{center}
\end{figure*}

Figure~\ref{fig_lfi_beams} shows three examples of measured beams compared with
calculations performed with the GRASP9\footnote{\url{http://www.ticra.com}} software. 
The calculated beams have been smeared appropriately to take into account satellite rotation and sampling.  
The effect of sampling is evident already at $-3$\,dB and cannot be neglected (the typical effect on FWHM is 2\% at 70\,GHz).  
In the comparison of measured and calculated beams, the peak of the simulated beam was aligned with the peak of the 
measured beam (calculated from a Gaussian elliptical fit). The electromagnetic model of the design 
telescope \citep{sandri2010} was used as a reference  in the comparison. Ideal parameters have been assumed for 
the shape of the mirrors, the alignment of the telescope and focal plane unit, as well as for the pattern of the 
feed horns. The good agreement down to $-20$\,dB demonstrates that, to first order, the
overall LFI optical system is performing as expected.  Comparison with a more realistic telescope model, including 
the actual alignment, measured mirror shapes, and measured feedhorn patterns, will be considered in future work. 

\begin{table}                    
\begingroup
\newdimen\tblskip \tblskip=5pt
\caption{LFI beam FWHM and mean ellipticity measured in flight from the first Jupiter pass.}  
\label{tab_beam-parameters}
\nointerlineskip
\vskip -3mm
\footnotesize
\setbox\tablebox=\vbox{
   \newdimen\digitwidth 
   \setbox0=\hbox{\rm 0} 
   \digitwidth=\wd0 
   \catcode`*=\active 
   \def*{\kern\digitwidth}
   \newdimen\signwidth 
   \setbox0=\hbox{+} 
   \signwidth=\wd0 
   \catcode`!=\active 
   \def!{\kern\signwidth}
\halign{\hbox to 1.31in{#\leaderfil}\tabskip=1em&
        \hfil#\hfil&
        \hfil#\hfil&
        \hfil#\hfil\tabskip=0pt\cr                         
\noalign{\doubleline}
\omit&FWHM$^{\rm a}$&Uncertainty$^{\rm b}$\cr
\omit\hfil RCA\hfil&[\arcm]&[\arcm]&Ellipticity$^{\rm c}$\cr
\noalign{\vskip 3pt\hrule\vskip 5pt}
{\bf 70\,GHz average}&\bf\getsymbol{LFI:FWHM:70GHz}&&\bf1.27\cr
\noalign{\vskip 4pt}
\hglue 2em  LFI18 &\getsymbol{LFI:FWHM:LFI18}&\getsymbol{LFI:FWHM:uncertainty:LFI18}\cr
\hglue 2em  LFI19 &\getsymbol{LFI:FWHM:LFI19}&\getsymbol{LFI:FWHM:uncertainty:LFI19}\cr
\hglue 2em  LFI20 &\getsymbol{LFI:FWHM:LFI20}&\getsymbol{LFI:FWHM:uncertainty:LFI20}\cr
\hglue 2em  LFI21 &\getsymbol{LFI:FWHM:LFI21}&\getsymbol{LFI:FWHM:uncertainty:LFI21}\cr
\hglue 2em  LFI22 &\getsymbol{LFI:FWHM:LFI22}&\getsymbol{LFI:FWHM:uncertainty:LFI22}\cr
\hglue 2em  LFI23 &\getsymbol{LFI:FWHM:LFI23}&\getsymbol{LFI:FWHM:uncertainty:LFI23}\cr
\noalign{\vskip 5pt}
{\bf 44\,GHz average}&\bf\getsymbol{LFI:FWHM:44GHz}&&\bf1.26\cr
\noalign{\vskip 4pt}
\hglue 2em  LFI24&\getsymbol{LFI:FWHM:LFI24}&\getsymbol{LFI:FWHM:uncertainty:LFI24}\cr
\hglue 2em  LFI25&\getsymbol{LFI:FWHM:LFI25}&\getsymbol{LFI:FWHM:uncertainty:LFI25}\cr
\hglue 2em  LFI26&\getsymbol{LFI:FWHM:LFI26}&\getsymbol{LFI:FWHM:uncertainty:LFI26}\cr
\noalign{\vskip 5pt}
{\bf 30\,GHz average}&\bf\getsymbol{LFI:FWHM:30GHz}&&\bf1.38\cr
\noalign{\vskip 4pt}
\hglue 2em  LFI27&\getsymbol{LFI:FWHM:LFI27}&\getsymbol{LFI:FWHM:uncertainty:LFI27}\cr
\hglue 2em  LFI28&\getsymbol{LFI:FWHM:LFI28}&\getsymbol{LFI:FWHM:uncertainty:LFI28}\cr
\noalign{\vskip 5pt\hrule\vskip 3pt}}}
\endPlancktable
\tablenote a The square root of the product of the major axis and minor axis FWHMs of the individual RCA beams, averaged between {\tt M} and {\tt S} radiometers.\par
\tablenote b The standard deviation of the mean of the $1\sigma$ statistical uncertainties of the fit. Although a small difference between the {\tt M} and {\tt S} beams caused by optics and receiver non-idealities can be expected, for the purpose of point-source extraction relevant for this paper the beams are considered identical, as this effect is well within the statistical uncertainty.\par
\tablenote c Ratio of the major and minor axes of the fitted elliptical Gaussian.\par
\endgroup
\end{table}


\section{Stability and calibration}
\label{sec_stability_and_calibration}

   \subsection{Stability}
\label{sec_stability}

Thanks to its differential scheme, the LFI is insensitive to many effects in the total-power data caused by $1/f$ noise, thermal fluctuations, or electrical instabilities.
    
\begin{figure}
        \begin{center}
            \includegraphics[width=9cm]{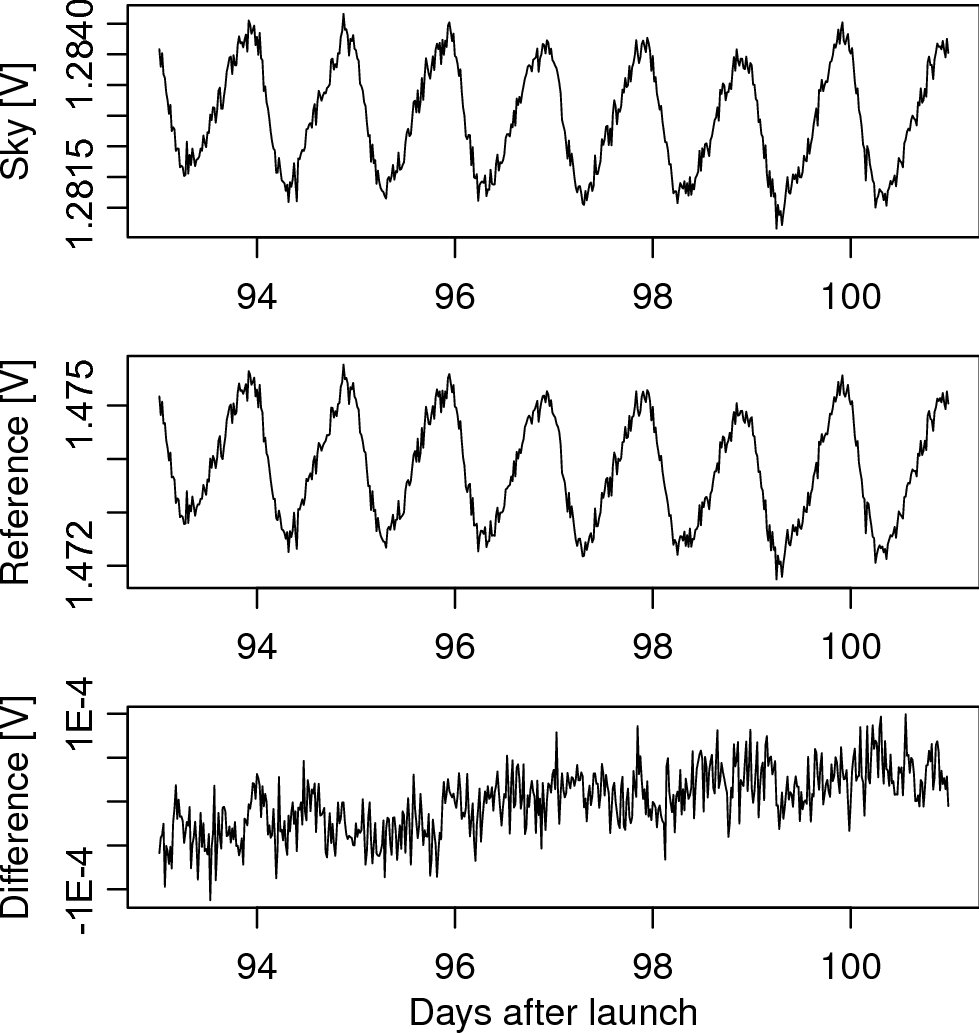}                                                                                                           
            \caption{Sky (top), reference (middle) and difference (bottom) signals from radiometer \texttt{LFI27M}. The modulation is due to the thermal effect induced by the satellite transponder being turned on and off for the daily downlink early in the mission.  The quasi-sinusoidal $\sim 0.1\%$ fluctuation in the sky and reference signals is almost completely removed in the differenced data.}
            \label{fig_dailyStability}
        \end{center}
\end{figure}

One effect detected during the first survey was the daily temperature fluctuation in the back-end unit induced by the transponder, which was turned on only for downlinks for the first 258 days of the mission (\citealt{planck2011-1.3}, \S\S\,2.4.1 and 5.1). Fig.~\ref{fig_dailyStability} shows the effect of these fluctuations on the radiometeric output of \texttt{LFI27M}.  As expected, the effect is highly correlated between the sky and reference load signals.  In the difference, the variation is reduced by a factor $\sim (1-r)$, where $r$ is the gain modulation factor defined in Eq.~(\ref{eq_r_v}).

\begin{figure*}
        \begin{center}
            \begin{tabular}{c c}
                \includegraphics[width=9cm]{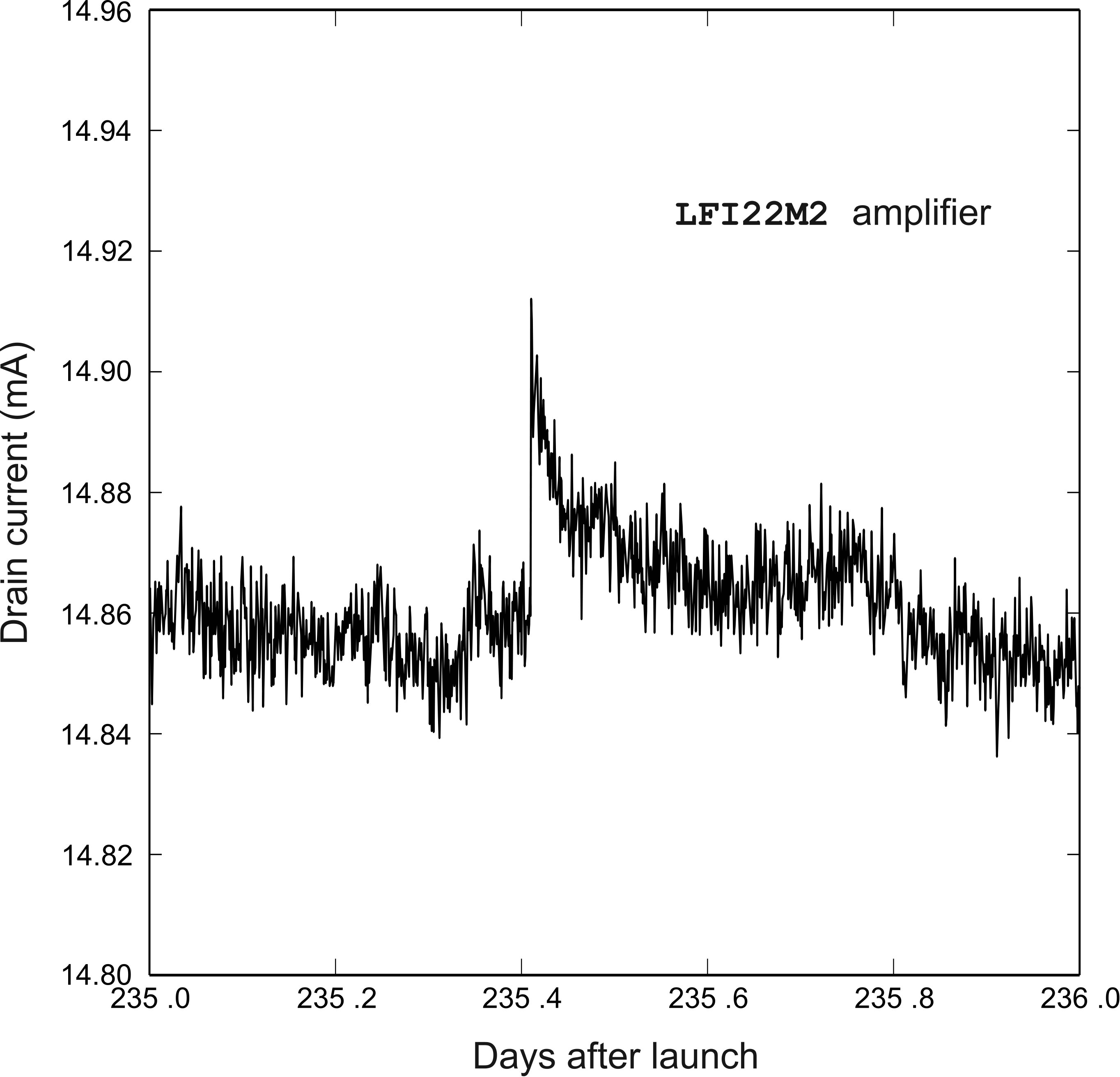} & 
                \includegraphics[width=9cm]{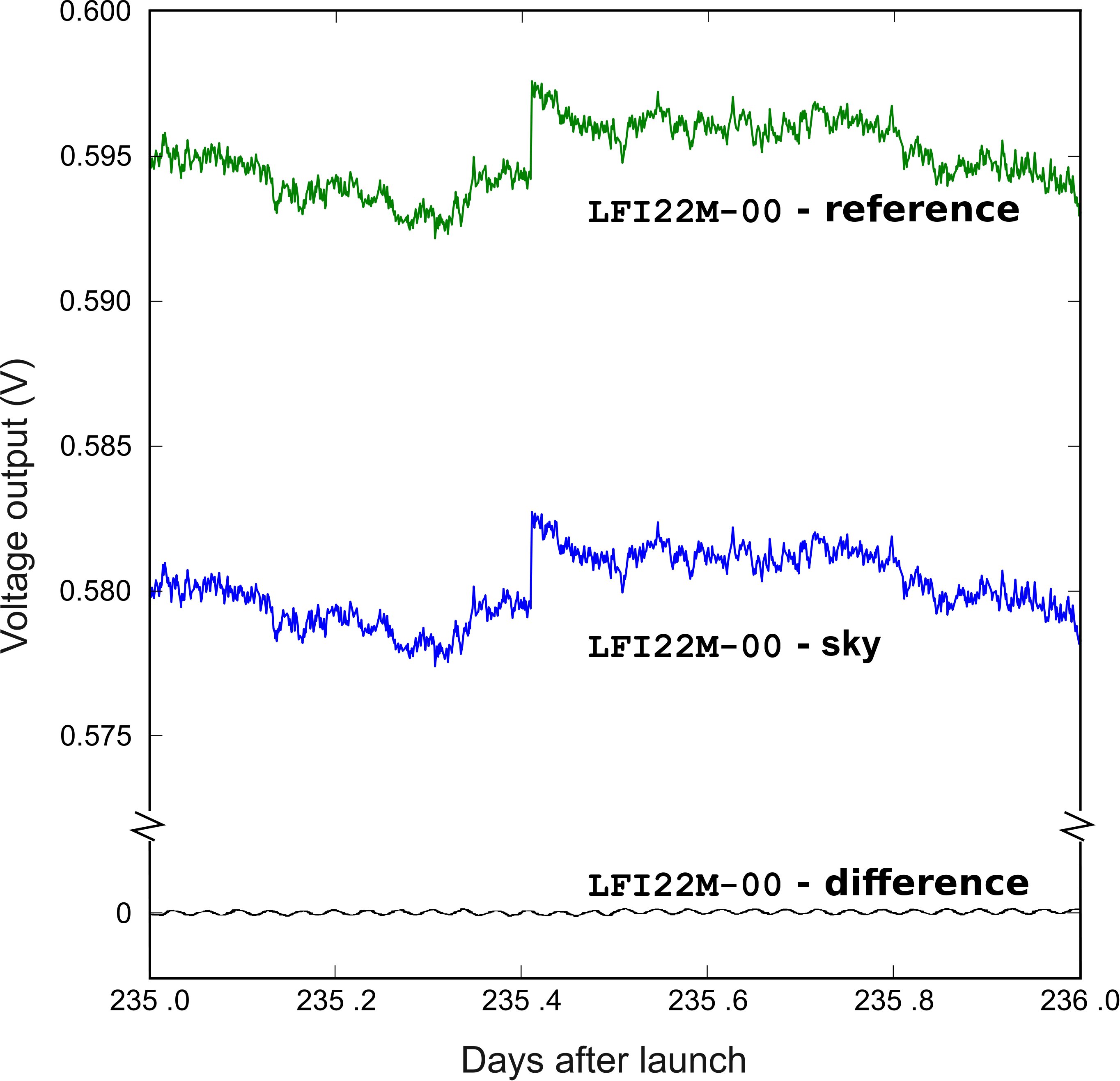}\\
            \end{tabular}
            \caption{Short spikes in the drain current (left) affect total power signals (right). 
            The jumps are strongly correlated in sky and reference signals, so that in the difference data 
            the effect essentially disappears.}
        \label{fig_Idjump}
        \end{center}
\end{figure*}

A  particular class of signal fluctuations occasionally observed during operations is represented by electrical instabilities that appear as abrupt increases in the measured drain current of the front-end amplifiers with a variable relaxation time from few seconds to some hundreds of seconds.  Typically, these events cause a change in both sky and reference load signals and disappear in the difference (see Fig.~\ref{fig_Idjump}).

Because these effects are essentially common-mode, their residual on the differenced data is negligible, and the data are suitable for science production.  In a few cases the residual fluctuation in the differential output was large enough (a few millikelvin in calibrated antenna temperature units) to be flagged, and the data are not used. The total amount of discarded data for all LFI channels until Operational Day 389 was about 2000\,s per detector, or 0.008\%.
        
A further peculiar effect appeared in the 44\,GHz detectors, where single 
isolated samples, either on the sky or the reference voltage output, were far from the rest.  
Over a reference period of four months, 15~occurrences of single-sample spikes (out of 
24 total anomaly events) were discarded, an insignificant loss of data.


\subsection{Calibration}
\label{sec_calibration}

Photometric calibration, i.e., conversion from voltage to antenna temperature, is performed for each radiometer after 
total power data have been cleaned of 1\,Hz frequency spikes (see \S~\ref{sec_systematic_effects} and \citet{planck2011-1.6}), and 
differenced.

Our calibrator is the well-known dipole signal induced by Earth and spacecraft motions
with respect to the CMB rest frame.  The largest calibration uncertainty comes from the presence of the Galaxy and of the CMB anisotropies 
in the measured signal. We therefore use an iterative calibration procedure in which the dipole is 
fit and subtracted, producing a sky map that is then subtracted from the original data to enhance the dipole signal 
for the next iteration.  Typically, convergence is obtained after few tens of iterations.

\begin{figure*}
        \begin{center}
            \includegraphics[width=18cm]{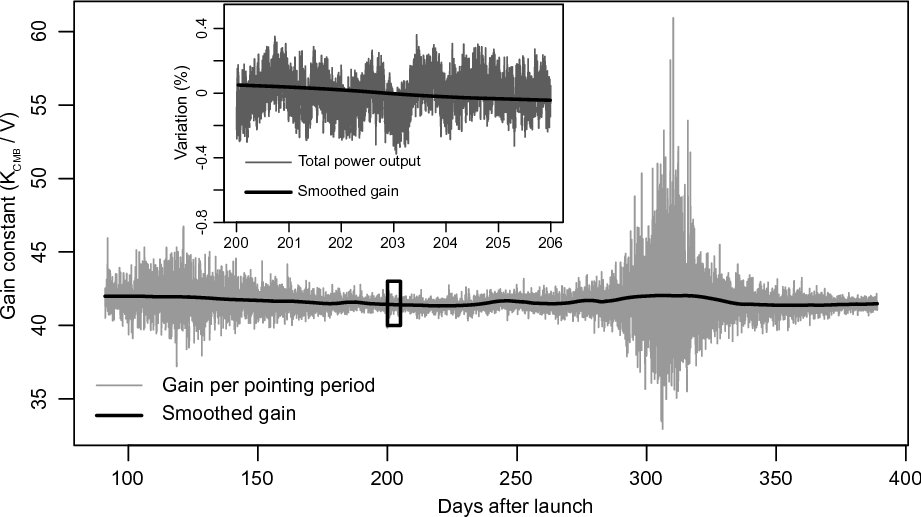}
            \caption{Reconstructed gain for radiometer \texttt{LFI21M}. Thin gray line: gain constants obtained for each pointing period by the iterative calibration procedure. The two regions showing large scatter in the reconstructed gain  correspond to regions with minimum amplitude in the CMB dipole.  Thick black line: gain constants obtained after smoothing - this is the actual curve used in the pipeline.  The inset shows a closer look at the region between Operational Days 200 and 205.  Here the relative variation (in \%) for the smoothed gain model is compared with the relative
            gain variation calculated from Eq.~(\ref{eq_deltav_over_v}) and shown in dark grey.
            }
            \label{fig_gain_constants}
        \end{center}
    \end{figure*}

The thin gray line in Fig.~\ref{fig_gain_constants} shows the result of this iterative process for the radiometer 
\texttt{LFI21M}, pointing period by pointing period. In the most stable regions the gain values display 
relative variations of $\sim 0.8\%$ rms and $\sim 7\%$ peak-to-peak. In the most unstable region, 
where the spacecraft spin was nearly aligned with the dipole and the dipole signal was weak, the relative variations 
are $\sim 4\%$ r.m.s. and $\sim 67\%$ peak-to-peak.  These variations reflect statistical uncertainties in the determination 
of the gain over a single pointing period, rather than actual changes in gain. 

We can put a limit on the true intrinsic gain variations by looking at the variation of the total power 
voltage output (cf. \S~\ref{sec_stability}).  The small variations in total power constrain intrinsic gain variations to be less than 1\%.
    
The gain solution based on individual pointing periods was therefore processed to improve its stability by implementing two 
running averages that have been further smoothed with wavelets: one with a 5-day 
window, used in the strong dipole regions, and the other with a 30-day window, 
used in the weak dipole regions.  In particular cases, where a true known instrument gain change had 
to be traced\footnote{e.g., when the satellite started using the transponder always ``on'' 
which caused a change in the instrument warm unit temperature.} we used a 
5-day un-smoothed window.  In Fig.~\ref{fig_gain_constants} the the thick black line gives the 
smoothed gain curve; variations are now $\sim 0.5\%$ rms and $\sim 1.6\%$ peak-to-peak. The inset in the figures 
compares the smoothed gain and the relative gain variation obtained from Eq.~(\ref{eq_deltav_over_v}), which represent 
true gain fluctuations in the instrument at the level of $\pm 0.2\%$. In the current 
gain model implementation these changes are neglected, but they will be incuded in future versions of our analysis.


\subsection{Calibration accuracy}
\label{sec_calibration_accuracy}

Following COBE and WMAP \citep{kogut1996,jarosik2010}, our main calibrator is the dipole modulation in the CMB. 
In our current calibration model we use as calibration signal the sum of the solar dipole 
$\Delta T_\mathrm{Sun}$ and the orbital dipole $\Delta T_\mathrm{orb}$, which is the 
contribution from \Planck\ orbital velocity around the Sun:
\begin{equation}
\Delta T = \bigl(\Delta T_\mathrm{Sun} + \Delta T_\mathrm{orb}\bigr) \sin \vartheta_\mathrm{axis},
\label{eq_absulute_calibration}
\end{equation}
where $\vartheta_\mathrm{axis}$ is the angle between the spacecraft axis and the overall dipole axis (solar + orbital).
    
In Eq.~(\ref{eq_absulute_calibration}), the absolute calibration uncertainty is dominated by 
the uncertainty in $\Delta T_\mathrm{Sun}$, which is known only to about 1\%. The modulation of the orbital dipole 
by the Earth's motion around the Sun is known with an uncertainty almost three orders of 
magnitude smaller; however, at least one complete \Planck\ orbit is needed for its measurement.  In the future, 
calibration based on the Earth orbital modulation will be significantly more accurate.  

The accuracy of our current calibration can be estimated by taking into account two components: 
1)~the statistical uncertainty in the regions of weak dipole; and 2)~the systematic 
uncertainty caused by neglecting gain fluctuations that occur on periods shorter than the 
smoothing window.  Neglecting the contribution from the wavelet filters and assuming, for 
simplicity, that the smoothing is done using plain averages, we can write the statistical uncertainty as:
\begin{equation}
    \left.\frac{\delta G}{G}\right|_\mathrm{stat} = \frac1{\sqrt{M}}\,\sqrt{\frac{\sum_N 
        \bigl(G_i - \left<G\right>\bigr)}{N - 1}},
\label{eq_gain_statistical_uncertainty}
\end{equation}
where $N$ is the overall number of pointings over the considered period and $M$ is the number of pointings 
for each smoothing region (5 or 30\,days). 
    
Table~\ref{tab_CalibrationErrors} lists the largest statistical uncertainties in four time windows 
(days\,100--140, 280--320, 205--245, 349--389), the first two corresponding to minimum and the second two to maximum dipole response.

\begin{table}                    
\begingroup
\newdimen\tblskip \tblskip=5pt
\caption{Worst-case relative calibration uncertainties.}
\label{tab_CalibrationErrors}
\nointerlineskip
\vskip -3mm
\footnotesize
\setbox\tablebox=\vbox{
   \newdimen\digitwidth 
   \setbox0=\hbox{\rm 0} 
   \digitwidth=\wd0 
   \catcode`*=\active 
   \def*{\kern\digitwidth}
   \newdimen\signwidth 
   \setbox0=\hbox{+} 
   \signwidth=\wd0 
   \catcode`!=\active 
   \def!{\kern\signwidth}
\halign{\hbox to 1.5in{#\leaderfil}\tabskip=2em&
        \hfil#\hfil\tabskip=3em&
        \hfil#\hfil\tabskip=0pt\cr                      
\noalign{\doubleline}
\omit&\multispan2\hfil $\left.\delta G/G\right|_\mathrm{stat}$ [\%]\hfil\cr
\noalign{\vskip -4pt}
\omit&\multispan2\hrulefill\cr
\omit\hfil RCA\hfil&\omit\hfil Rad. M\hfil&\omit\hfil Rad. S\hfil\cr
\noalign{\vskip 3pt\hrule\vskip 5pt}
\bf 70\,GHz average&\multispan2\hfil\bf 0.12\hfil\cr
\noalign{\vskip 4pt}
\hglue 2em\texttt{LFI18}&0.28&0.24\cr
\hglue 2em\texttt{LFI19}&0.14&0.23\cr
\hglue 2em\texttt{LFI20}&0.25&0.26\cr
\hglue 2em\texttt{LFI21}&0.37&1.04\cr
\hglue 2em\texttt{LFI22}&0.37&0.18\cr
\hglue 2em\texttt{LFI23}&0.29&0.38\cr
\noalign{\vskip 8pt}
\bf 44\,GHz average&\multispan2\hfil\bf 0.07\hfil\cr
\noalign{\vskip 4pt}
\hglue 2em\texttt{LFI24}&0.35&0.31\cr
\hglue 2em\texttt{LFI25}&0.36&0.30\cr
\hglue 2em\texttt{LFI26}&0.40&0.41\cr
\noalign{\vskip 8pt}
\bf 30\,GHz average&\multispan2\hfil\bf 0.05\hfil\cr
\hglue 2em\texttt{LFI27}&0.31&0.35\cr
\hglue 2em\texttt{LFI28}&0.23&0.38\cr
\noalign{\vskip 5pt\hrule\vskip 3pt}}}
\endPlancktable
\endgroup
\end{table}

Our current calibration scheme neglects systematic gain variations caused by thermal fluctuations, 
which introduce spurious signal fluctuations on timescales ranging from 1 to 24\,hours.  Being 
much slower than the satellite spin period, these fluctuations are  well-removed during map-making by the destriping algorithm
\citep{Keihanen2005,Keihanen2010,planck2011-1.6}, so that the rms systematic uncertainty 
per pixel in the final maps due to imperfect calibration is at sub-microkelvin levels (see \S~\ref{sec_thermal_fluctuations}).  
However, we expect that a more accurate description of the gain variations over short timescales 
will improve our calibration accuracy even at the level of time ordered data. If we use 
the dipole signature to estimate an average value $G_0$ for the gain over long time 
periods (e.g., several months), we can write the gain versus time as
    \begin{equation}
        \label{eqGainCorrections}
        G(t) = G_0 \times \Bigl(1 + \xi (t)\Bigr),
    \end{equation}
    where $\xi(t)\equiv \delta G/\langle G\rangle$ is the relative gain variation.
    
The simplest and most direct measurement of gain changes is through the monitoring of the stability of the reference 
load total power voltage output, $V_{\rm ref} = K\left(T_{\rm ref} + T_{\rm noise}\right )$, where $K = 1/G$. The 
relative output voltage variation is:
\begin{eqnarray}
        \frac{\delta V_{\rm ref}}{\langle V_{\rm ref}\rangle} &&= \left[ \left(\frac{\delta K}{\langle K\rangle}\right)^2 +\right.      \\
         && + \left.\left(\frac{\delta T_{\rm noise}}{\langle T_{\rm ref} +T_{\rm noise}\rangle}\right)^2 +
        \left(\frac{\delta T_{\rm ref}}{\langle T_{\rm ref}+T_{\rm noise}\rangle }\right)^2\right]^{1/2}.\nonumber
        \label{eq_ref_variation}
    \end{eqnarray}

If the fractional variations in the noise temperature and in the reference load signal are negligible compared to gain 
variations, then we have that:
\begin{equation}
        \frac{\delta V_{\rm ref}}{\langle V_{\rm ref}\rangle} = \frac{\delta K}{\langle K 
           \rangle} = -\xi(t).
\label{eq_deltav_over_v}
\end{equation}

Considering that $T_{\rm noise}$ is in the range 10--30\,K and that the reference load 
temperature is about 4.5\,K with fluctuations of $\lesssim 2$\,mK, we have 
$\frac{\delta T_{\rm ref}}{\langle T_{\rm ref} + T_{\rm noise}\rangle}< 2\times 10^{-4}$.  Fluctuations in the noise 
temperature due to thermal changes can be estimated by assuming a coupling of 0.5\,K\,K\mo\ and taking 
a typical temperature fluctuation of 1\,mK in the front-end 
unit.  We have $\frac{\delta T_{\rm noise}}{\langle T_{\rm ref} + T_{\rm noise}\rangle}< 5\times 10^{-4}$, which 
implies that any variation in the total power signal that is larger than $\sim 6\times 10^{-4}$ is due to actual changes in the gain.
    
\begin{figure}
\begin{center}
\includegraphics[width=9cm]{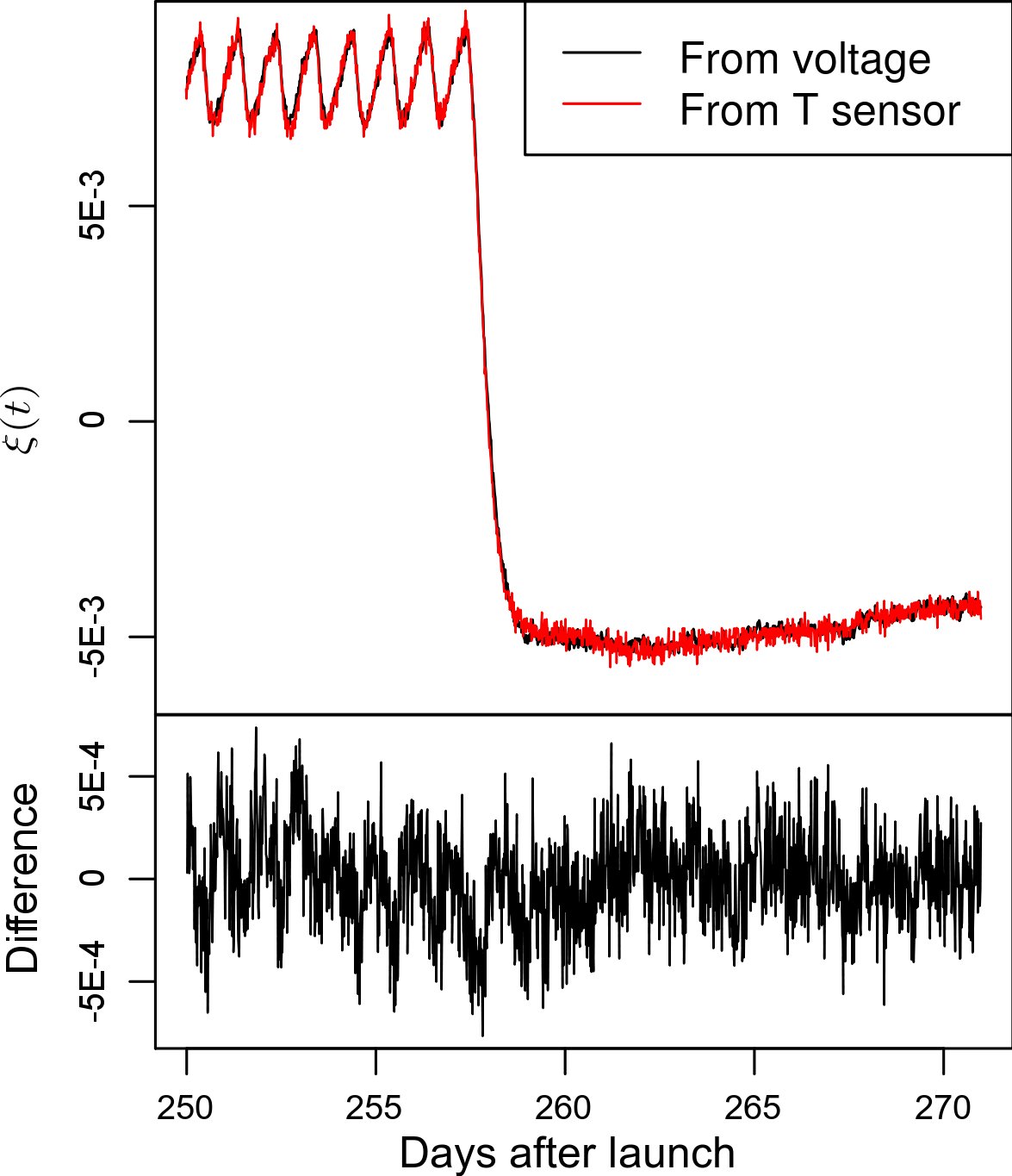}
\caption{Comparison between two gain models for radiometer \texttt{LFI26S}. Each model uses the average gain 
calculated using the dipole, but modulates it over time using either Eq.~(\ref{eq_deltav_over_v}) (black line) or one of the temperature 
sensors in the LFI back-end (red line). The bottom 
plot shows the difference between the two gain models.}
\label{fig_calibration_model}
\end{center}
\end{figure}

Figure~\ref{fig_calibration_model} shows two estimates for $\xi(t)$ based on 
the temperature sensors in the radiometer back-end and on the relative variation of the reference load voltage output (black line).

In future versions of the LFI gain model, we plan to use the iterative solution as a starting point and then trace gain changes down to short time scales both by using housekeeping information and by monitoring the relative variation of the total power radiometric output (Eq.~(\ref{eq_deltav_over_v})).


\section{Noise properties}
\label{sec_signal_noise_properties}

    The noise characteristics of the LFI datastreams are closely reproduced by a simple (white + $1/f$) noise model:
\begin{equation}
    P(f) = \sigma^2\left[1+\left(\frac{f}{f_{\rm knee}}\right)^\alpha\right],
    \label{eq_noise_model}
\end{equation}
where $P(f)$ is the power spectrum and $\alpha \approx -1$.

In this model, noise properties are characterised by three parameters, 
the white noise limit $\sigma$, 
the knee frequency\footnote{The frequency at which the $1/f$ and white noise contribute equally in power.} 
$f_{\rm knee}$, and the exponent of the $1/f$ component $\alpha$, also referred to as ``slope.''  In this section we 
show how the LFI noise has been characterised in flight, and we give noise performace estimates based on  one year of operations.

\subsection{Method}
\label{sec_sensitivity_method}

Noise properties have been calculated following two different and complementary approaches: 
1)~fitting Eq.~(\ref{eq_noise_model}) to time-ordered data for each radiometer; 
and 2)~building normalised noise maps by differencing data from the first half of the pointing period with 
data from the second half of the pointing period to remove the sky signal (``jackknife'' data sets).
    
\subsubsection{Noise power spectrum estimation in frequency domain}
\label{sec_noise_fitting}

Estimation of noise power spectra from time ordered data is part of the iterative approach 
in the {\it ROMA\/} map-making suite \citep{prunet2001,natoli2001,degasperis2005}.  
We produced joint estimates for both signal (i.e., maps) and noise, and 
then fitted the resulting noise power spectra to Eq.~(\ref{eq_noise_model}) to estimate the three basic noise 
parameters $\sigma$, $f_{\rm knee}$, and $\alpha$. All details relative to the pipeline are provided in \citet{planck2011-1.6}.
    
\begin{figure}
    \begin{center}
        \includegraphics[width=9cm]{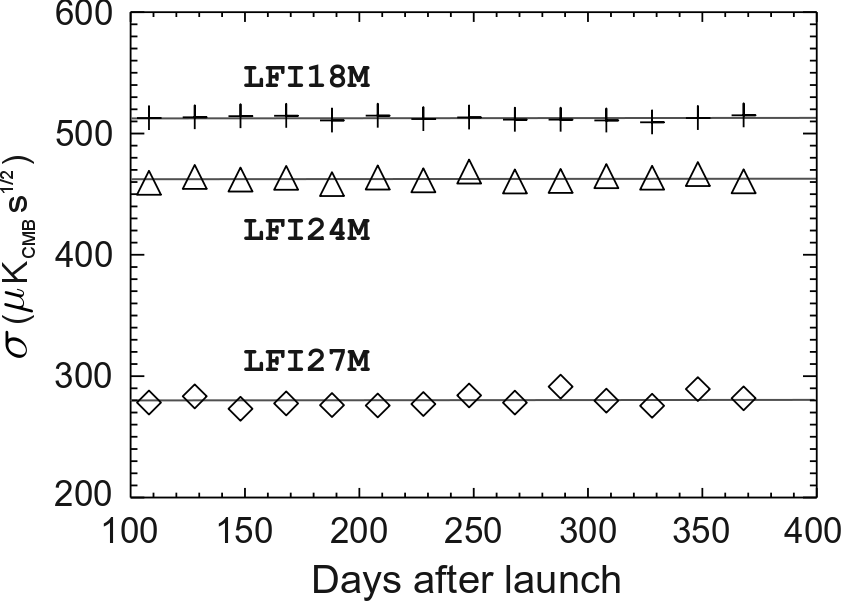}
    \end{center}
    \caption{White noise level versus time for three radiometers: \texttt{LFI18M}, \texttt{LFI24M}, and \texttt{LFI27M}.}
    \label{fig_whitenoise_vs_time}
\end{figure}

To verify the stability of the instrument, we produced noise estimates from five-day chunks of 
data separated by 20\,days each.  Figure~\ref{fig_whitenoise_vs_time} gives examples for three radiometers.  
No significant systematic variations were observed in any of the three parameters, so that 
final values and uncertainties could be obtained by taking the average and the standard deviation of the five-day values.

\begin{figure*}
        \begin{center}
            \includegraphics[width=4.5cm]{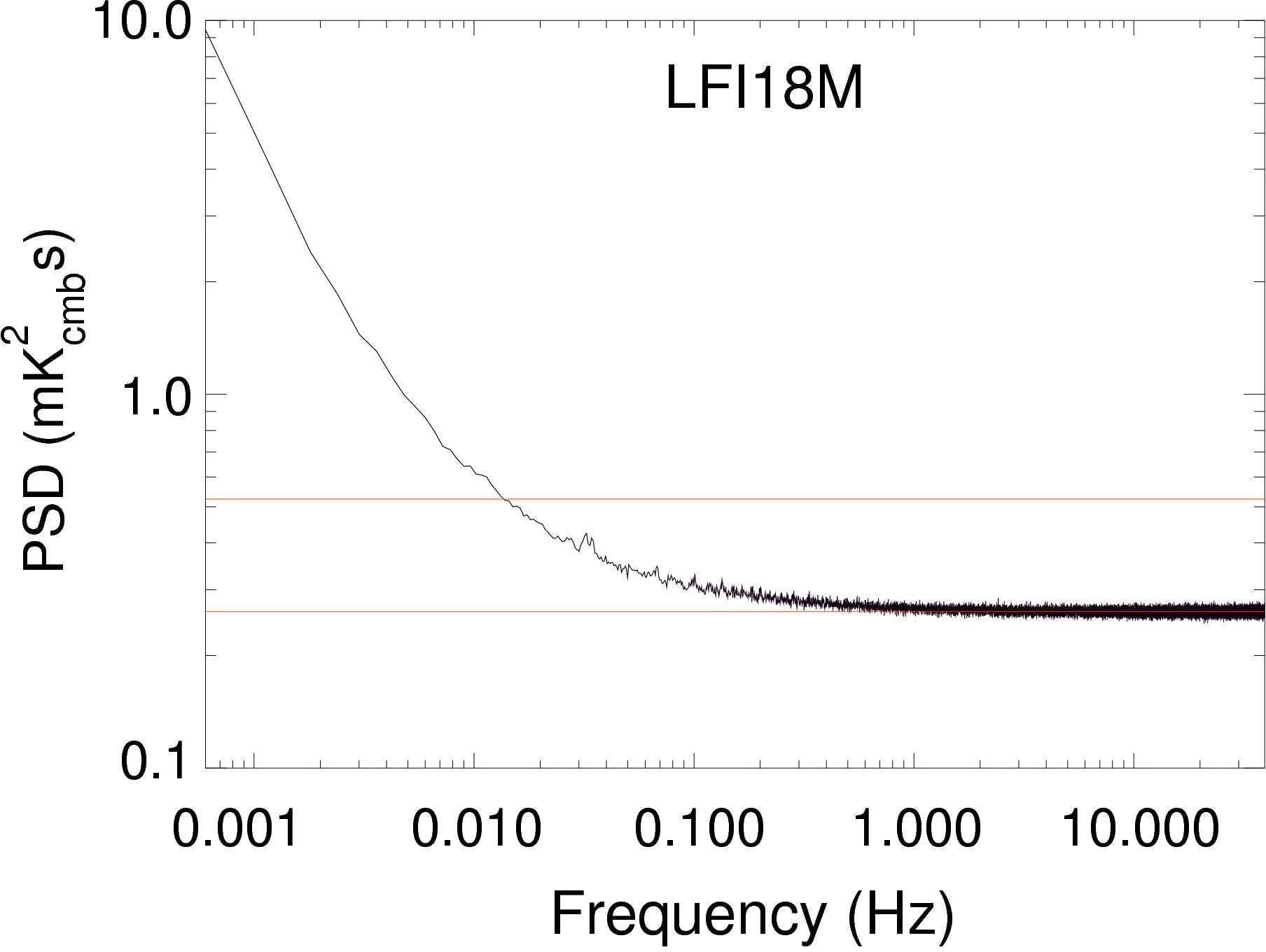}
            \includegraphics[width=4.5cm]{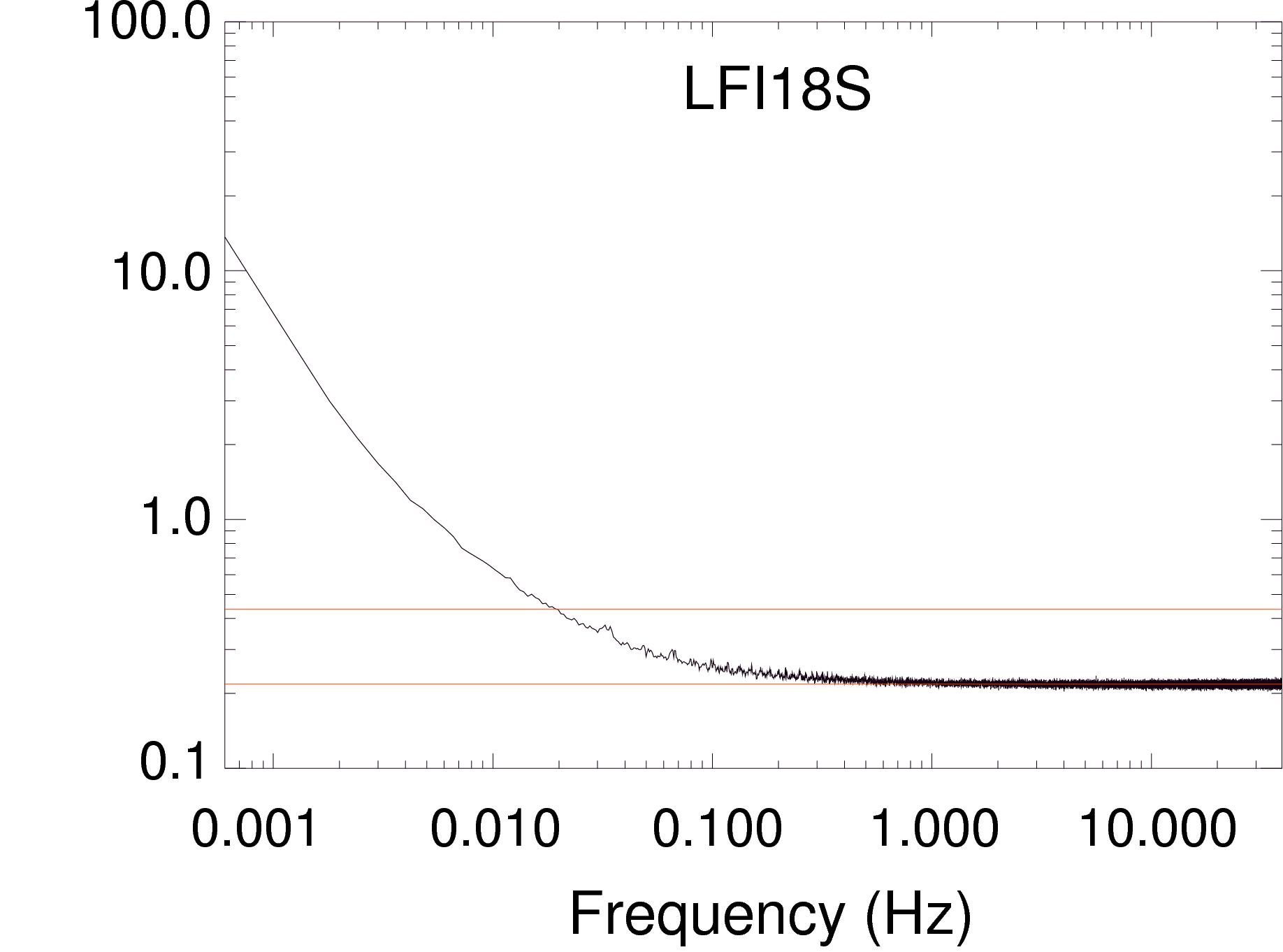}
            \includegraphics[width=4.5cm]{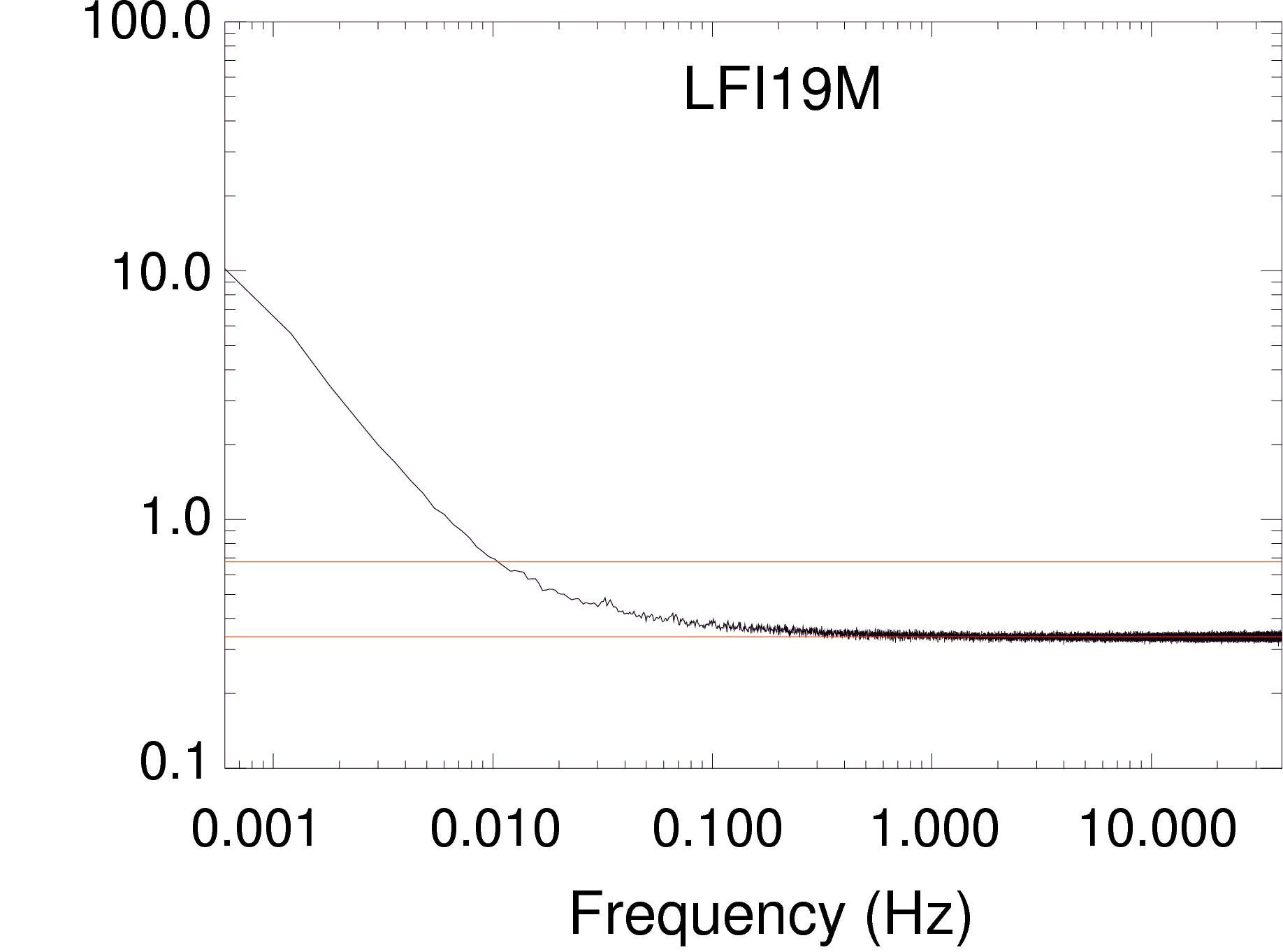}
            \includegraphics[width=4.5cm]{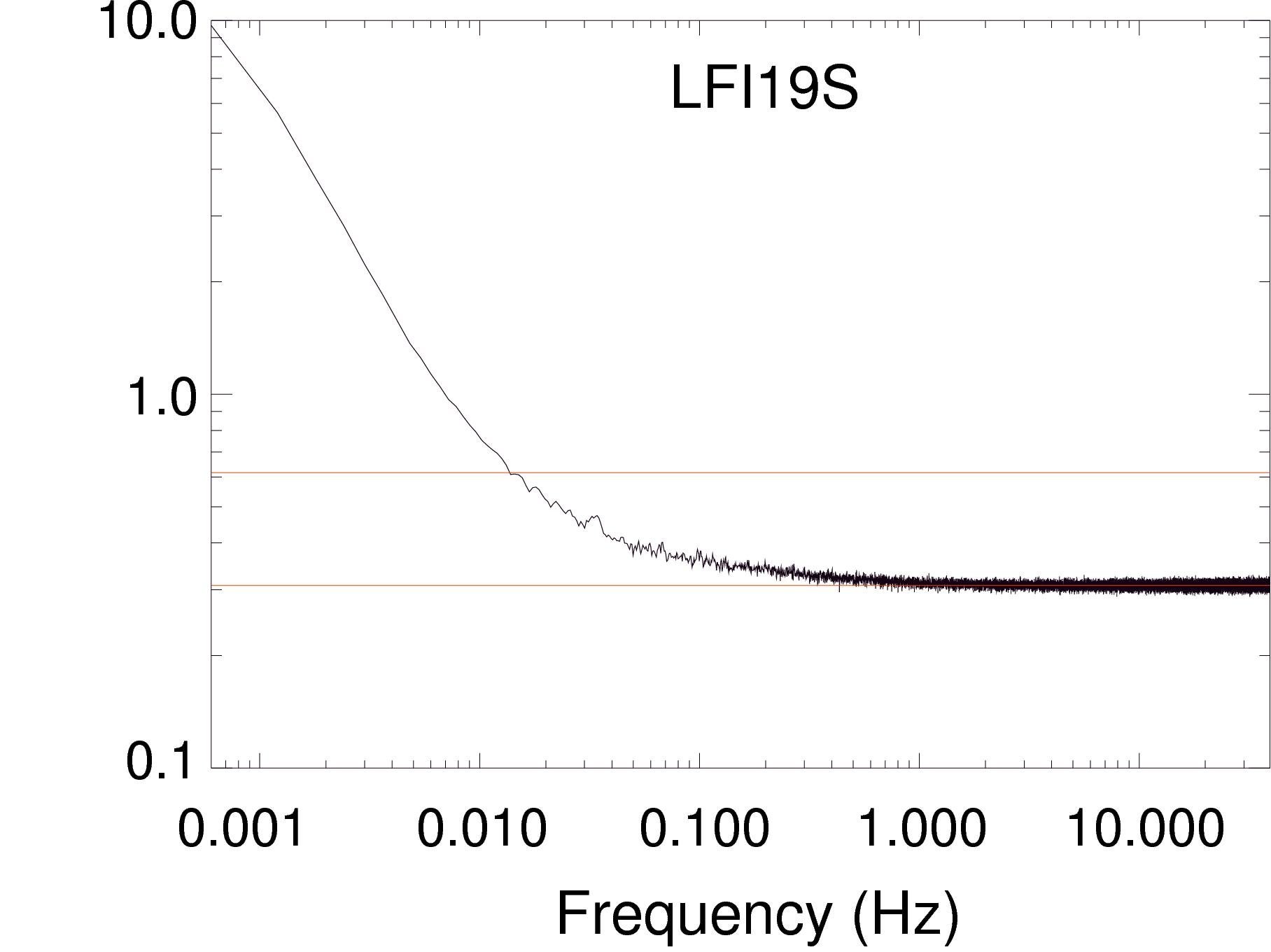}\\
            \vspace{0.3cm}
            \includegraphics[width=4.5cm]{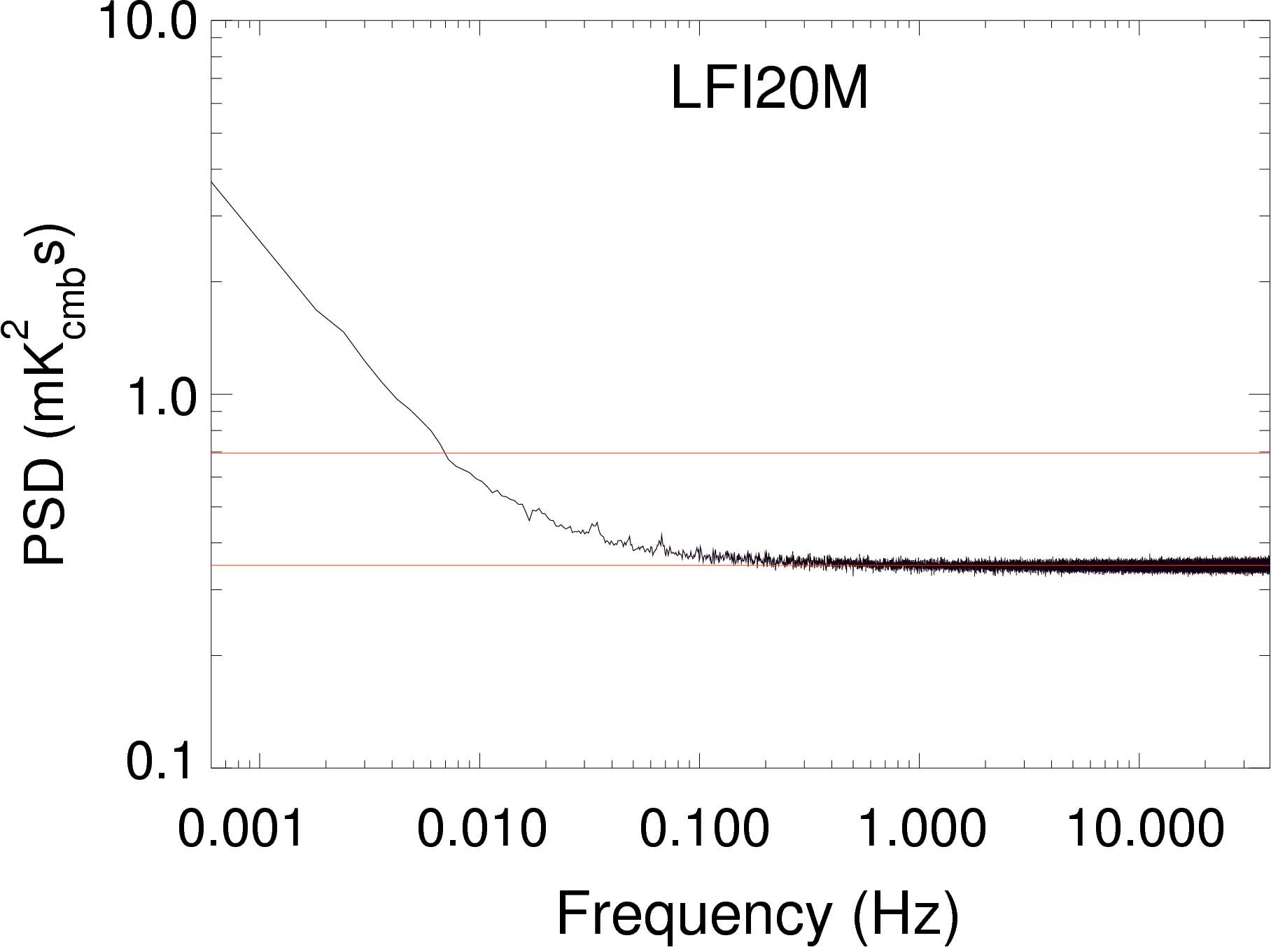}
            \includegraphics[width=4.5cm]{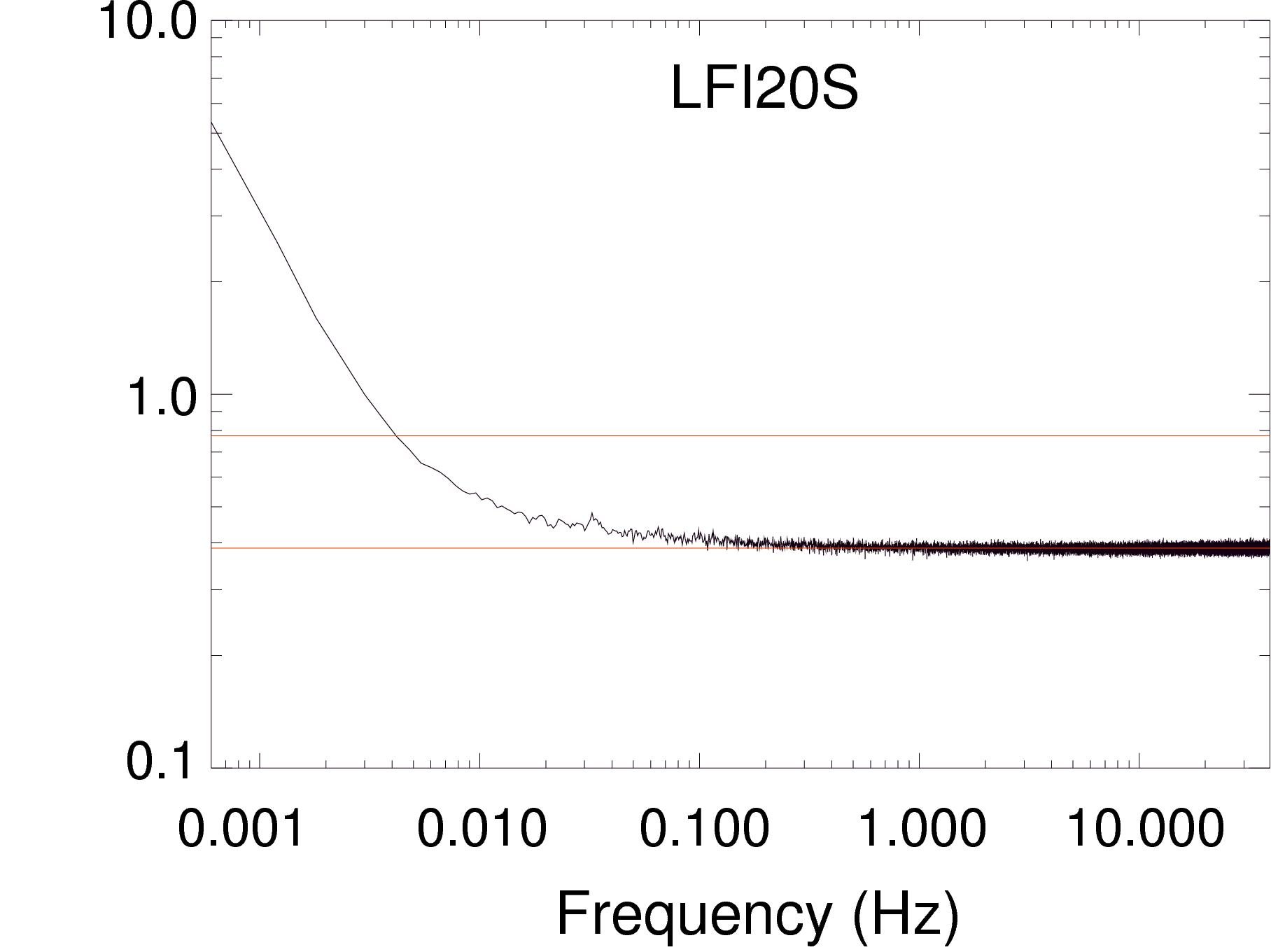}
            \includegraphics[width=4.5cm]{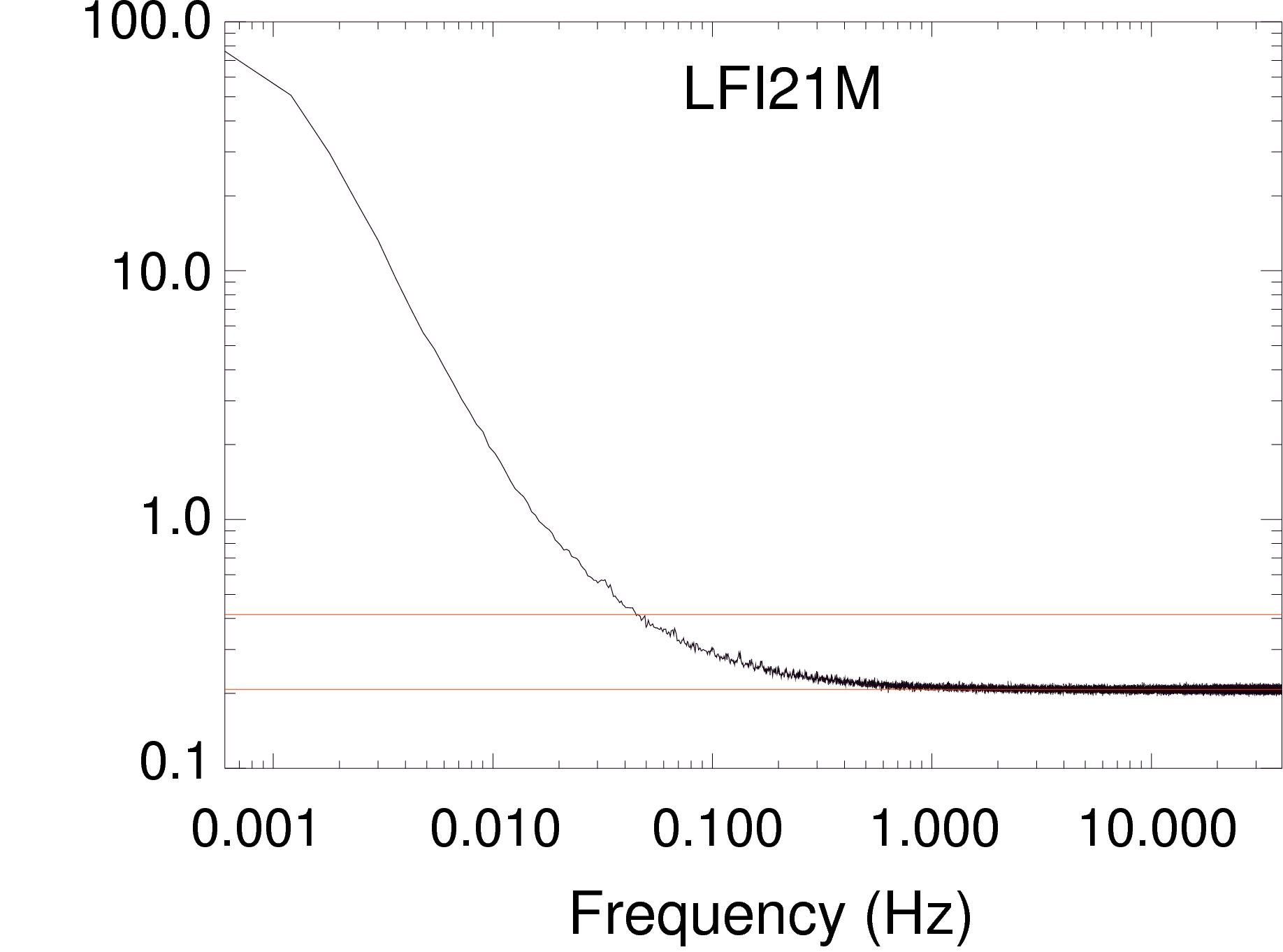}
            \includegraphics[width=4.5cm]{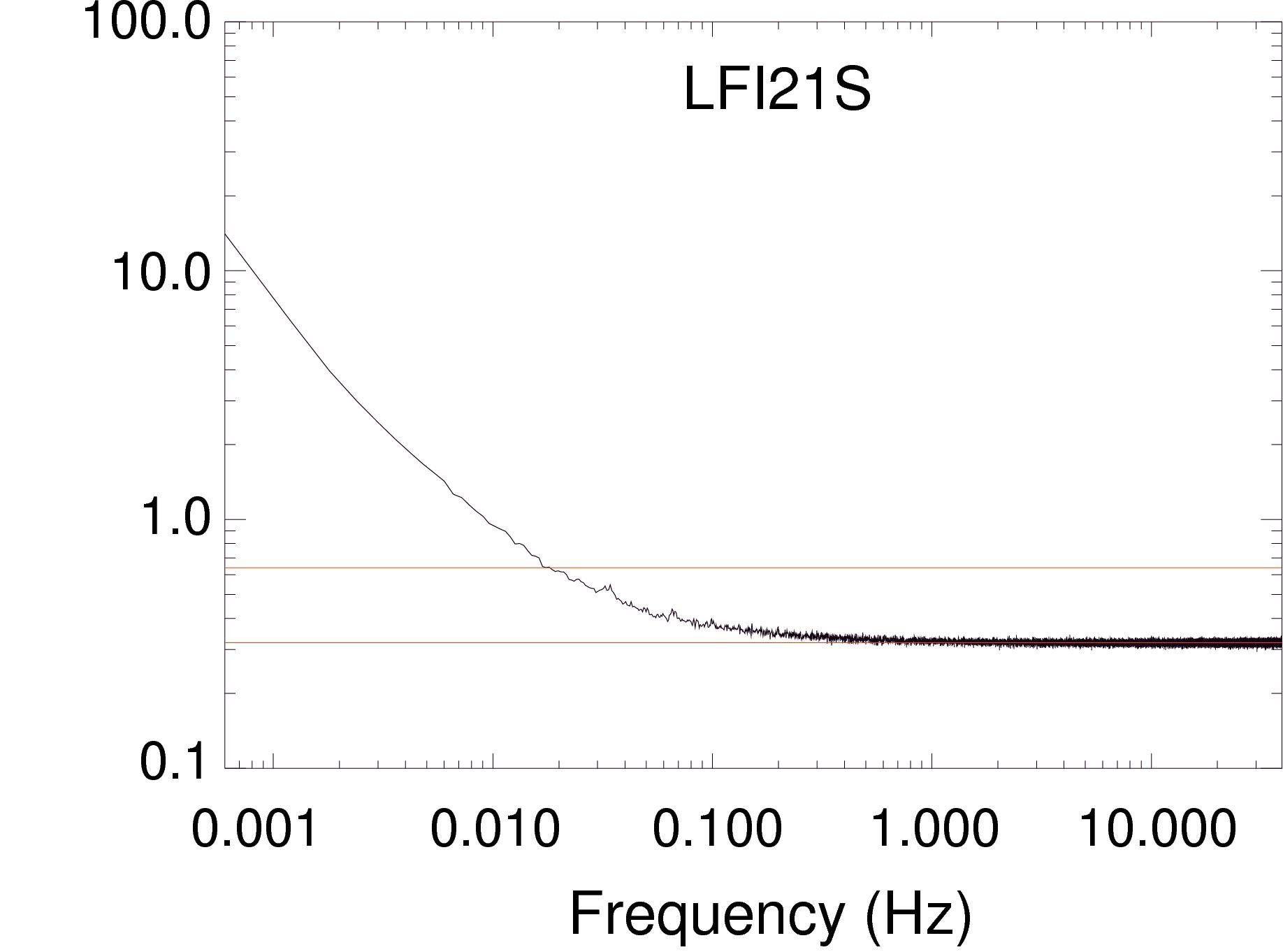}\\
            \vspace{0.3cm}
            \includegraphics[width=4.5cm]{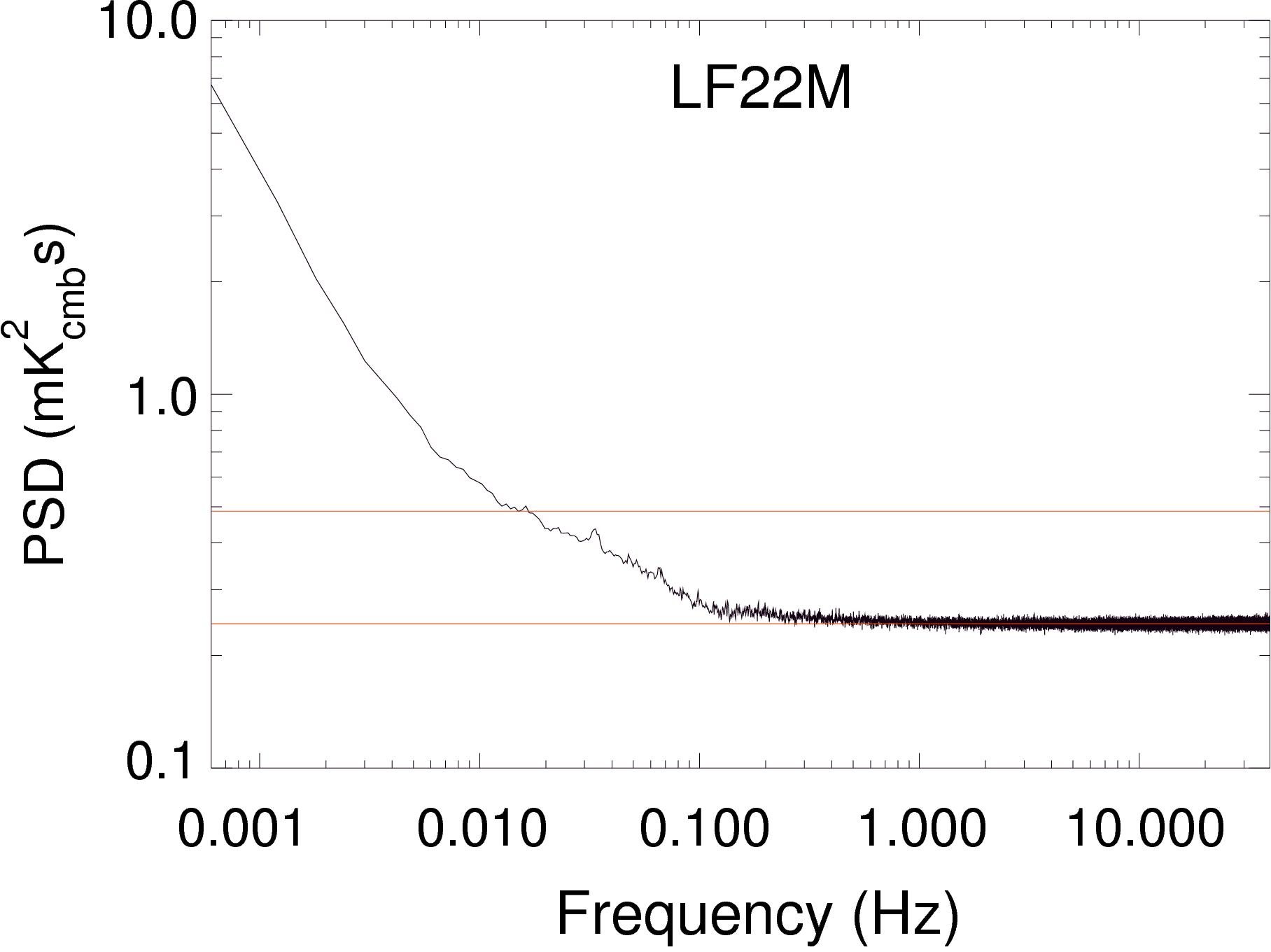}
            \includegraphics[width=4.5cm]{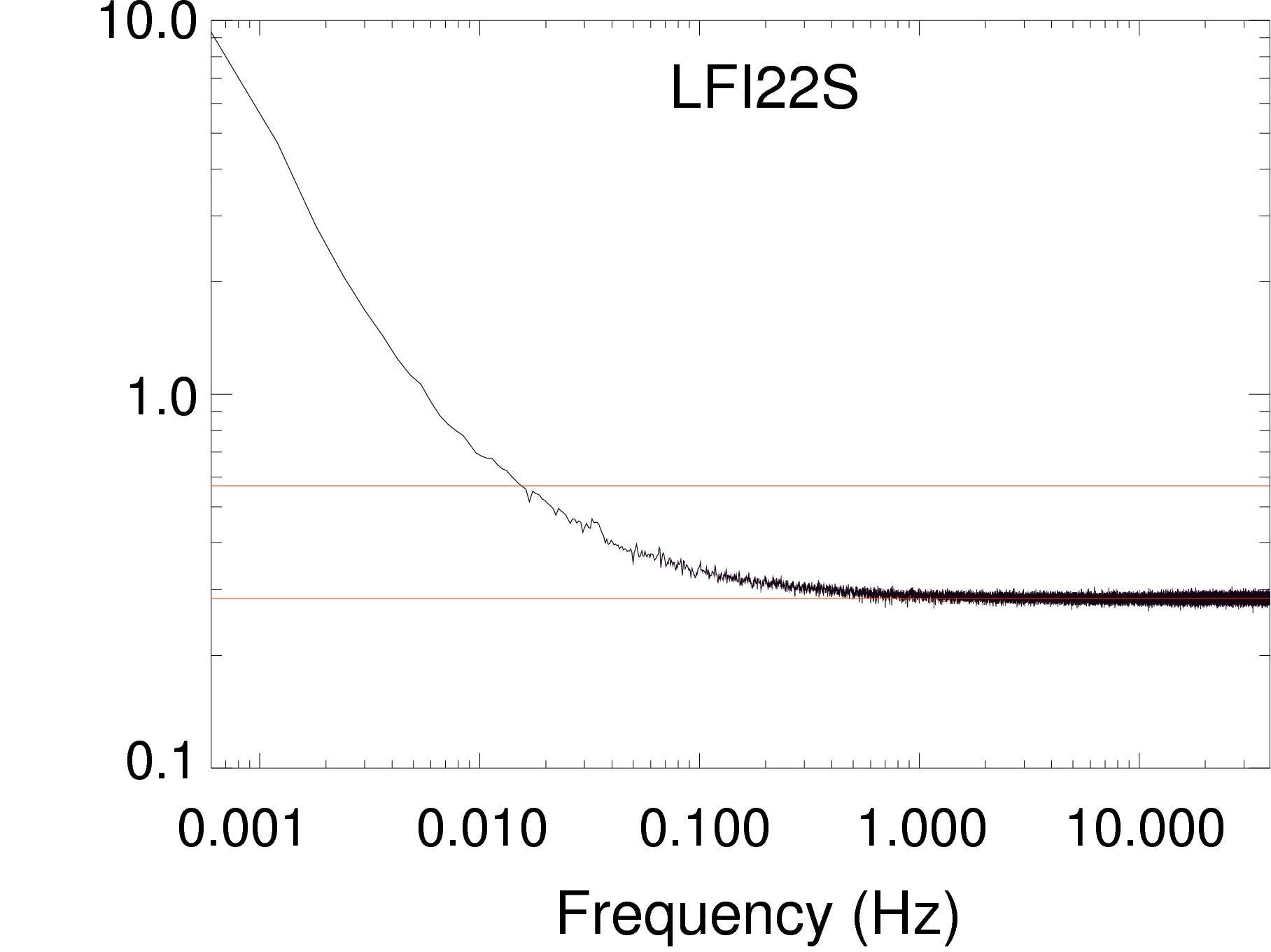}
            \includegraphics[width=4.5cm]{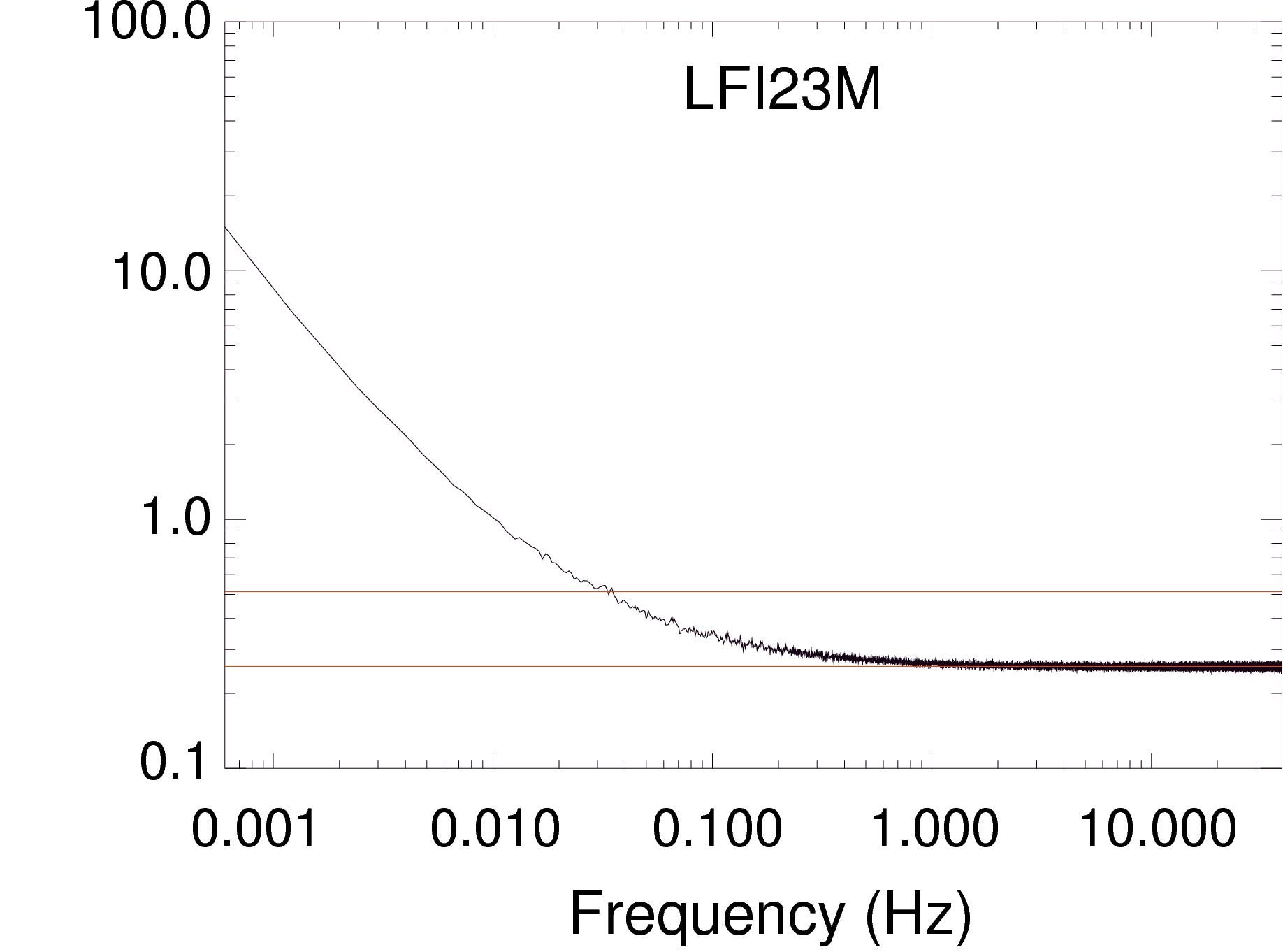}
            \includegraphics[width=4.5cm]{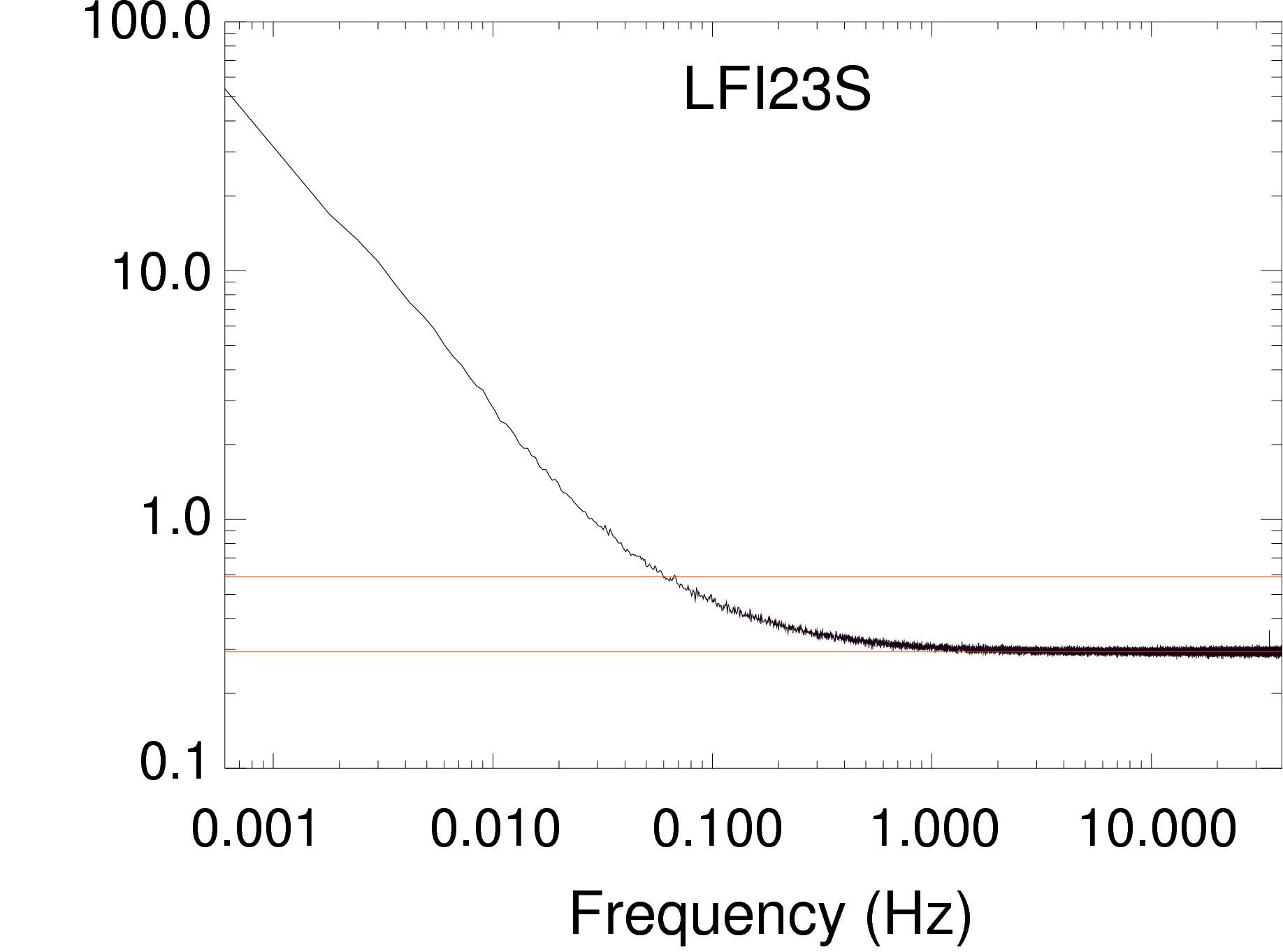}\\
            \vspace{0.3cm}
            \includegraphics[width=4.5cm]{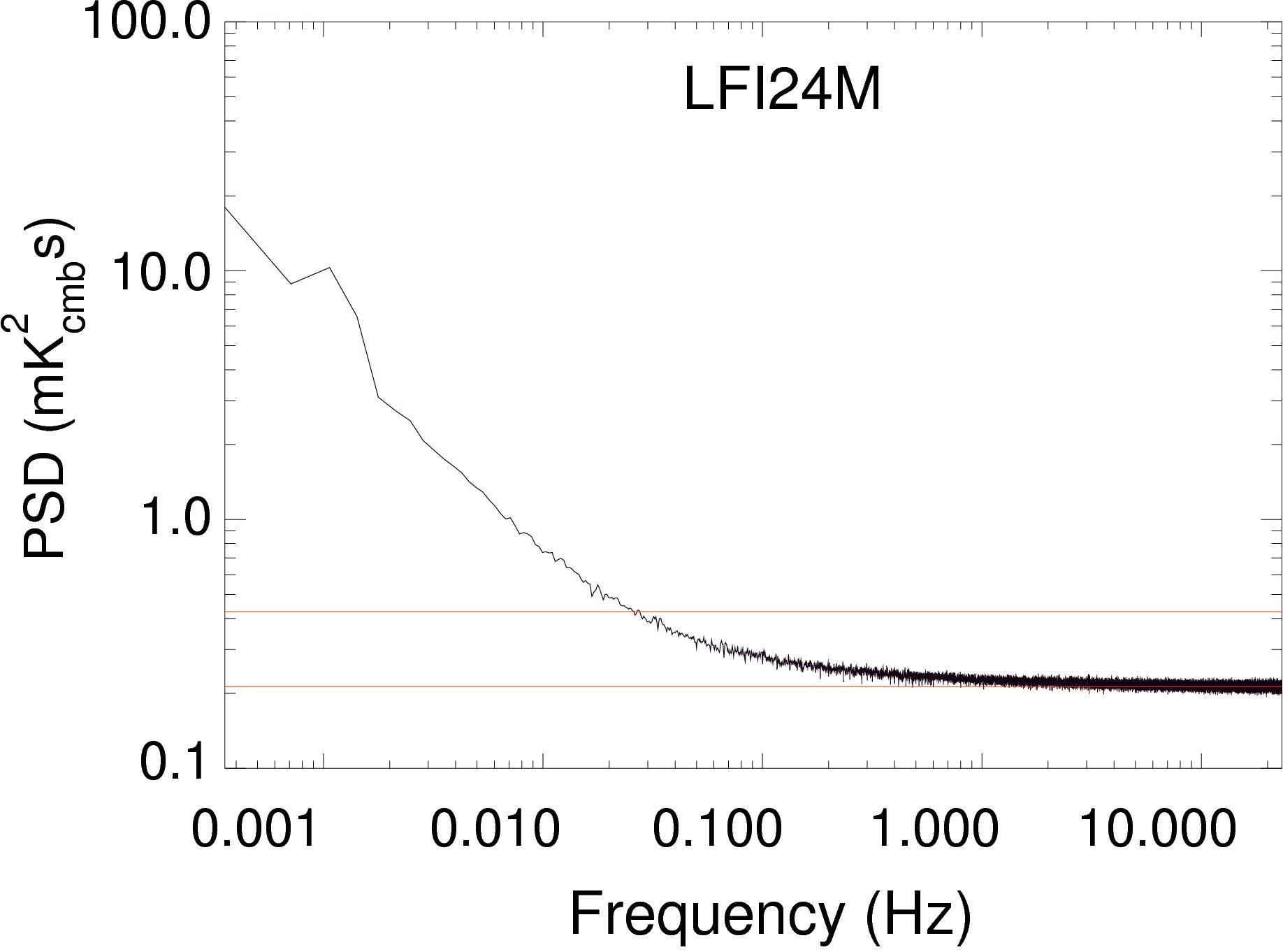}
            \includegraphics[width=4.5cm]{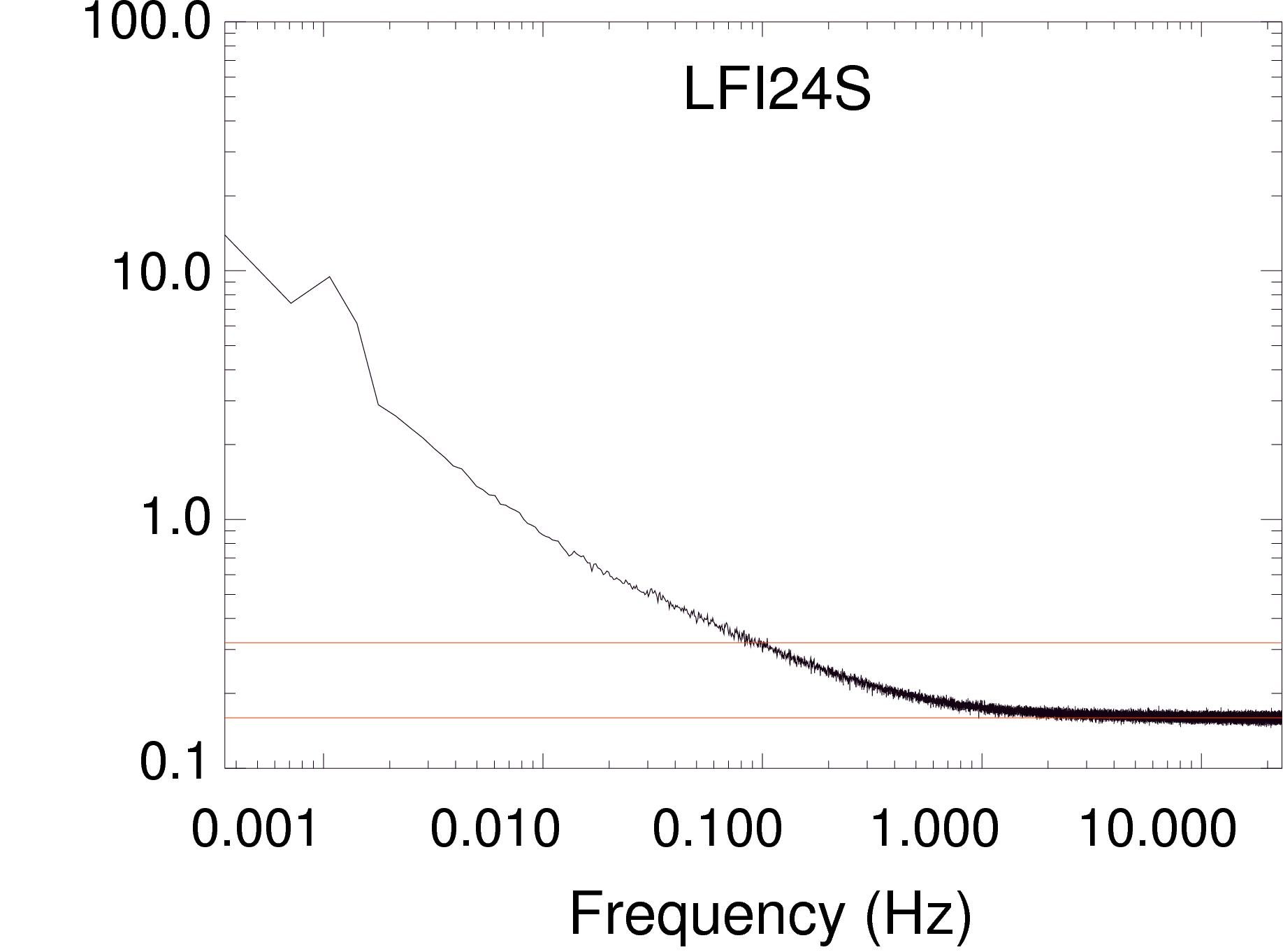}
            \includegraphics[width=4.5cm]{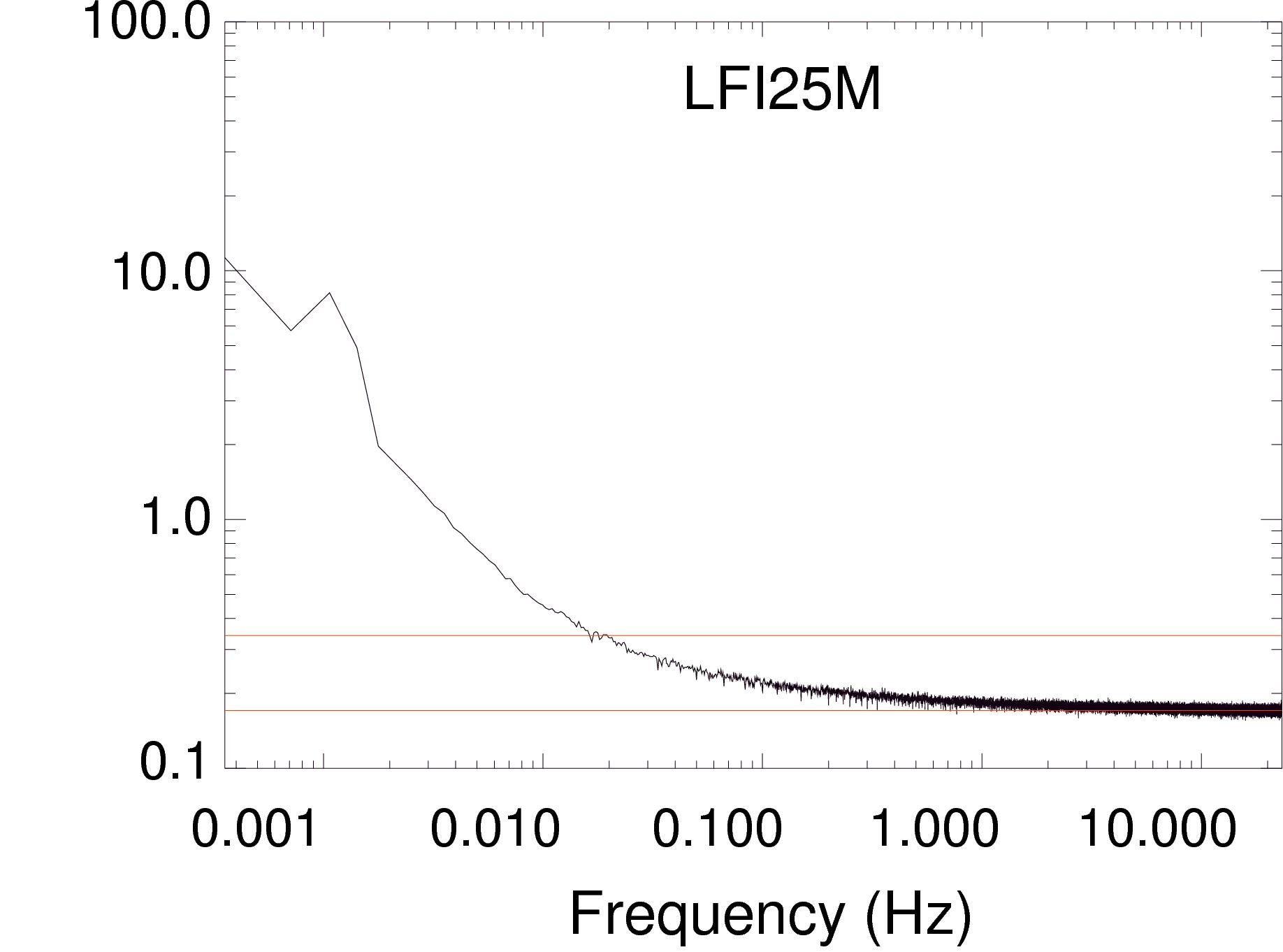}
            \includegraphics[width=4.5cm]{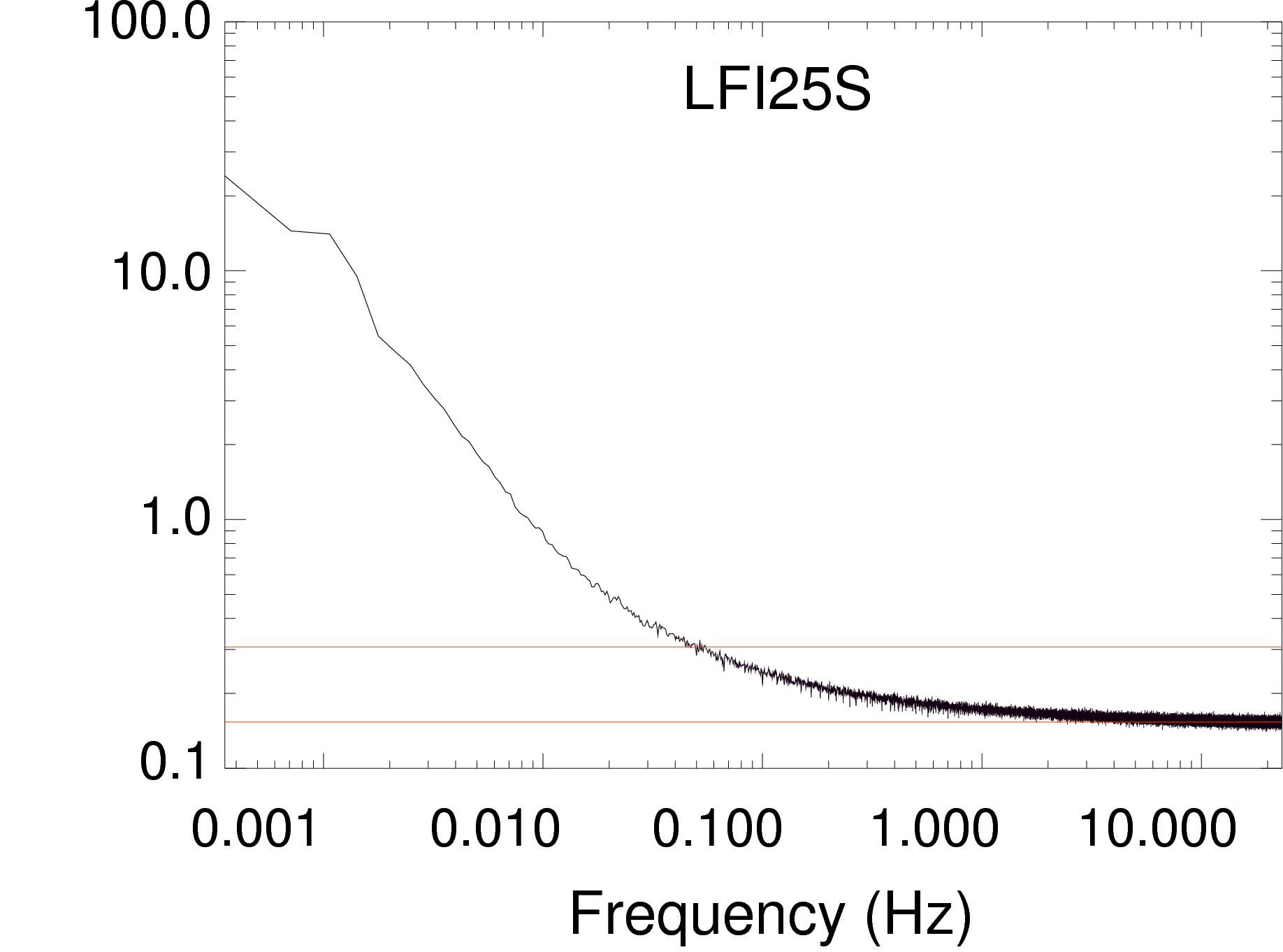}\\
            \vspace{0.3cm}
            \includegraphics[width=4.5cm]{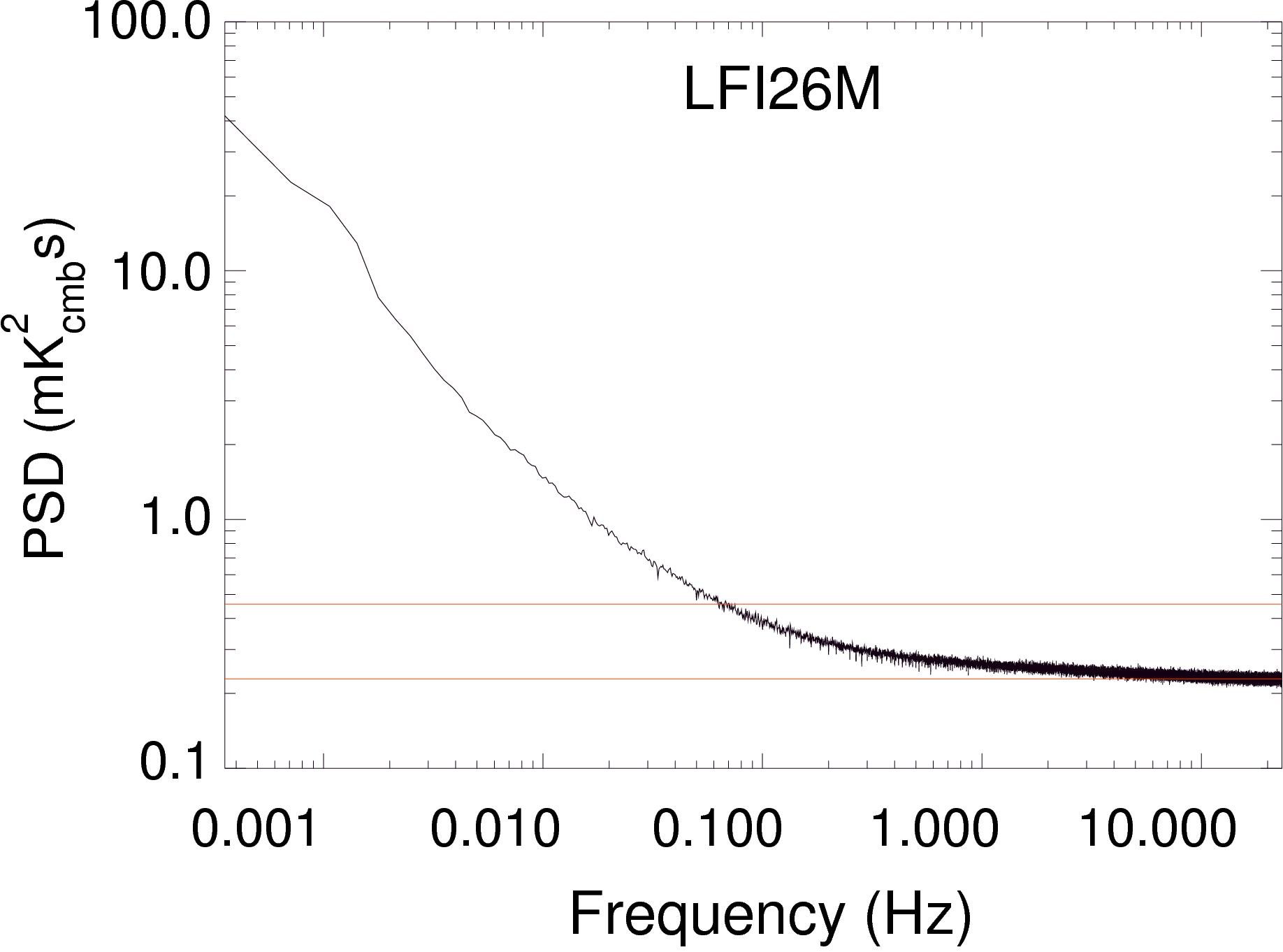}
            \includegraphics[width=4.5cm]{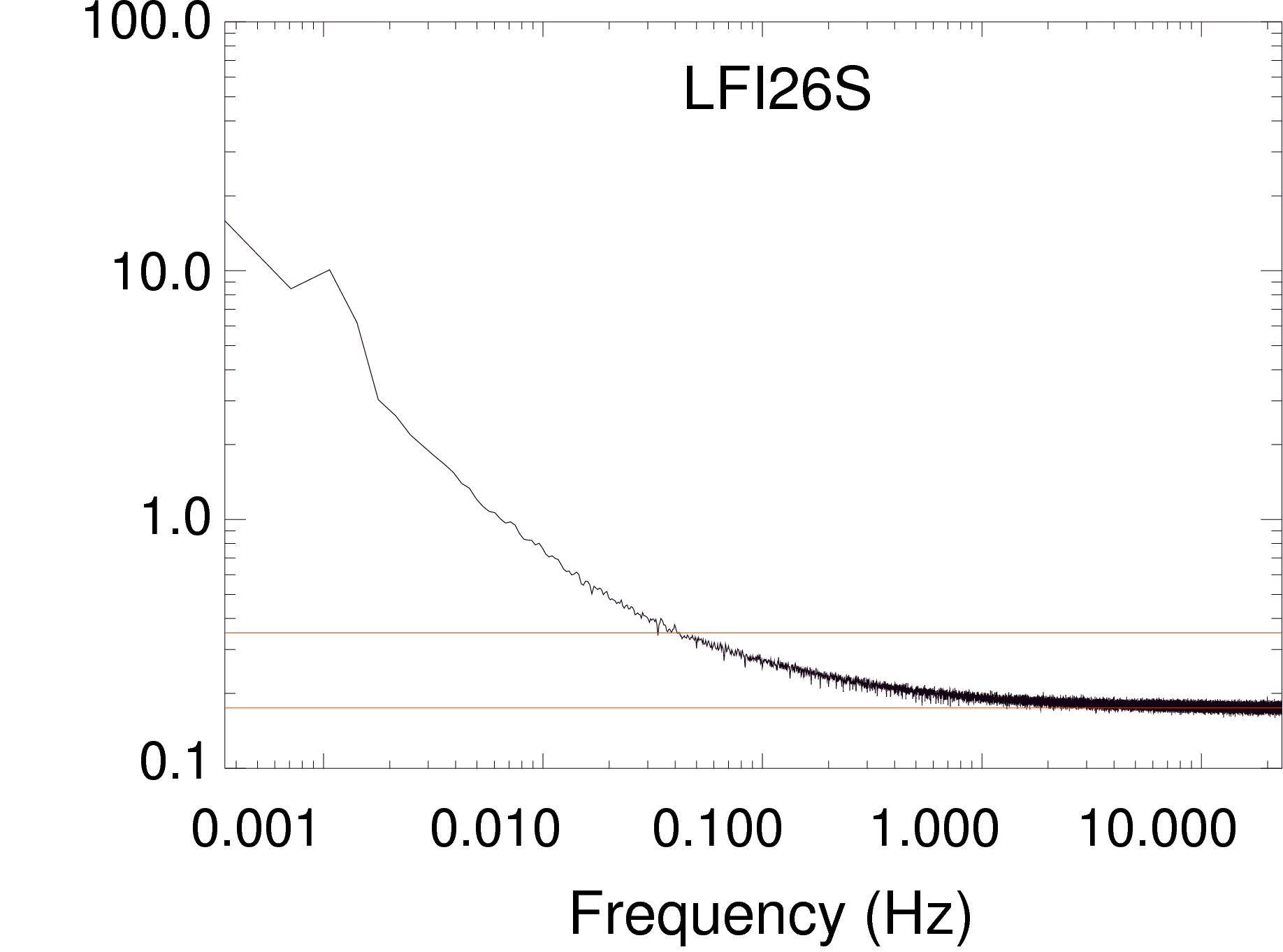}
            \includegraphics[width=4.5cm]{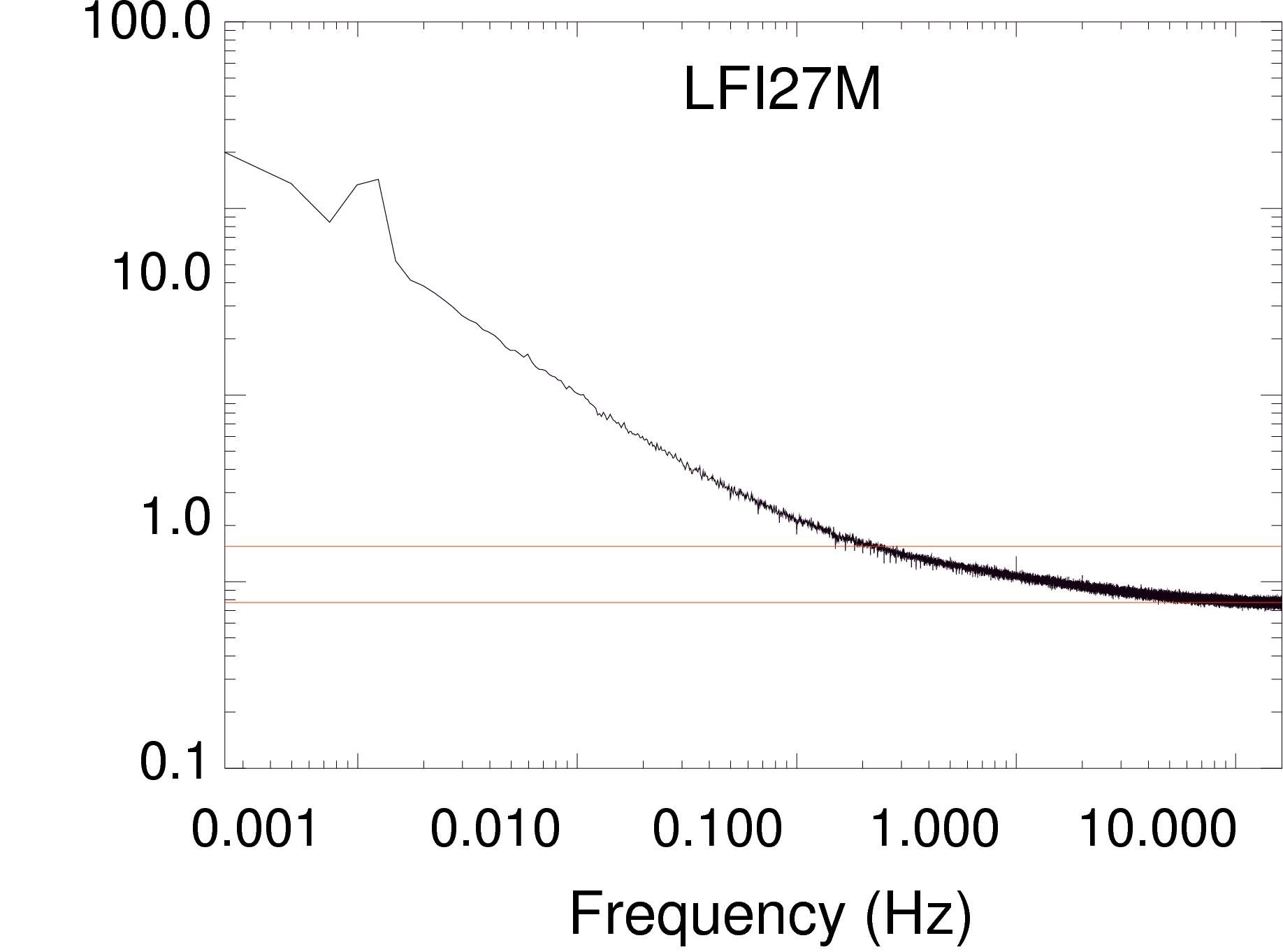}
            \includegraphics[width=4.5cm]{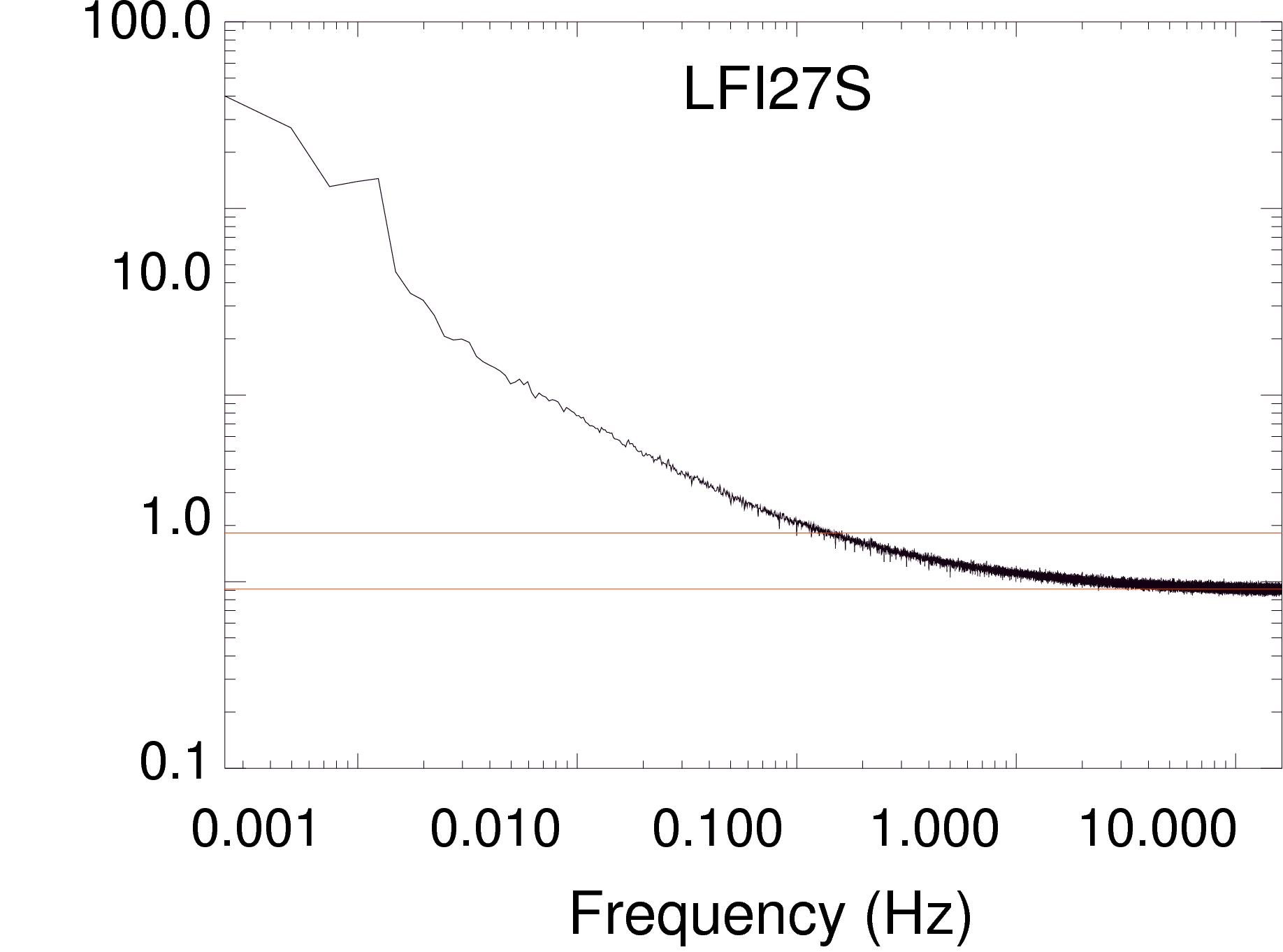}\\
            \vspace{0.3cm}
            \includegraphics[width=4.5cm]{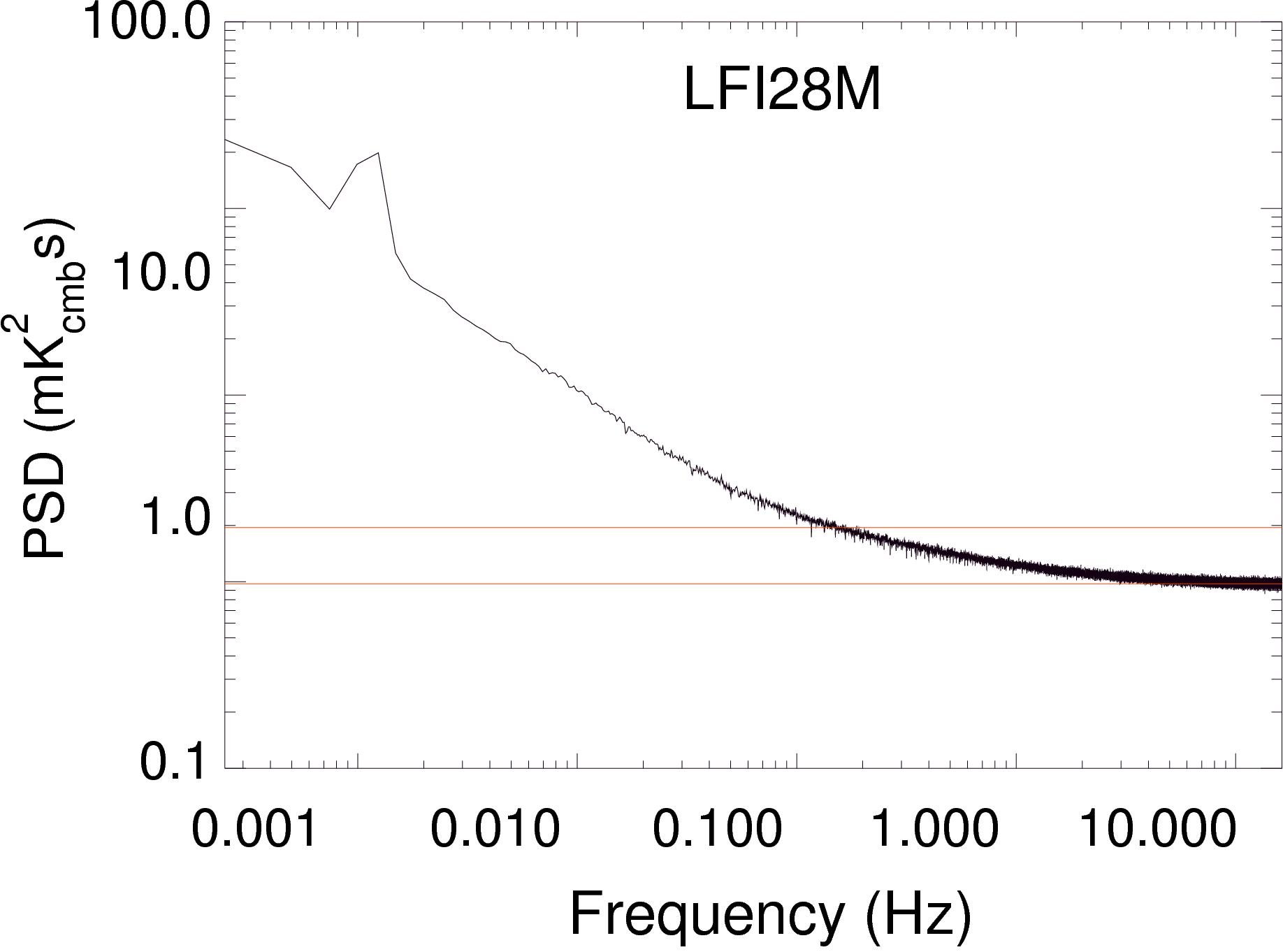}
            \includegraphics[width=4.5cm]{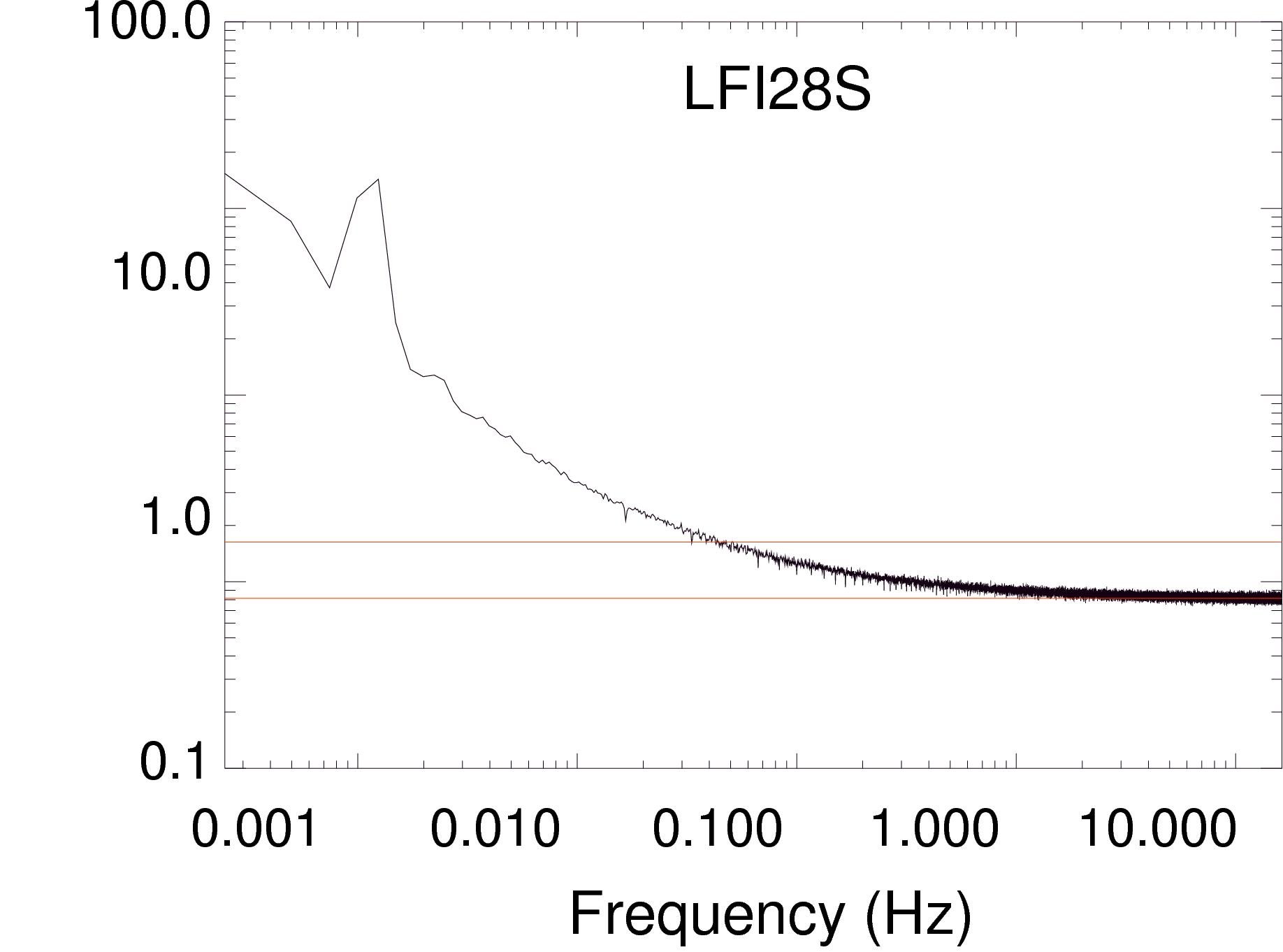}\\

            \caption{Power spectral densities (PSDs) for the LFI radiometers. The lower red line marks the white noise level.  
            The upper red line marks the level of equal contribution from white and $1/f$ noise, so the intercept 
            with the power spectrum marks the knee frequecy, $f_{\rm knee}$.}
            \label{fig_noiseps_all0}
        \end{center}
    \end{figure*}

Results for white noise $\sigma$, knee frequency $f_{\rm knee}$, and slope $\alpha$ are reported in 
Tables~\ref{tab_white_noise_per_radiometer} and \ref{tab_one_over_f_noise_per_radiometer}.   Typical uncertainties are of the 
order of $\sim 0.5\%$ for the white noise, between 5 and 10\% for the slope, and 10 and 20\% for the knee frequency.  
Comparing the noise power spectra for all the 22 LFI radiometers with their best-fit model, 
we found good agreement as shown in Fig.~\ref{fig_noiseps_all0}. In the figure, the horizontal red lines represent 
the white noise level (lower line) and the level of equal contribution from white and $1/f$ noise. 
The frequency corresponding to the intercept of the upper red line with the power spectrum is the knee frequecy $f_{\rm knee}$.

\begin{table}                    
\begingroup
\newdimen\tblskip \tblskip=5pt
\caption{White noise sensitivities for the 22 LFI radiometers.}  
\label{tab_white_noise_per_radiometer}
\nointerlineskip
\vskip -3mm
\footnotesize
\setbox\tablebox=\vbox{
   \newdimen\digitwidth 
   \setbox0=\hbox{\rm 0} 
   \digitwidth=\wd0 
   \catcode`*=\active 
   \def*{\kern\digitwidth}
   \newdimen\signwidth 
   \setbox0=\hbox{+} 
   \signwidth=\wd0 
   \catcode`!=\active 
   \def!{\kern\signwidth}
\halign{\hbox to 1.3in{#\leaderfil}\tabskip=2em&
        \hfil#\hfil&
        \hfil#\hfil\tabskip=0pt\cr 
\noalign{\doubleline}
\omit&\multispan2\hfil W{\sc hite} N{\sc oise} S{\sc ensitivity}\hfil\cr
\noalign{\vskip -4pt}
\omit&\multispan2\hrulefill\cr
\omit&Radiometer M&Radiometer S\cr
\omit\hfil RCA\hfil&[$\,\mu\mathrm{K}_{\rm CMB}\, \mathrm{s}^{1/2}$]&[$\,\mu\mathrm{K}_{\rm CMB}\, \mathrm{s}^{1/2}$]\cr
\noalign{\vskip 3pt\hrule\vskip 5pt}
\omit{\bf 70\,GHz}\hfil\cr
\noalign{\vskip 4pt}
\hglue 2em LFI18   &\getsymbol{LFI:white:noise:sensitivity:LFI18:Rad:M}$\,\pm 1.8$ &\getsymbol{LFI:white:noise:sensitivity:LFI18:Rad:S}$\,\pm 1.7$\cr
\hglue 2em LFI19   &\getsymbol{LFI:white:noise:sensitivity:LFI19:Rad:M}$\,\pm 1.6$ &\getsymbol{LFI:white:noise:sensitivity:LFI19:Rad:S}$\,\pm 2.4$\cr
\hglue 2em LFI20   &\getsymbol{LFI:white:noise:sensitivity:LFI20:Rad:M}$\,\pm 1.7$ &\getsymbol{LFI:white:noise:sensitivity:LFI20:Rad:S}$\,\pm 1.5$\cr
\hglue 2em LFI21   &\getsymbol{LFI:white:noise:sensitivity:LFI21:Rad:M}$\,\pm 2.8$ &\getsymbol{LFI:white:noise:sensitivity:LFI21:Rad:S}$\,\pm 6.2$\cr
\hglue 2em LFI22   &\getsymbol{LFI:white:noise:sensitivity:LFI22:Rad:M}$\,\pm 2.3$ &\getsymbol{LFI:white:noise:sensitivity:LFI22:Rad:S}$\,\pm 2.8$\cr
\hglue 2em LFI23   &\getsymbol{LFI:white:noise:sensitivity:LFI23:Rad:M}$\,\pm 1.7$ &\getsymbol{LFI:white:noise:sensitivity:LFI23:Rad:S}$\,\pm 2.6$\cr
\noalign{\vskip 5pt}
\omit{\bf 44\,GHz}\hfil\cr
\noalign{\vskip 4pt}
\hglue 2em LFI24   &\getsymbol{LFI:white:noise:sensitivity:LFI24:Rad:M}$\,\pm 1.9$ &\getsymbol{LFI:white:noise:sensitivity:LFI24:Rad:S}$\,\pm 1.3$\cr
\hglue 2em LFI25   &\getsymbol{LFI:white:noise:sensitivity:LFI25:Rad:M}$\,\pm 3.6$ &\getsymbol{LFI:white:noise:sensitivity:LFI25:Rad:S}$\,\pm 1.6$\cr
\hglue 2em LFI26   &\getsymbol{LFI:white:noise:sensitivity:LFI26:Rad:M}$\,\pm 3.1$ &\getsymbol{LFI:white:noise:sensitivity:LFI26:Rad:S}$\,\pm 4.2$\cr
\noalign{\vskip 5pt}
\omit{\bf 30\,GHz}\hfil\cr
\noalign{\vskip 4pt}
\hglue 2em LFI27   &\getsymbol{LFI:white:noise:sensitivity:LFI27:Rad:M}$\,\pm 2.1$ &\getsymbol{LFI:white:noise:sensitivity:LFI27:Rad:S}$\,\pm 1.6$\cr
\hglue 2em LFI28   &\getsymbol{LFI:white:noise:sensitivity:LFI28:Rad:M}$\,\pm 1.7$ &\getsymbol{LFI:white:noise:sensitivity:LFI28:Rad:S}$\,\pm 1.4$\cr
\noalign{\vskip 5pt\hrule\vskip 3pt}}}
\endPlancktable
\endgroup
\end{table}

\begin{table*}                    
\begingroup
\newdimen\tblskip \tblskip=5pt
\caption{Knee frequency and slope for the 22 LFI radiometers.}
\label{tab_one_over_f_noise_per_radiometer}
\nointerlineskip
\vskip -3mm
\footnotesize
\setbox\tablebox=\vbox{
   \newdimen\digitwidth 
   \setbox0=\hbox{\rm 0} 
   \digitwidth=\wd0 
   \catcode`*=\active 
   \def*{\kern\digitwidth}
   \newdimen\signwidth 
   \setbox0=\hbox{+} 
   \signwidth=\wd0 
   \catcode`!=\active 
   \def!{\kern\signwidth}
\halign{\hbox to 1.3in{#\leaderfil}\tabskip=2em&
        \hfil#\hfil&
        \hfil#\hfil&
        \hfil#\hfil&
        \hfil#\hfil\tabskip=0pt\cr 
\noalign{\doubleline}
\omit&\multispan2\hfil K{\sc nee} F{\sc requency}\hfil&\multispan2\hfil S{\sc lope}\hfil\cr
\noalign{\vskip -4pt}
\omit&\multispan2\hrulefill&\multispan2\hrulefill\cr
\omit&Radiometer M&Radiometer S&Radiometer M&Radiometer S\cr
\omit\hfil RCA\hfil&[mHz]&[mHz]\cr
\noalign{\vskip 3pt\hrule\vskip 5pt}
\omit{\bf 70\,GHz}\hfil\cr
\noalign{\vskip 4pt}
\hglue 2em LFI18   &*\getsymbol{LFI:knee:frequency:LFI18:Rad:M} $\pm *3.9$& *\getsymbol{LFI:knee:frequency:LFI18:Rad:S} $\pm *2.5$&\getsymbol{LFI:slope:LFI18:Rad:M} $\pm$ 0.08&\getsymbol{LFI:slope:LFI18:Rad:S} $\pm$ 0.07\cr
\hglue 2em LFI19   &*\getsymbol{LFI:knee:frequency:LFI19:Rad:M} $\pm *2.2$& *\getsymbol{LFI:knee:frequency:LFI19:Rad:S} $\pm *5.0$&\getsymbol{LFI:slope:LFI19:Rad:M} $\pm$ 0.05&\getsymbol{LFI:slope:LFI19:Rad:S} $\pm$ 0.08\cr
\hglue 2em LFI20   &*\getsymbol{LFI:knee:frequency:LFI20:Rad:M} $\pm *5.3$&**\getsymbol{LFI:knee:frequency:LFI20:Rad:S} $\pm *1.6$&\getsymbol{LFI:slope:LFI20:Rad:M} $\pm$ 0.07&\getsymbol{LFI:slope:LFI20:Rad:S} $\pm$ 0.09\cr
\hglue 2em LFI21   &*\getsymbol{LFI:knee:frequency:LFI21:Rad:M} $\pm *6.1$& *\getsymbol{LFI:knee:frequency:LFI21:Rad:S} $\pm *7.9$&\getsymbol{LFI:slope:LFI21:Rad:M} $\pm$ 0.06&\getsymbol{LFI:slope:LFI21:Rad:S} $\pm$ 0.12\cr
\hglue 2em LFI22   &*\getsymbol{LFI:knee:frequency:LFI22:Rad:M} $\pm 19.7$& *\getsymbol{LFI:knee:frequency:LFI22:Rad:S} $\pm *6.2$&\getsymbol{LFI:slope:LFI22:Rad:M} $\pm$ 0.15&\getsymbol{LFI:slope:LFI22:Rad:S} $\pm$ 0.18\cr
\hglue 2em LFI23   &*\getsymbol{LFI:knee:frequency:LFI23:Rad:M} $\pm *4.8$&  \getsymbol{LFI:knee:frequency:LFI23:Rad:S} $\pm *6.9$&\getsymbol{LFI:slope:LFI23:Rad:M} $\pm$ 0.04&\getsymbol{LFI:slope:LFI23:Rad:S} $\pm$ 0.06\cr
\noalign{\vskip 5pt}
\omit{\bf 44\,GHz}\hfil\cr
\noalign{\vskip 4pt}
\hglue 2em LFI24   &*\getsymbol{LFI:knee:frequency:LFI24:Rad:M} $\pm*3.9$&  \getsymbol{LFI:knee:frequency:LFI24:Rad:S} $\pm *8.3$&\getsymbol{LFI:slope:LFI24:Rad:M} $\pm$ 0.04&\getsymbol{LFI:slope:LFI24:Rad:S} $\pm$ 0.03\cr
\hglue 2em LFI25   &*\getsymbol{LFI:knee:frequency:LFI25:Rad:M} $\pm*3.7$& *\getsymbol{LFI:knee:frequency:LFI25:Rad:S} $\pm *1.6$&\getsymbol{LFI:slope:LFI25:Rad:M} $\pm$ 0.01&\getsymbol{LFI:slope:LFI25:Rad:S} $\pm$ 0.01\cr
\hglue 2em LFI26   &*\getsymbol{LFI:knee:frequency:LFI26:Rad:M} $\pm*2.2$& *\getsymbol{LFI:knee:frequency:LFI26:Rad:S} $\pm *4.4$&\getsymbol{LFI:slope:LFI26:Rad:M} $\pm$ 0.01&\getsymbol{LFI:slope:LFI26:Rad:S} $\pm$ 0.02\cr
\noalign{\vskip 5pt}
\omit{\bf 30\,GHz}\hfil\cr
\noalign{\vskip 4pt}
\hglue 2em LFI27   &\getsymbol{LFI:knee:frequency:LFI27:Rad:M} $\pm 29.5$&  \getsymbol{LFI:knee:frequency:LFI27:Rad:S} $\pm 26.1$&\getsymbol{LFI:slope:LFI27:Rad:M} $\pm$ 0.05&\getsymbol{LFI:slope:LFI27:Rad:S} $\pm$ 0.14\cr
\hglue 2em LFI28   &\getsymbol{LFI:knee:frequency:LFI28:Rad:M} $\pm 10.2$& *\getsymbol{LFI:knee:frequency:LFI28:Rad:S} $\pm *7.2$&\getsymbol{LFI:slope:LFI28:Rad:M} $\pm$ 0.05&\getsymbol{LFI:slope:LFI28:Rad:S} $\pm$ 0.13\cr
\noalign{\vskip 5pt\hrule\vskip 3pt}}}
\endPlancktable
\endgroup
\end{table*}

\subsubsection{Jackknife noise maps}
\label{sec_jacknoise}

\begin{figure*}
            \begin{center}
                \includegraphics[width=18cm]{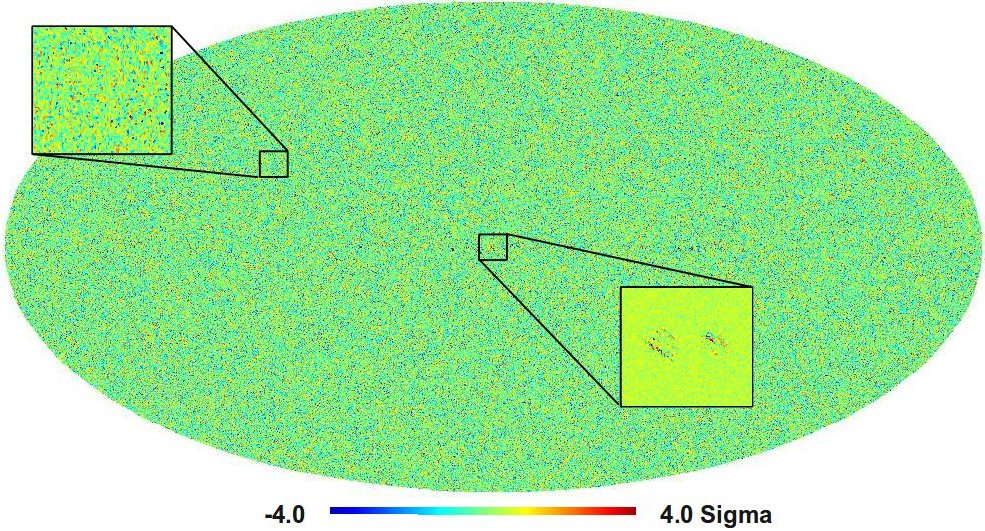}
            \end{center}
            \caption{30\,GHz normalised jackknife noise map. We also highlight a region in the Galactic plane
            affected by leakage from strong foreground emission. The inset at high
            galactic latitude highlights the white nature of the noise.}
            \label{fig_jacknoise_map}
        \end{figure*}

\begin{figure}
            \begin{center}
                \includegraphics[width=9cm]{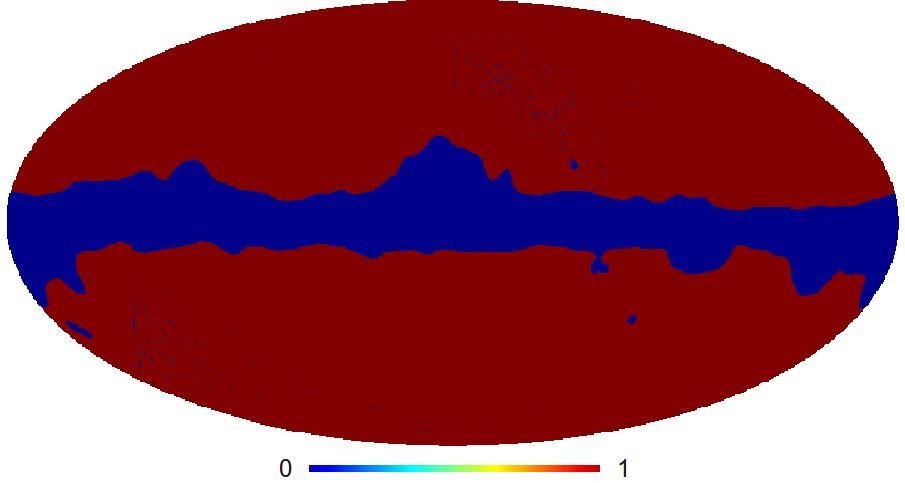}
            \end{center}
            \caption{30 GHz mask used to extract noise parameters from the corresponding noise map (masked points are shown in blue).}
            \label{fig_mask_30GHz}
        \end{figure}

During each pointing period \Planck\ repeatedly scans essentially the same circle in the sky.   This provides 
a powerful method to remove the sky signal and produce maps containing only the instrumental noise 
(both uncorrelated and correlated on timescales shorter than about 20 minutes).  The details of 
this method are given in \citet{planck2011-1.6}.  These ``jackknife'' 
noise maps can be normalised to the white noise estimate at each pixel obtained from the white 
noise covariance matrix, so that a perfectly white noise map would be Gaussian and isotropic with 
unit variance.  Figure~\ref{fig_jacknoise_map} shows an example of such normalised noise maps 
for the 30\,GHz LFI frequency channel.  The figure shows a generally structureless 
map, as expected, apart from a few regions in the Galactic plane where the temperature maps have 
large gradients over the pixel scale, causing the sky signal to leak into the noise map.  
In subsequent analysis steps we applied a $\sim 20\%$ mask to remove these 
regions (see Fig.~\ref{fig_mask_30GHz}).  Table~\ref{tab_noise_maps_rms} gives the standard 
deviation of normalised noise maps obtained for the three LFI frequencies.  Deviations from unity trace 
the contribution of residual $1/f$ noise in the 
final maps, which ranges from $0.2\%$ at 70\,GHz to $4\%$ at 30\,GHz.  A further Gaussianity test performed 
by calculating skewness and kurtosis of the normalised map after masking the galactic plane yielded null values within 
two standard deviations of 500 Gaussian simulations.

        \begin{table}
            \caption{Standard deviation of normalised noise maps.}
            \label{tab_noise_maps_rms}
            \begin{center}
                \begin{tabular}{c c c}
                    \hline
                    \hline
                    30 GHz & 44 GHz & 70 GHz \\
                    \hline
                    1.039  & 1.016  & 1.002 \\
                    \hline
                \end{tabular}
            \end{center}
        \end{table}

\subsection{Sensitivity}
\label{sec_sensitivity}

\begin{figure*} 
    \begin{center}
            \includegraphics[width=9cm]{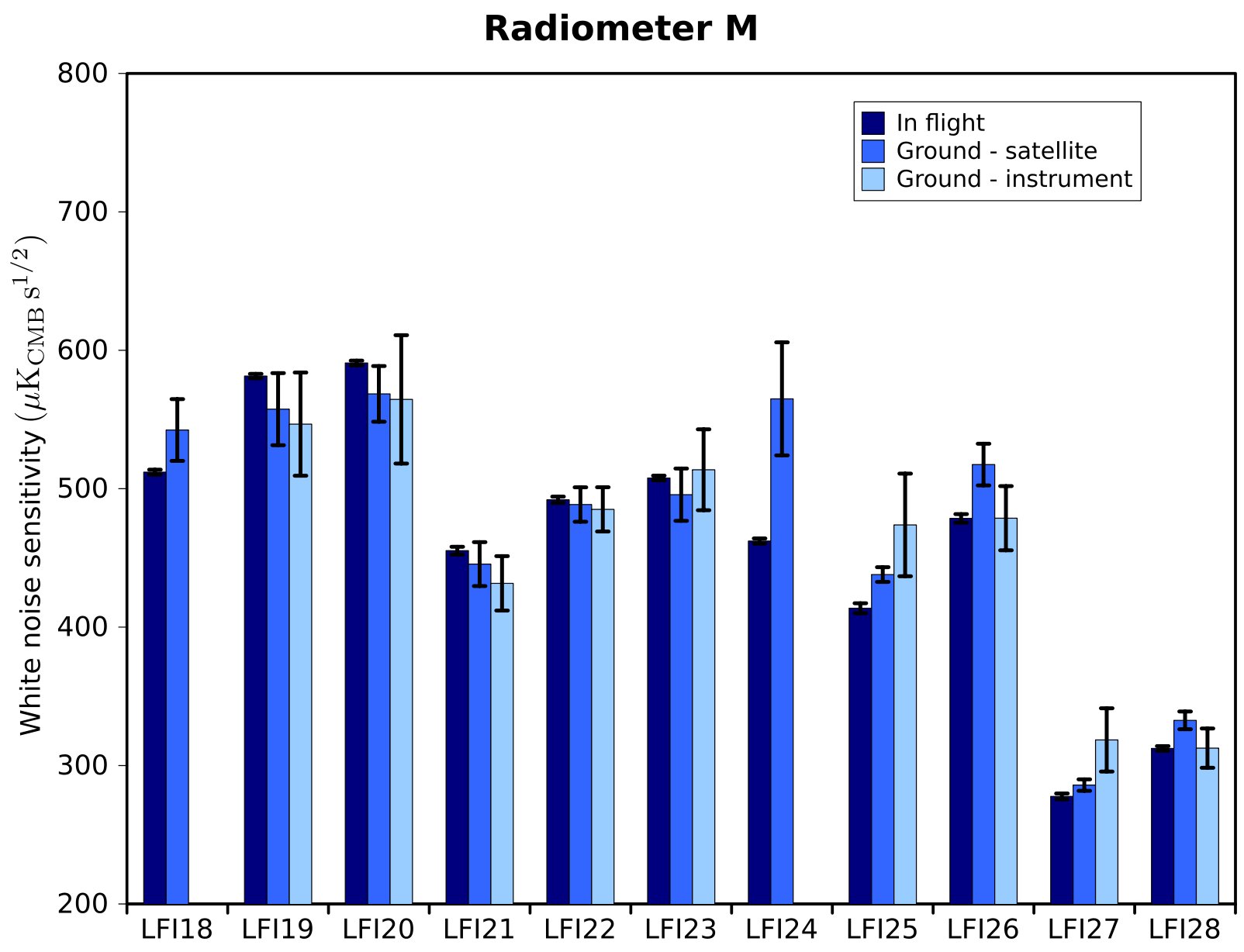}
            \includegraphics[width=9cm]{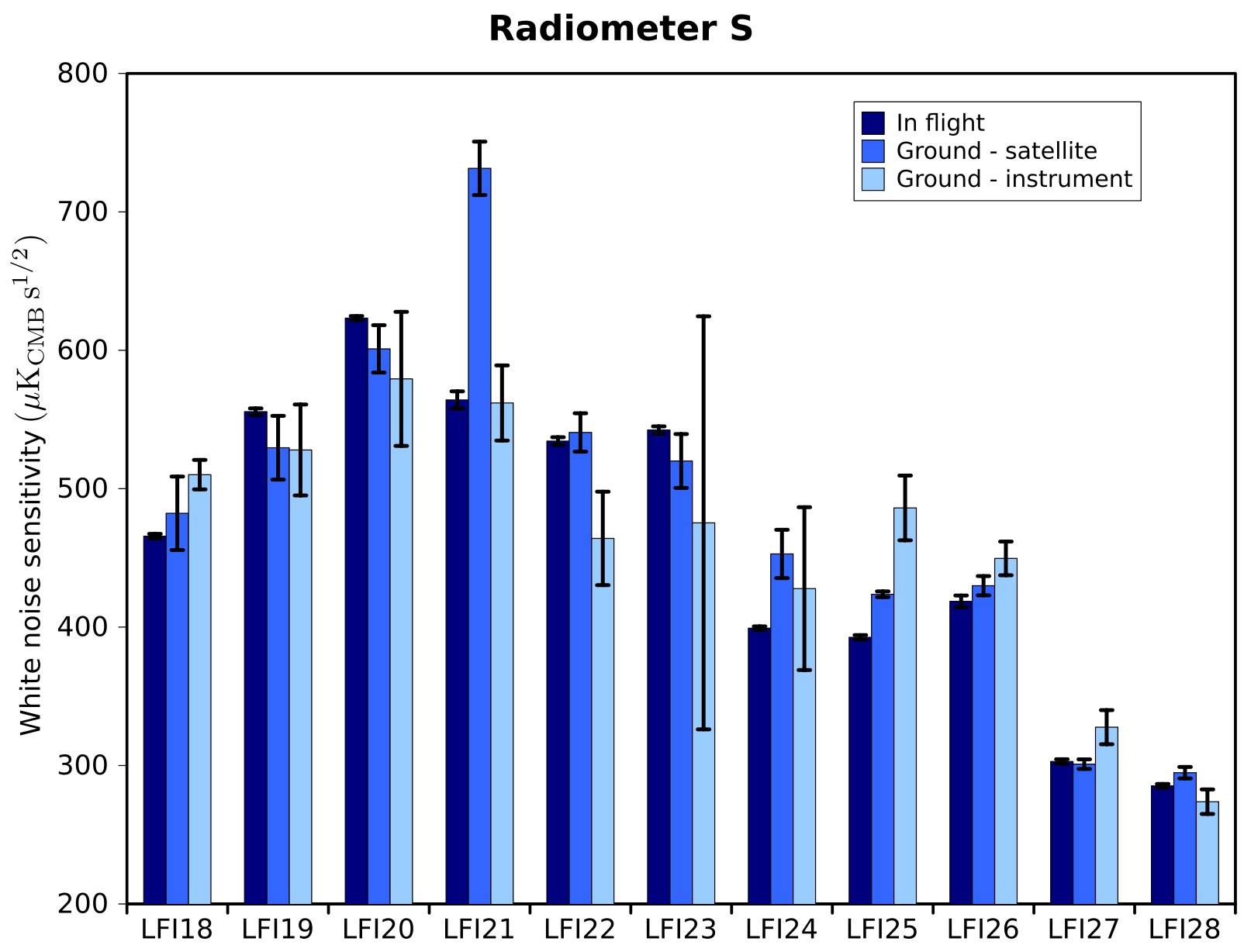}
        \end{center}
        \caption{Comparison of white noise sensitivity values calculated in flight and on ground during 
        the satellite-level and instrument-level test campaigns. Values for \texttt{LFI18M} and \texttt{LFI24M} measured 
        at instrument-level are not reported because these two radiometers failed before instrument delivery and were replaced and 
        repaired, respectively.}
        \label{fig_white_noise_sensitivity_comparison}
\end{figure*}

Fig.~\ref{fig_white_noise_sensitivity_comparison} compares the calibrated white noise sensitivity for the 
22 LFI radiometers calculated from flight data and during ground tests performed both at instrument and satellite 
level. Error bars reported for values measured in flight and during satellite tests are statistical uncertainties derived 
from calculations performed on several 
data chunks\footnote{About 20 one-hour chunks for on-ground satellite data, one year of operations for in-flight data.}. 
Error bars reported for values measured at instrument level, instead, represent the uncertainty between two 
different methods used to extrapolate the white noise sensitivity measured at test conditions 
($\sim 20$\,K input and $\sim 26$\,K front-end temperature) to flight conditions ($\sim 2$\,K input 
and $\sim 20$\,K front-end temperature).  Further details about this extrapolation are reported in 
\citet{mennella2010}.  

Figure~\ref{fig_white_noise_sensitivity_comparison} shows general agreement in the sensitivity calculated in 
various test campaigns, with two outstanding exceptions, radiometers 
\texttt{LFI24M} and \texttt{LFI21S}, which displayed significantly improved noise levels in-flight. This resulted 
from an incorrect bias setting of these radiometers during ground tests \citep[see][]{mennella2010} 
and was resolved during in-flight tuning.  
    
In Fig.~\ref{fig_white_noise_sensitivity_comparison} values for \texttt{LFI18M} and \texttt{LFI24M} measured 
at instrument-level are not reported because these two radiometers failed.  \texttt{LFI18M} was replaced 
with a spare unit and \texttt{LFI24M} was repaired before instrument  delivery \citep[see][]{mennella2010}.
    
Table~\ref{tab_white_noise_per_frequency} summarizes the sensitivity numbers calculated during the first 
year of operations using methods and procedures outlined in \S~\ref{sec_sensitivity_method} and described 
in detail in \citet{planck2011-1.6}  compared with scientific requirements. The measured sensitivity in
very good agreement with pre-launch expectations. While the white noise
moderately exceeds the design specification, this performance is fully in
line with the LFI science objectives.

    \begin{table}
        \caption{White noise sensitivities for the LFI frequency
        channels compared with requirements.}
        \label{tab_white_noise_per_frequency}
        \begin{center}
            \begin{tabular}{l c c}
                \hline
                \hline
                Channel  &Value  &Requir.\\
                \hline
                70 GHz   &152.6$\,\mu\mathrm{K}_{\rm CMB}\, \mathrm{s}^{1/2}$& 119$\,\mu\mathrm{K}_{\rm CMB}\, \mathrm{s}^{1/2}$\\
                44 GHz   &173.1$\,\mu\mathrm{K}_{\rm CMB}\, \mathrm{s}^{1/2}$& 119$\,\mu\mathrm{K}_{\rm CMB}\, \mathrm{s}^{1/2}$ \\
                30 GHz   &146.8$\,\mu\mathrm{K}_{\rm CMB}\, \mathrm{s}^{1/2}$& 119$\,\mu\mathrm{K}_{\rm CMB}\, \mathrm{s}^{1/2}$ \\
                \hline
            \end{tabular}
        \end{center}
    \end{table}
    
The consistency of the noise estimates outlined in Table~\ref{tab_white_noise_per_radiometer} was tested  
by comparing the jackknife noise maps, the white noise covariance matrices, and 101 
noise Monte Carlo realisations \citep[see][for details]{planck2011-1.6}.   In fact, 
noise covariance matrices and Monte Carlo simulations are dependent on estimated noise parameters, while the jackknife 
noise maps describe directly the noise in the final maps.  

\begin{figure*}
        \begin{center}
            \includegraphics[width=18cm]{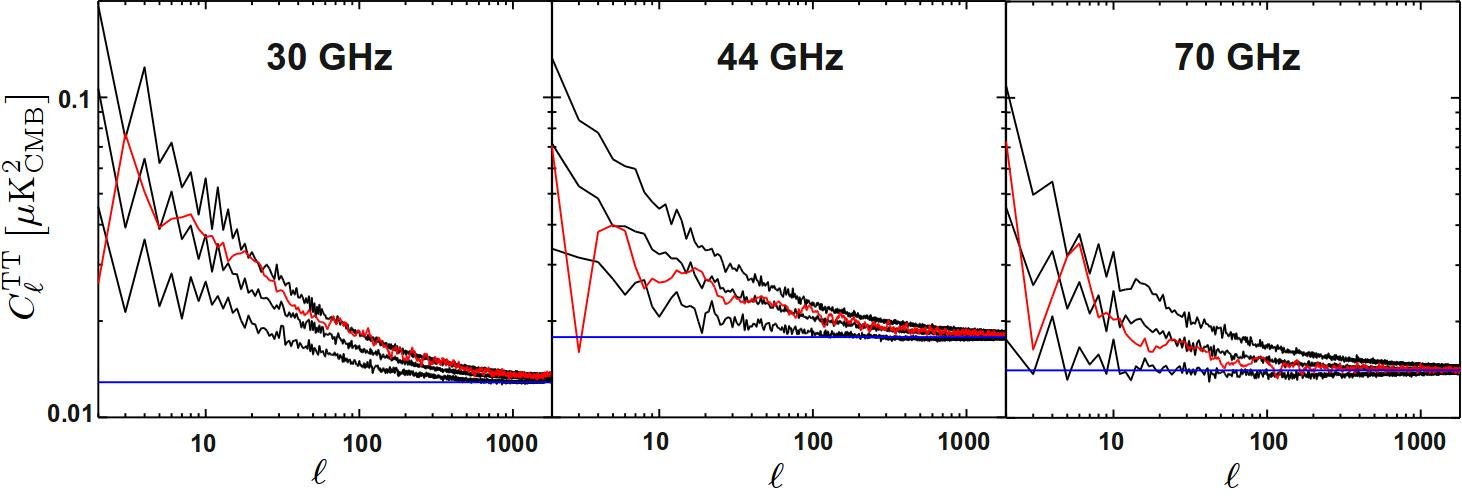}
        \end{center}
        \caption{The noise angular power spectra.
            Black: noise Monte Carlo (median, 16\% quantile, and 84\%
            quantile). Red: jackknife noise
            map pseudo-$C_\ell$. Blue: values from white noise covariance matrices.}
        \label{fig_Clspectra}
    \end{figure*}

We produced pseudo-$C_{\ell}$ spectra from the jackknife noise maps and noise Monte Carlo 
maps by {\tt anafast} (see  Fig.~\ref{fig_Clspectra}), and compared high multipole 
tails ($1150<\ell<1800$) to the predictions from the white noise covariance  matrices.  
In the analysis we masked out a 20\% region of the galactic plane and unsolved pixels  
(pixels that have HEALPix\footnote{\url{http://healpix.jpl.nasa.gov}} bad pixel value in noise maps), 
leaving us with sky fraction  $f_{\rm sky} \approx 0.8$.
    
\begin{figure}
        \begin{center}
            \includegraphics[width=9cm]{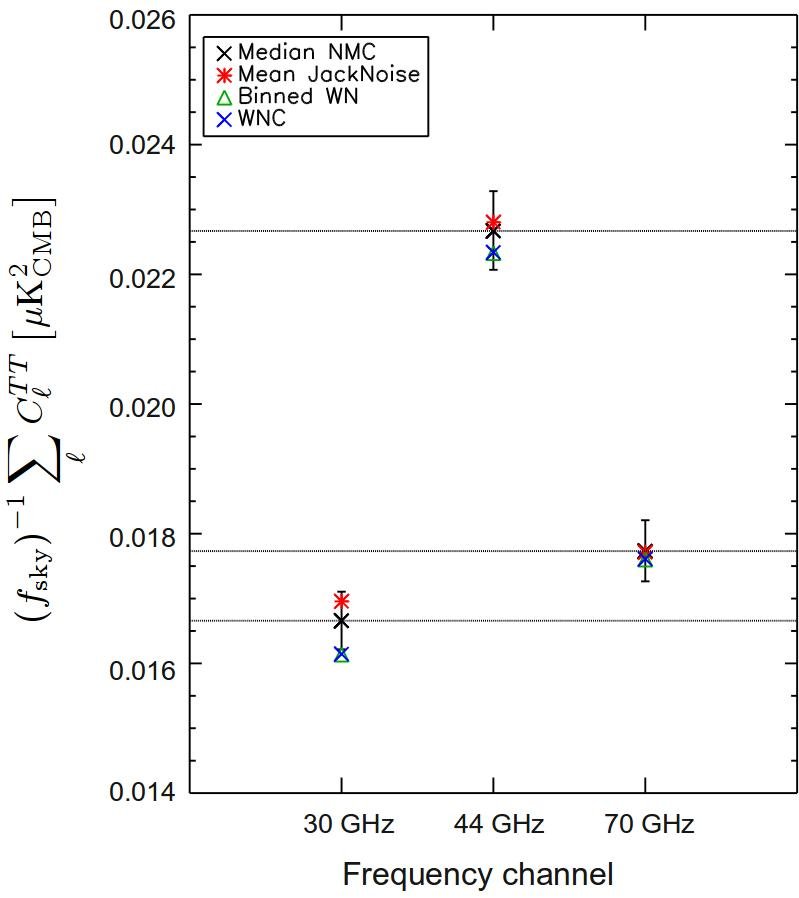}
            \caption{Comparison of noise estimates from different methods. Black: noise
            from high-$\ell$ means of noise Monte Carlo statistics.
            Red: noise from the high-$\ell$ mean of the jackknife noise
            map pseudo-$C_{\ell}$ spectra. Green: noise from the binned white noise 
            Monte Carlo maps. Blue: noise from white noise covariance matrices. Error bars are relative
            to values from Monte Carlo simulations. We have weighted all
            the values with the inverse of the analysed sky fraction, $f^{-1}_{\mbox{\scriptsize sky}}$, to
            represent the noise levels of temperature $C_{\ell}$ spectra that would
            result from using full sky maps.}
            \label{fig_MeanCls}
        \end{center}
    \end{figure}

Figure~\ref{fig_MeanCls} shows a comparison of the noise estimates obtained with the various 
methods.  In the figure the green triangles refer to noise obtained from \textit{binned} pure 
white noise Monte Carlo maps, i.e., maps containing no residual  $1/f$ noise.  At all frequencies the 
high-$\ell$ mean of the jackknife noise map pseudo-$C_{\ell}$ spectra weighted 
by the inverse sky coverage, $(f_{\mbox{\scriptsize sky}})^{-1}$ (shown in red), 
are in the 68\% range of the noise Monte Carlos   (represented with the back error bars). The point from the 
30\,GHz jackknife noise map is almost 1-$\sigma$ higher than the median value for the noise MC 
due to residual gradient leakage (due to point sources etc.), which is strongest for 
the 30\,GHz channel.  The noise power from the white noise covariance matrices and binned 
maps is always lower than the noise power from the full noise Monte Carlo and jackknife 
noise maps, as expected, due to residual $1/f$ noise present even at high multipoles.  This effect is especially pronounced 
in the 30\,GHz channels, which have larger knee frequencies (see \S~\ref{sec_1_over_f_noise_properties}).

\subsection{$1/f$ Noise properties}
\label{sec_1_over_f_noise_properties}

\begin{figure*}
        \begin{center}
            \includegraphics[width=9cm]{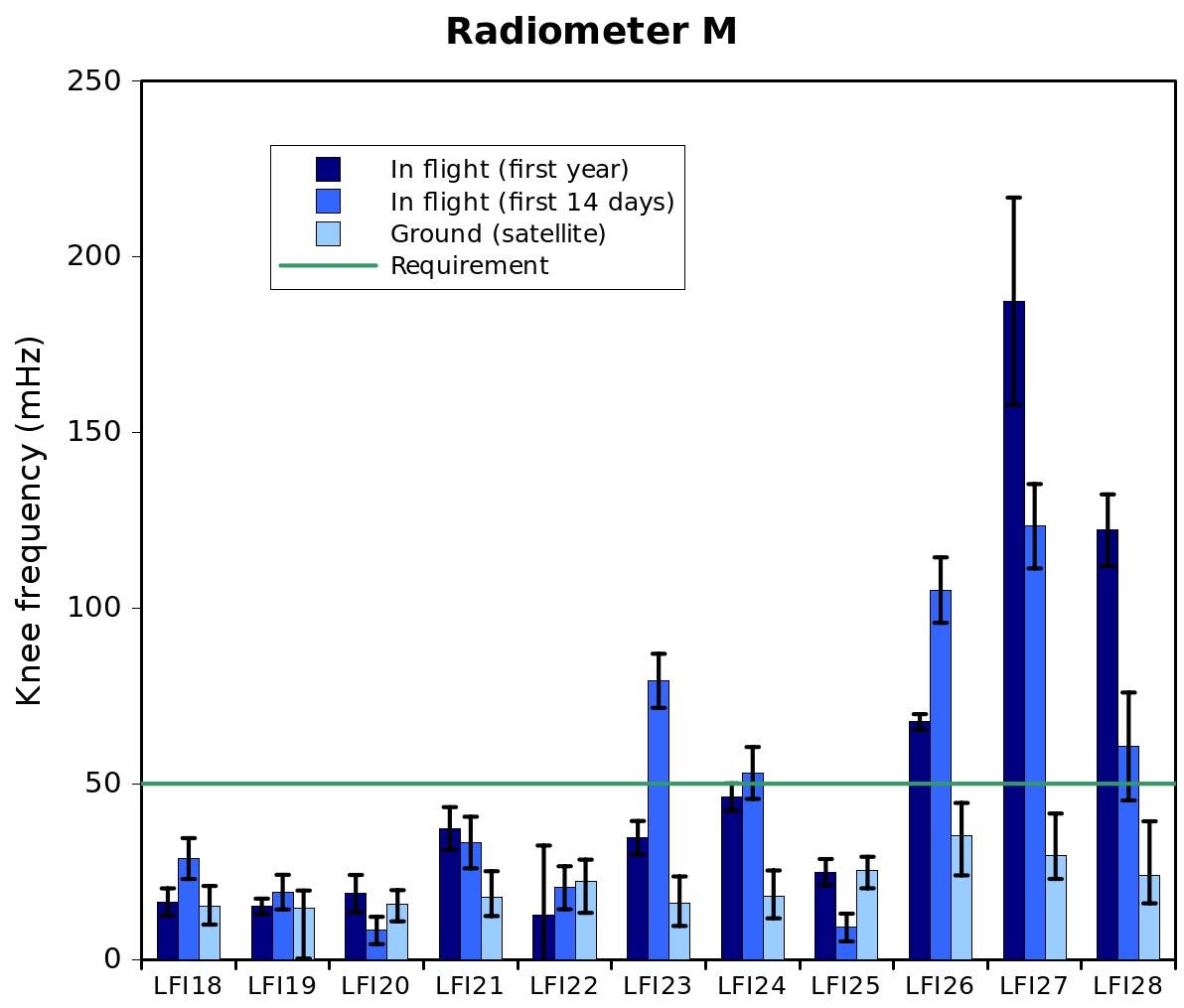}
            \includegraphics[width=9cm]{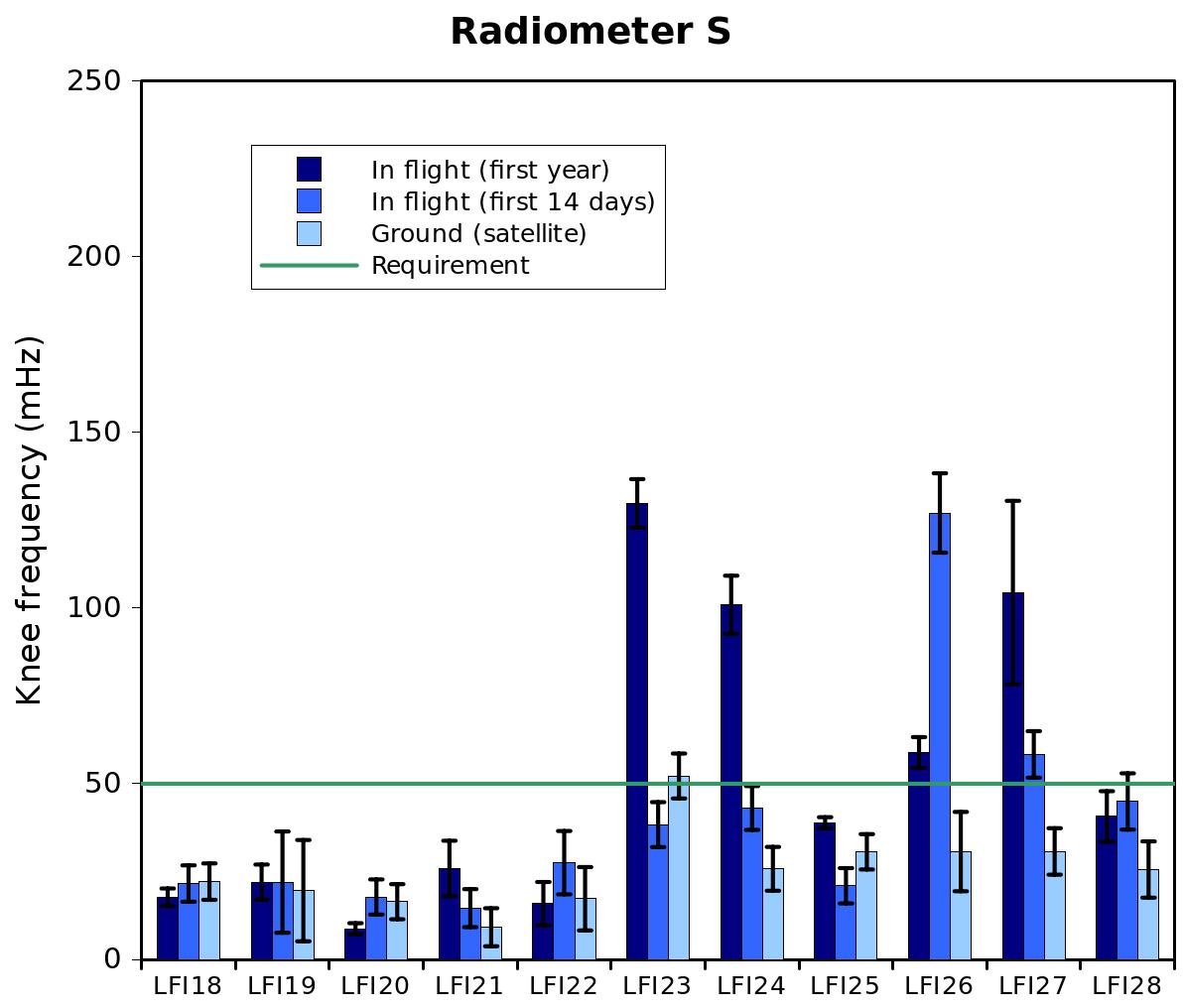}
            \caption{Comparison of  $1/f$  knee frequencies measured in flight and on ground during the satellite-level
            test campaign.}
            \label{fig_fk_comparison}
        \end{center}
    \end{figure*}

Figure~\ref{fig_fk_comparison} shows a comparison of the knee frequency for the 22~LFI radiometers estimated 
from flight data and ground satellite calibration. Knee frequency and slope after one year of operations 
for all LFI radiometers were given earlier in Table~\ref{tab_one_over_f_noise_per_radiometer}.  With some   exceptions 
the values are below or compatible with requirements and are comparable within the error  bars among different measurements. 
    
There are a few cases (\texttt{LFI23S}, \texttt{LFI24S}, \texttt{LFI27M}, \texttt{LFI28M}, \texttt{LFI27S}) in which 
the knee frequency measured in flight is higher than that measured on ground.  The 
cause is still under investigation, but the impact on the scientific quality of the data is small.  
Destriping effectively removes correlated structures, limiting the impact of the high knee frequencies to 
a small ($< 4\%$) increase of the noise variance  (see Table~\ref{tab_noise_maps_rms} and Fig.~\ref{fig_jacknoise_map}).
        
As shown in the low-$\ell$ part of the angular power spectra in Fig.~\ref{fig_Clspectra}, the noise parameters estimated 
in flight are a good representation of the noise in the actual maps.  This means that for cosmological 
purposes we cannot rely on simple white noise covariance, and a full timeline-to-map Monte Carlo is needed.  On the 
other hand, this also means that there is no need for a more complex functional noise model with respect to 
the one reported in Eq.~(\ref{eq_noise_model}).


\section{First assessment of systematic effects}
\label{sec_systematic_effects}

    In this section we discuss the most relevant systematic effects in the first year data. 
The LFI design was driven by the need to suppress systematic effects well below instrument white noise. The receiver differential scheme, 
with reference loads cooled to 4\,K, greatly minimises the effect of $1/f$ noise and common-mode fluctuations, such as thermal 
perturbations in the 20\,K LFI focal plane. The use of a gain modulation factor (see Eq.~(\ref{eq_r_v})) largely compensates 
for spurious contributions from input offsets.  Furthermore, diode averaging (Eq.~(\ref{eq_vout_radiometer})) allows us to cancel 
second-order correlations such as those originating from phase switch imbalances.

We have developed an error budget for systematic effects \citep{bersanelli2010} as a reference for both instrument design and data analysis.  Our goal is to ensure that each systematic effect is rejected to the specified level, either by design or by robust removal in software.  At this stage, the following effects proved to be relevant: 
\begin{itemize}
    \item $1/f$ noise (discussed in \S~\ref{sec_1_over_f_noise_properties}),
    \item 1\,Hz frequency spikes, 
    \item thermal fluctuations in the back-end modules driven by temperature oscillations from the transponder during the first survey, 
    \item thermal fluctuations in the 20\,K focal plane, 
    \item thermal fluctuations of the 4\,K reference loads.
\end{itemize}

\begin{figure*}
    \begin{center}
        \includegraphics[width=18cm]{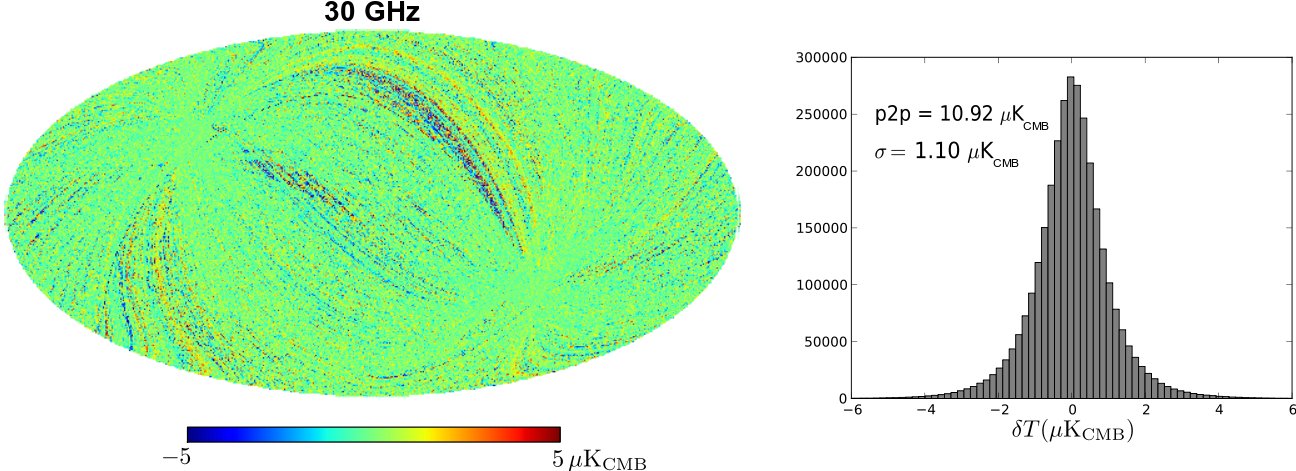}
        \includegraphics[width=18cm]{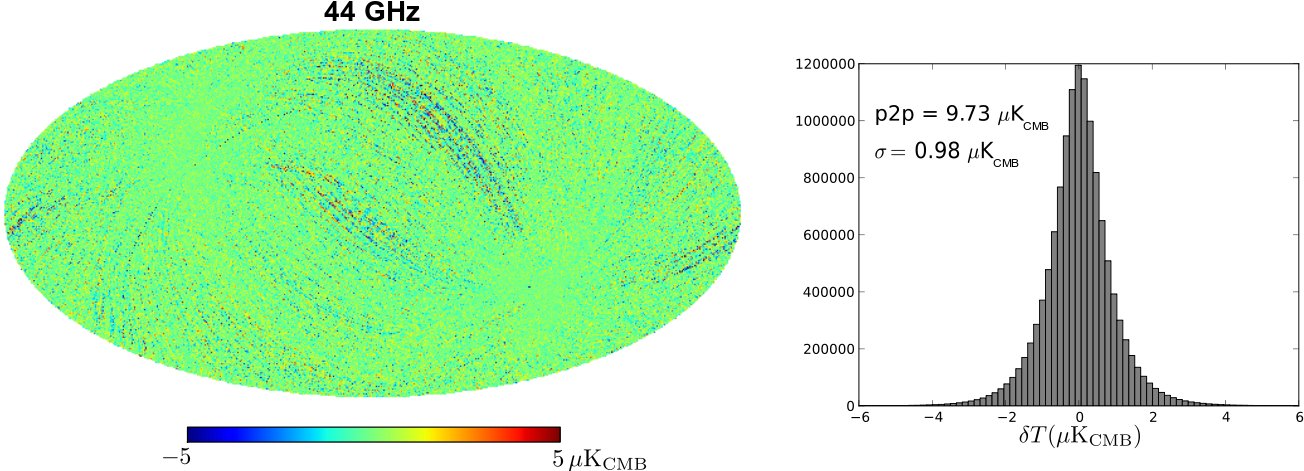}
        \includegraphics[width=18cm]{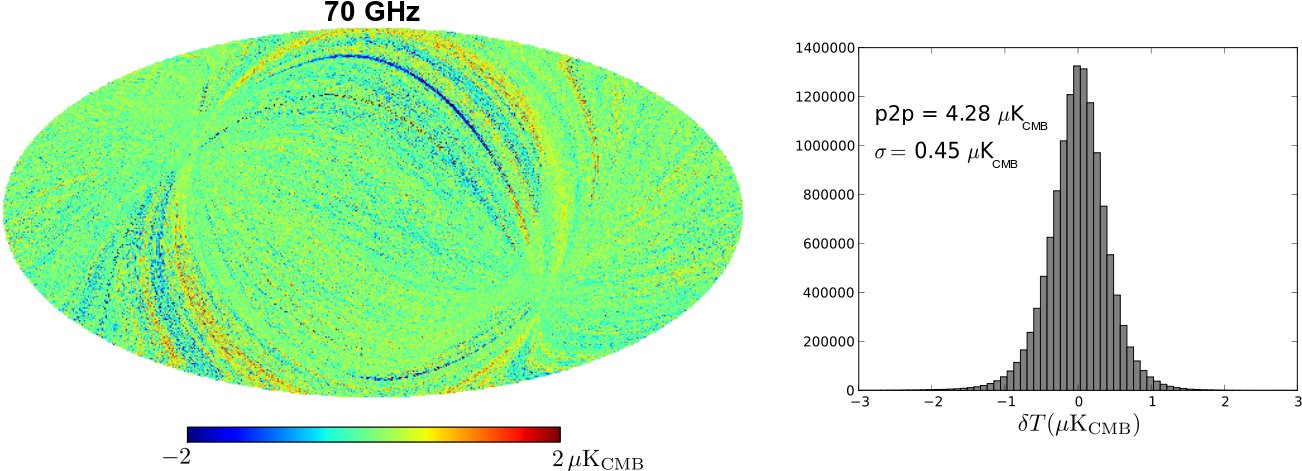}
    \end{center}
    \caption{Maps of the combined systematic effects from frequency spikes and thermal fluctuations.}
    \label{fig_total_maps}
\end{figure*}

For each of these effects we used flight data and information from ground tests to build timelines, maps, 
and angular power spectra that represent our best knowledge of their impact on the scientific analysis. 
Figure~\ref{fig_total_maps} shows maps and histograms of the combination of all the above 
mentioned systematic effects for the three LFI channels.  The figure shows that the 30 and 
44\,GHz channels display a residual level of spurious signals about twice larger 
than the 70\,GHz 
channel. As shown in the following sections, this is caused by temperature fluctuations of 
4\,K reference loads that are significantly larger at 30 and 44\,GHz than at 70\,GHz.

\begin{figure}[h!]
    \includegraphics[width=9cm]{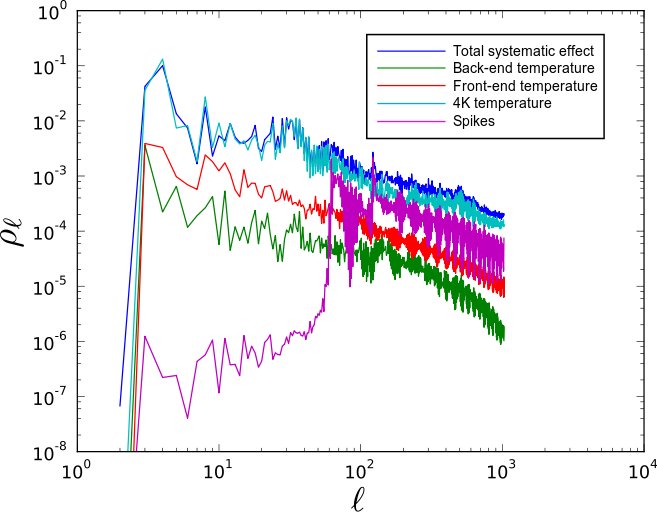}
    \includegraphics[width=9cm]{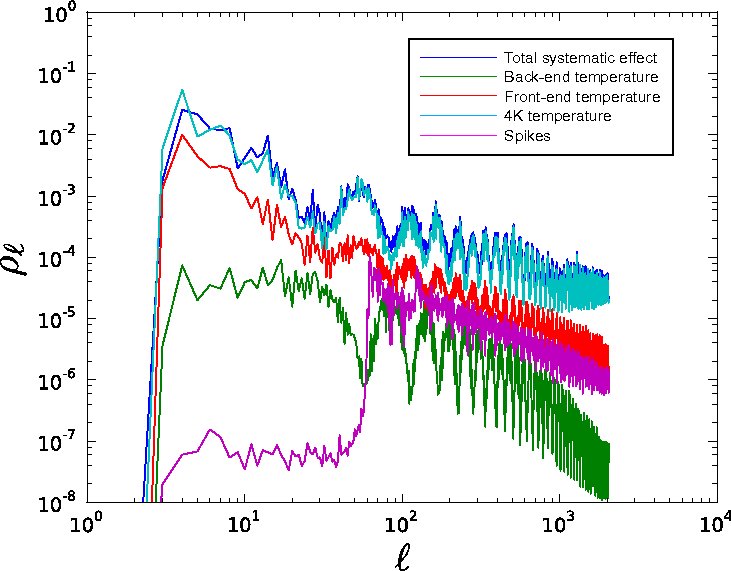}
    \includegraphics[width=9cm]{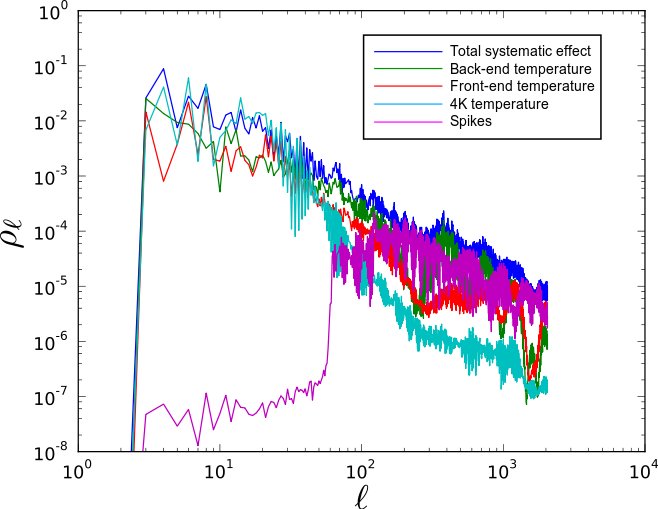}
    \caption{Ratio $\rho_\ell = C_\ell^{\rm syst} / C_\ell^{\rm noise}$ for the three LFI frequency channels (top: 30\, GHz;
    middle: 44\,GHz; bottom: 70\,GHz).  In each panel 
    we have plotted both the global spectrum obtained from the maps in Fig.~\ref{fig_total_maps} 
    and the individual spectra calculated from the single-component maps.}
    \label{fig_condensed_ps}
\end{figure}

A convenient way to assess the effect on angular power spectra is to calculate the ratio 
$\rho_\ell = C_\ell^{\rm syst} / C_\ell^{\rm noise}$, where $C_\ell^{\rm syst}$ is the 
angular power spectrum of the systematic effect map and $C_\ell^{\rm noise}$ is the angular power spectrum 
of the instrument noise contribution including the residual $1/f$ component remaining after destriping.  
Figure~\ref{fig_condensed_ps} shows $\rho_\ell$ for the three LFI frequency channels. In each panel we 
have plotted the spectrum obtained from the global maps in Fig.~\ref{fig_total_maps} and 
the spectra derived from the single-component maps (see next sections).

Figure~\ref{fig_condensed_ps} shows that $\rho_\ell$ is in the range $10^{-1}$--$10^{-2}$ for $\ell$ in 
the range 2--20, and $\rho_\ell < 10^{-2}$ for $\ell > 10$. The figure also shows that at 30 and 44\,GHz the largest 
contribution to systematic effects is determined by temperature fluctuations of the 4\,K reference loads, 
while the residual systmatic uncertainty in the 70\,GHz channel is mainly caused by back-end temperature 
fluctuations and, at small angular scales, by frequency spikes. 
Table~\ref{tab_residual_systematics} gives an overview of our current assessment of residual peak-to-peak and rms systematic 
effects per pixel on LFI temperature maps.  Maps were made with $N_{\rm side}=512$ at 30\,GHz 
and $N_{\rm side}=1024$ at 44 and 70\,GHz. Corresponding pixel sizes are $\sim 6\parcm8$ and $\sim 3\parcm4$.

Further advances in the data analysis pipeline will be aimed at removing spikes from the 70\,GHz data, improving relative 
calibration to account for thermally driven fluctuations, and further suppressing spurious fluctuations caused by 
4\,K temperature instabilities at 30 and 44\,GHz.

\begin{table}
    \begin{center}
        \caption{Summary of residual effects on maps in $\mu \mathrm{K}_{\rm CMB}$ from main systematic effects.}
        \label{tab_residual_systematics}
        \begin{tabular}{l c c c c c c c}
            \hline
            \hline
            & \multicolumn{2}{c}{30\,GHz}& \multicolumn{2}{c}{44\,GHz}& \multicolumn{2}{c}{70\,GHz}\\
            \hline
            & p-p & r.m.s.& p-p & r.m.s.& p-p & r.m.s.\\
            \hline
            1-Hz spikes  & 4.00 & 0.45 & 1.51 & 0.15 & 2.56 & 0.30   \\
            BEM T fluct. & 1.27 & 0.11 & 0.63 & 0.05 & 2.70 & 0.24 \\
            FEM T fluct. & 1.05 & 0.23 & 1.15 & 0.22 & 1.12 & 0.21  \\
            4\,K T fluct. & 9.76 & 0.98 & 9.73 & 0.98 & 1.30 & 0.16  \\
            \hline
            Total\tablefootmark{a}  & 10.92 & 1.10 & 9.73 & 0.98 & 4.28 & 0.45 \\
            \hline
        \end{tabular}
    \end{center}
    \tablefoottext{a}{The total has been estimated from the maps combining all systematic effects
    (Fig.~\ref{fig_total_maps}).}
\end{table}


\subsection{Frequency spikes}
\label{sec_spikes}

Spikes are seen in the radiometer outputs in the frequency domain at multiples of 1\,Hz.  These  ``frequency spikes'' were 
first detected during ground tests, and are caused by pickup from the clock of the housekeeping electronics \citep{meinhold2009,mennella2010}.  
The pickup occurs between the detector diodes and the DAE gain stage. Frequency spikes are present at some level in 
the output from all detectors, but affect the 44\,GHz data most strongly because 
of the low voltage output and high DAE gain values in that channel. 
In this section we provide estimates of the residual systematic effect after frequency spike removal at 44\,GHz, 
and the effect caused by the spikes without any removal at 30 and 70\,GHz.  

In the time domain, the frequency domain spikes comprise a one second square wave 
with a rising edge near 0.5\,s and a falling edge near 0.75\,s in on-board time. During the 
first year of operations we did not observe any deviation in either the phase or the shape of these signals.  

Templates of these spurious square waves obtained from the output voltages have been used to remove this effect 
from the data before differentiation (procedures and algorithms are described in \citet{planck2011-1.6}).  
Although we could in principle apply the removal process to all data, we decided to 
remove the spikes only from the 44\,GHz data, which are the most affected by this effect. 

To estimate the residual in the spike removal process caused by slow variations in the 
square wave amplitude, we used a simple $\chi^2$ minimisation procedure to estimate the amplitude of 
the template for each hour of data, then smoothed the amplitude by simple 
binning with a 10-day window size.   We took this smoothed amplitude as our estimate of the 
true time variation of the spike signal, and calculated the residual error that would result if we approximated 
this time-varying signal with a constant template in the spike removal process.  Although we could use, in principle, 
the time-varying template to remove the effect, we decided to use the simplest and most robust approach of a constant-amplitude
template. The possibility to implement of a model that accounts for slow spike amplitude drifts will be considered for future
data releases.

\begin{figure*}
        \begin{center}
            \includegraphics[width=16.5cm]{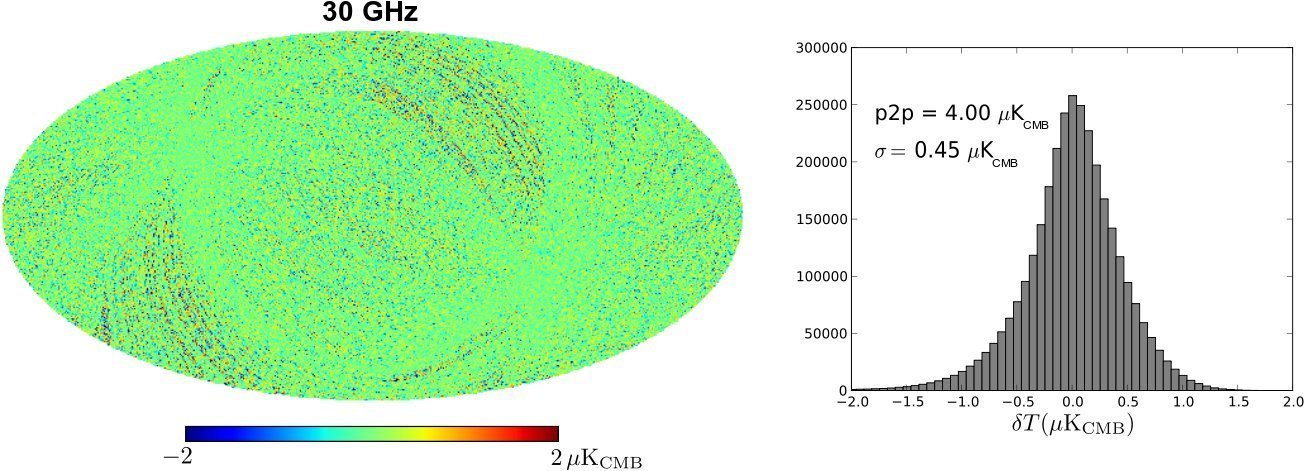}
            \includegraphics[width=16.5cm]{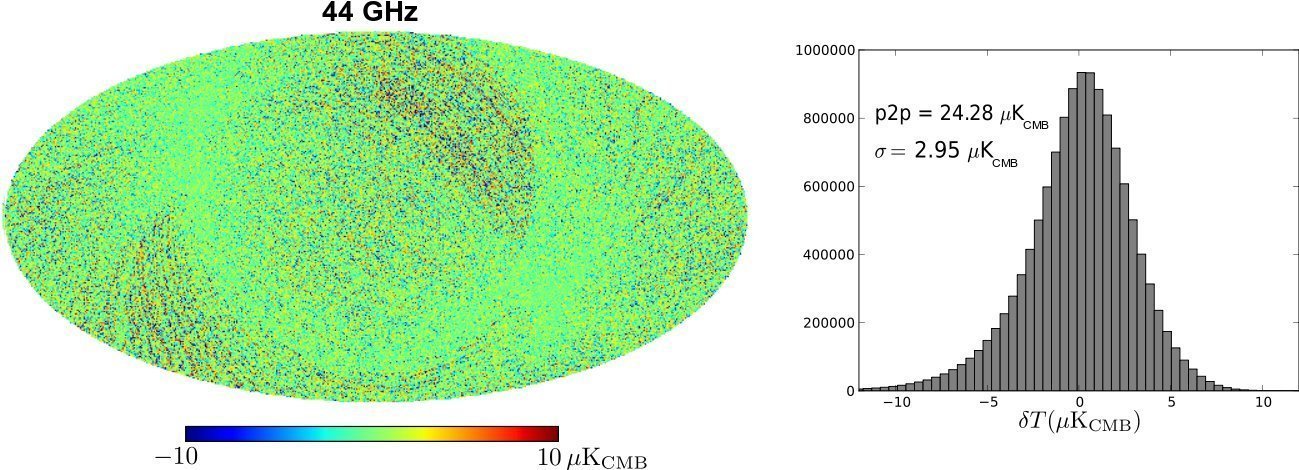}
            \includegraphics[width=16.5cm]{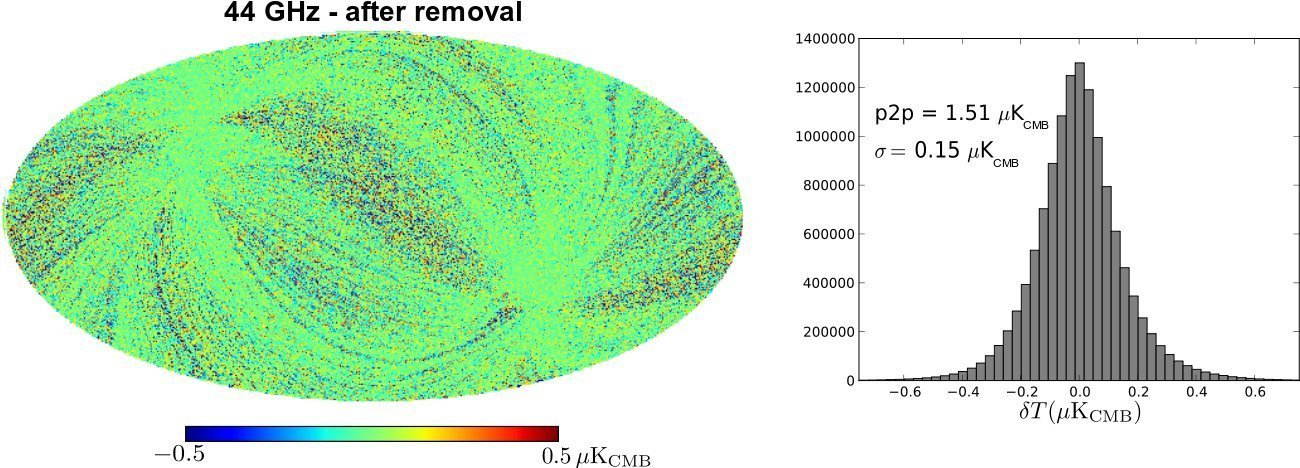}
            \includegraphics[width=16.5cm]{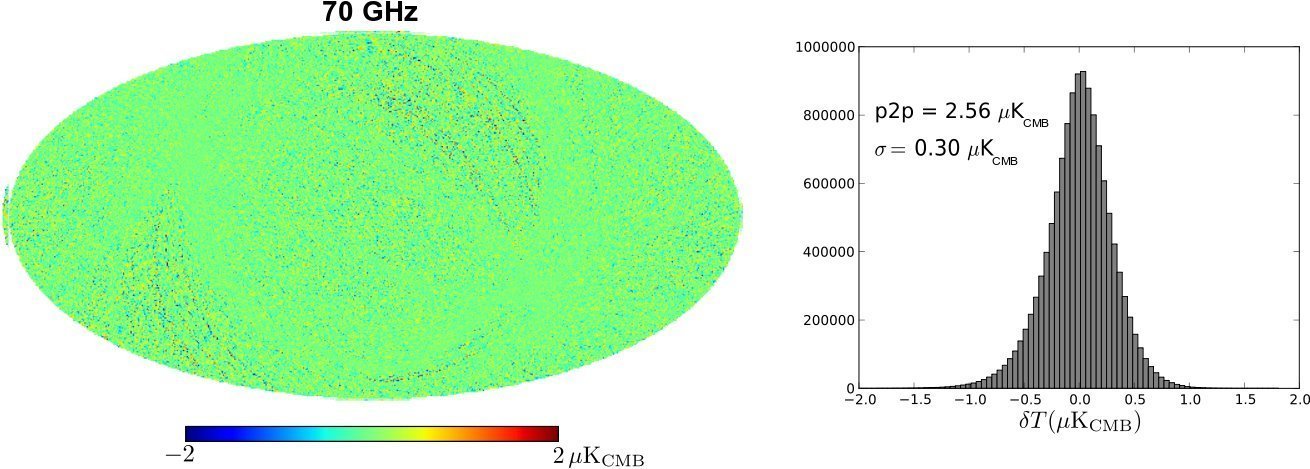}
        \end{center}
        \caption{Simulated spike maps for each channel, and the residual after removal
        for the 44\,GHz channel. Maps have been obtained using constant-amplitude templates
        to generate spike TODs. The datastream used to generate the residual map 
        has been obtained by subtracting a spike TOD with time-varying amplitude from a constant-amplitude
        spike TOD.}
        \label{fig_spike_maps}
    \end{figure*}

Templates were also used to generate maps for each frequency channel to determine the impact of the spike signal on the science results.
These maps and the corresponding histograms are shown in Fig.~\ref{fig_spike_maps}. For the 44\,GHz channel
we also show the residual effect after removing the constant template, showing that after removal the residual
effect is reduced by a factor of about 20.

\subsection{Thermal fluctuations}
\label{sec_thermal_fluctuations}

The LFI is sensitive to temperature fluctuations of the warm back-end unit, the 20\,K focal plane, 
and the 4\,K reference loads.  In the first two, temperature variations impact the sky and reference load signals at 
a similar level, so that in the differential radiometric output the residual spurious variations are reduced by more than 
one order of magnitude. Fluctuations at the level of the 4\,K reference loads, 
instead, transfer directly to the radiometric output, and may represent a more critical source of systematic errors.
    
The effects of low frequency thermal fluctuations are strongly suppressed by the spacecraft spin itself and 
the destriping map-making codes \citep{planck2011-1.6}.  High frequency thermal fluctuations are strongly damped by the 
thermo-mechanical structure of the spacecraft and instruments \citep{tomasi2009}.  The dominant effects, 
then, come from fluctuations at frequencies in a range around the spin frequency.
    
In this section we present an overview of the impact on the LFI science caused by the 
residual level of systematic effects in the final maps using a dedicated pipeline that follows the procedure outlined below. 
    
    \begin{itemize}
        \item Start from a time ordered datastream of a temperature sensor representative of the temperature
        behaviour of the considered thermal stage;
        \item low-pass filter to remove high-frequency sensor noise;
        \item apply thermal transfer functions where appropriate to obtain the physical temperature behaviour
        at the level of the receiver components sensitive to the fluctuation;
        \item apply the radiometric transfer function to convert the physical temperature fluctuation into
        antenna temperature fluctuation;
        \item resample at scientific sampling frequency, calibrate and build differenced time-ordered data;
        \item build maps with \texttt{Madam}.
    \end{itemize}

These maps have been generated combining actual flight housekeeping data with thermal and radiometric transfer functions obtained 
from flight and ground-test data \citep{terenzi2009b}. They represent, therefore, our current best estimate of 
the impact of the individual effects on the science in flight conditions.
    
As explained in \S~\ref{sec_calibration_accuracy} further developments of this work will aim at including thermal fluctuations 
from the back-end unit in the calibration model which will reduce the need to remove the spurious signal 
from the time ordered data during map-making.


    \subsubsection{Back-end temperature fluctuations}
    \label{sec_bem_fluctuations}

        As mentioned in \S\,\ref{sec_stability}, for the first 258 days of the mission the satellite transponder was 
        switched on only for the daily telecommunication period.  This induced quasi-sinusoidal fluctuations in the 
        \Planck\ service module temperature that propagated to the LFI  back-end unit, 
        and were recorded by the instrument housekeeping sensors at the level of $\pm 0.2$\,K/day.
        
\begin{figure}
            \begin{center}
                \includegraphics[width=9cm]{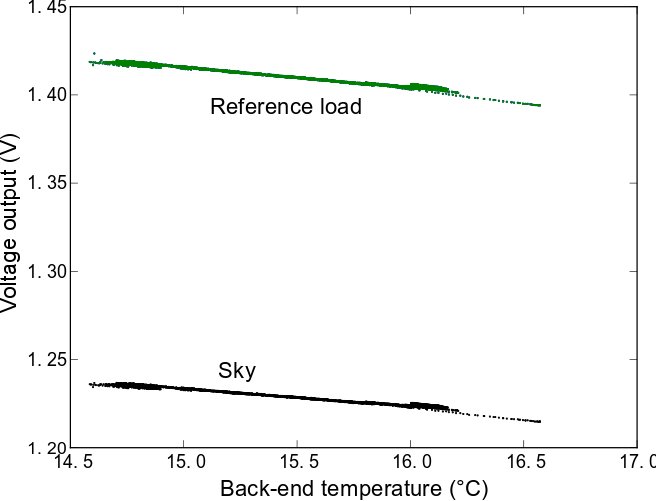}
                \caption{Correlation between back-end temperature and total power voltage output
                relative to the LFI detector \texttt{LFI27M-00}.}
                \label{fig_beu_corr}
            \end{center}
        \end{figure}

        This temperature variation drives variation in the total power voltage output of both sky and 
        reference detectors, which largely disappears in the difference, shown in Fig.~\ref{fig_dailyStability}.  
        The effect on the undifferenced total power data, however, was useful in calculating correlation 
        coefficients between temperature and voltage outputs. These coefficients have been used to estimate the 
        residual effect on time-ordered data and maps.  Figure~\ref{fig_beu_corr} 
        shows this correlation for the detector \texttt{LFI27M-00}. 

\begin{figure*}
            \begin{center}
                \includegraphics[width=18cm]{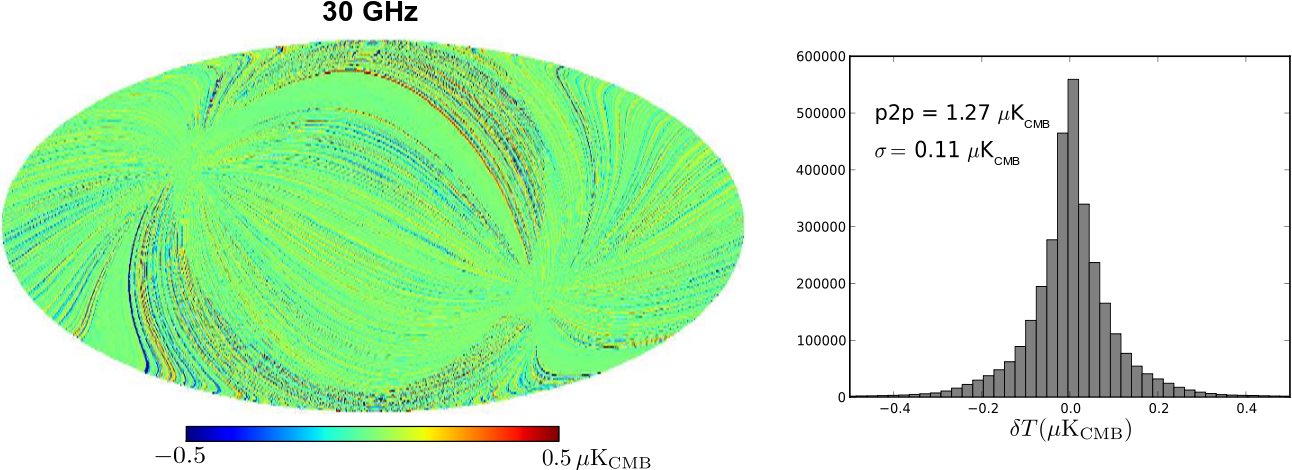}
                \includegraphics[width=18cm]{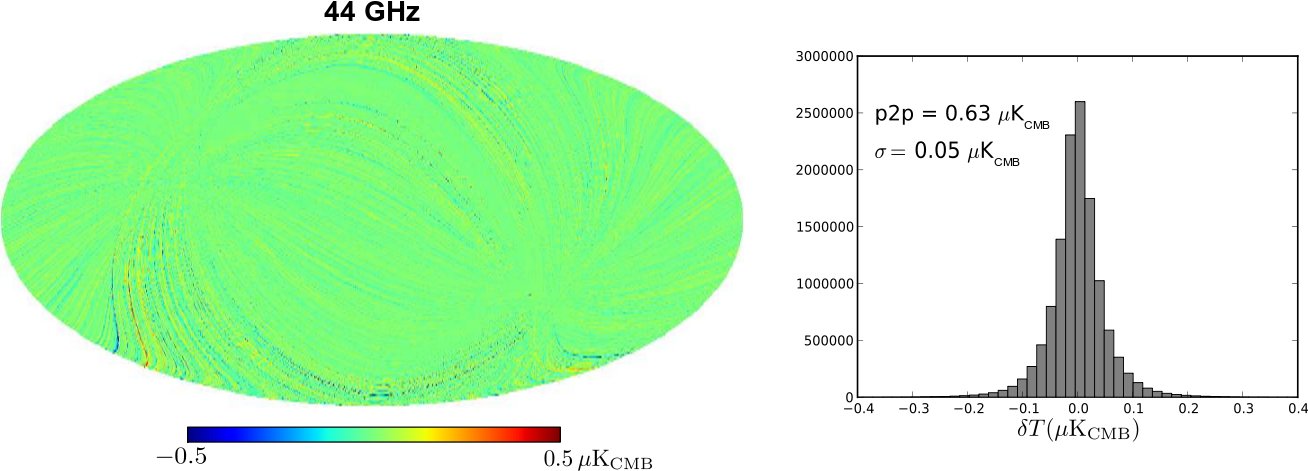}
                \includegraphics[width=18cm]{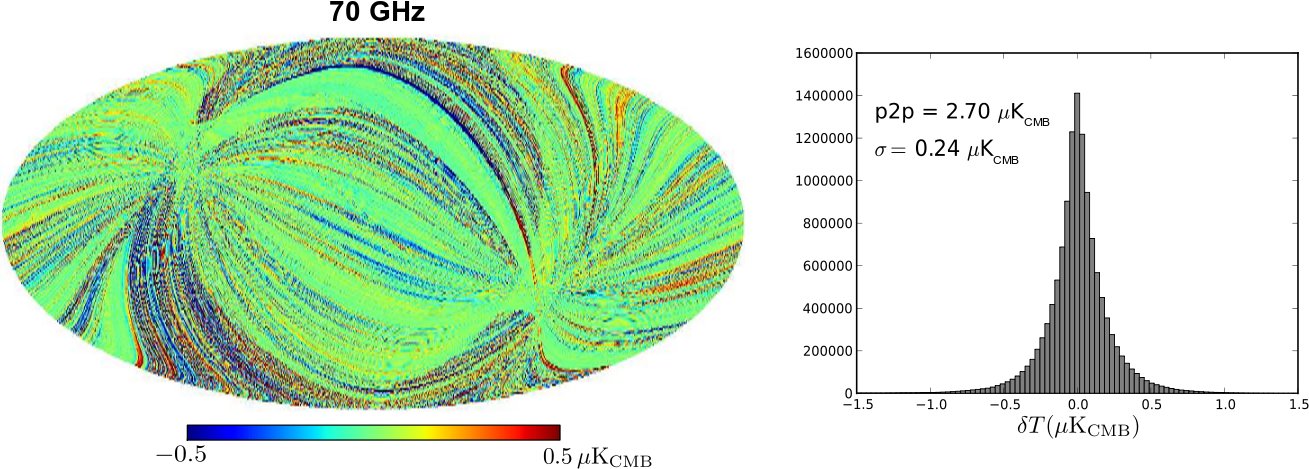}
            \end{center}
            \caption{Simulation of the systematic effects of temperature fluctuations of the LFI warm back-end 
            unit. Maps are $N_{\rm side}=512$ at 30\,GHz, and $N_{\rm side}=1024$ at 44 and 70\,GHz.}
            \label{fig_bem_maps}
        \end{figure*}

        Figure~\ref{fig_bem_maps} shows the expected systematic effects of temperature fluctuations of 
        the back-end unit during the first year of operations. The peak-to-peak effect is $\sim 1\,\mu \mathrm{K_{CMB}}$, 
        and the rms is well below $1\, \mu\mathrm{K_{CMB}}$. 
        
    \subsubsection{Front-end temperature fluctuations}
    \label{sec_fem_fluctuations}

        Another source of temperature fluctuations is the \textit{Planck} sorption cooler
        \citep{morgante2009,planck2011-1.3}, which displays temperature variations of the order of $\sim 0.5$\,K 
        peak-to-peak at the cold end attached to the LFI focal plane unit.  These fluctuations are 
        damped by a thermal stabilisation assembly (TSA; \citealt{planck2011-1.3}) down to about 100\,mK peak-to-peak at low frequency.
        
\begin{figure*}
            \begin{center}
                \includegraphics[width=18cm]{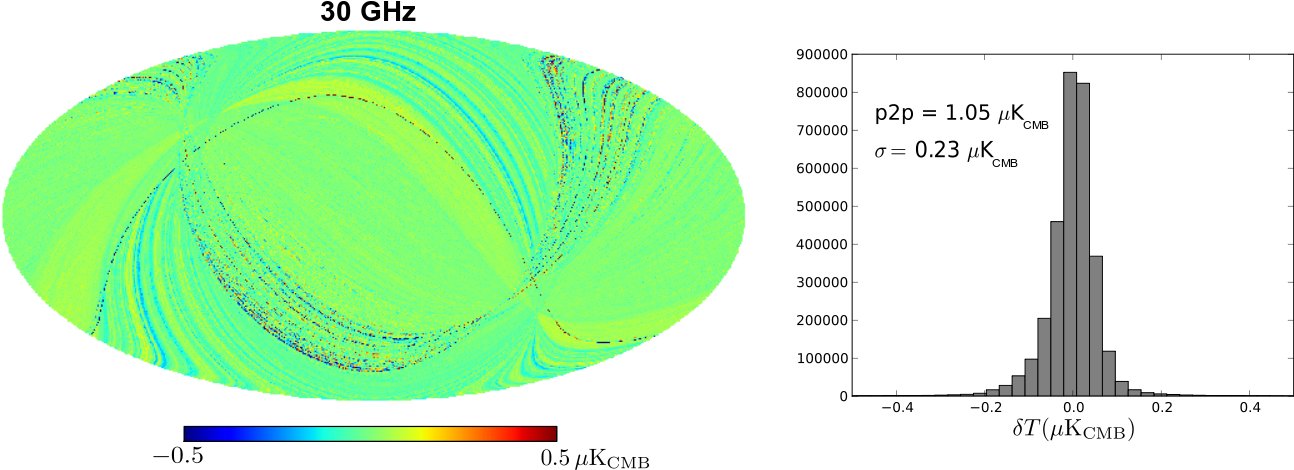}
                \includegraphics[width=18cm]{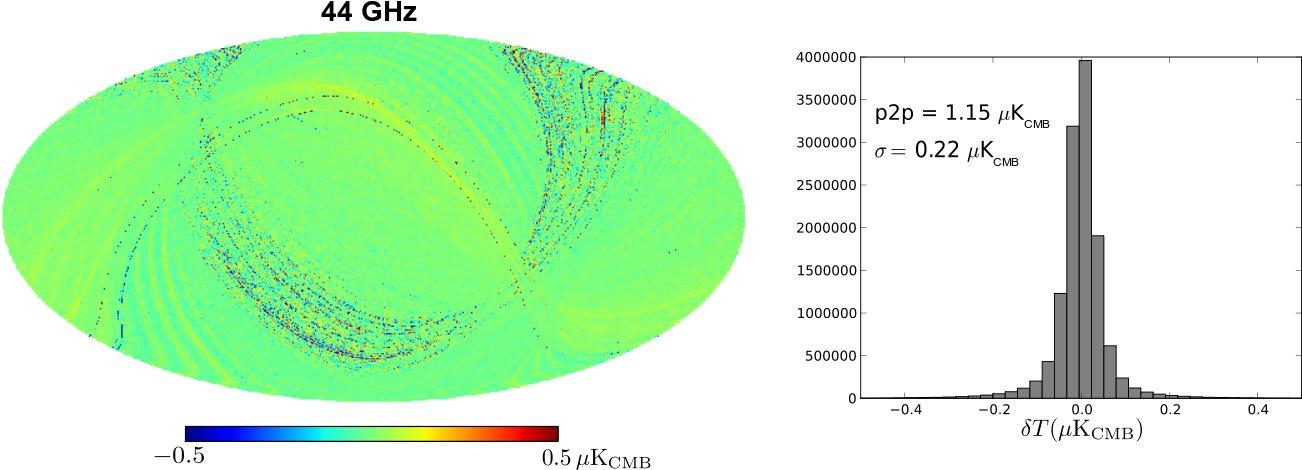}
                \includegraphics[width=18cm]{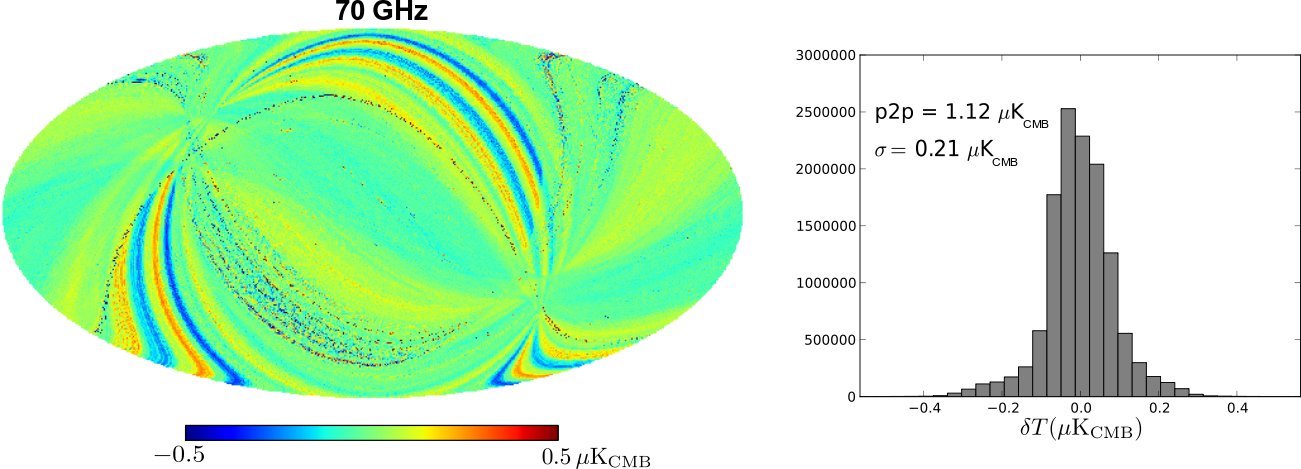}
            \end{center}
            \caption{Simulation of the systematic effects of temperature fluctuations of the LFI focal 
            plane unit. Maps are $N_{\rm side}=512$ at 30\,GHz, and $N_{\rm side}=1024$ at 44 and 70\,GHz.}
            \label{fig_fem_maps}
        \end{figure*}

        During the second half of the first year of operations the expected cooler degradation led to an increase of 
        the cold end temperature and the low-frequency fluctuations.   At the radiometers the temperature increased 
        by about 0.2\,K, with fluctuations running about 30\,mK peak-to-peak.  We 
        estimated the effect of these fluctuations on the science data using ground-measured transfer functions 
        that include the effect of temperature on both gain and noise.  
        Figure~\ref{fig_fem_maps} shows the calculated effects for the three LFI 
        frequency channels.  As in the case of back-end temperature fluctuations, 
        the overall effect is $\sim1\,\mu\mathrm{K_{CMB}}$ peak-to-peak and $0.2\,\mu\mathrm{K_{CMB}}$ rms.

\subsubsection{4\,K reference load temperature fluctuations}
\label{sec_4k_fluctuations}

    The temperature stability of the 4\,K reference loads, attached to the HFI 4\,K mechanical box, 
    is a key factor in the LFI systematic effects budget.  Temperature fluctuations of the loads are driven primarily 
    by fluctuations at the 20\,K cold-end interface, which propagate to the LFI reference loads by conduction 
    through the HFI mounting struts and radiation from the \hbox{LFI}.  See \cite{planck2011-1.3} for details. 
        
    The 70\,GHz reference loads are mounted close to the HFI 4\,K plate and are stabilised 
    by proportional-integral-derivative (PID) active controllers that remove almost completely low frequency fluctuations. 
    The loads at 30 and 44\,GHz are farther from the HFI 4\,K plate and closer to 
    the HFI mounting struts; their temperature is less stable. Details of temperature sensor measurements are provided 
    in \citet{planck2011-1.3}.

\begin{figure*}
        \begin{center}
            \includegraphics[width=18cm]{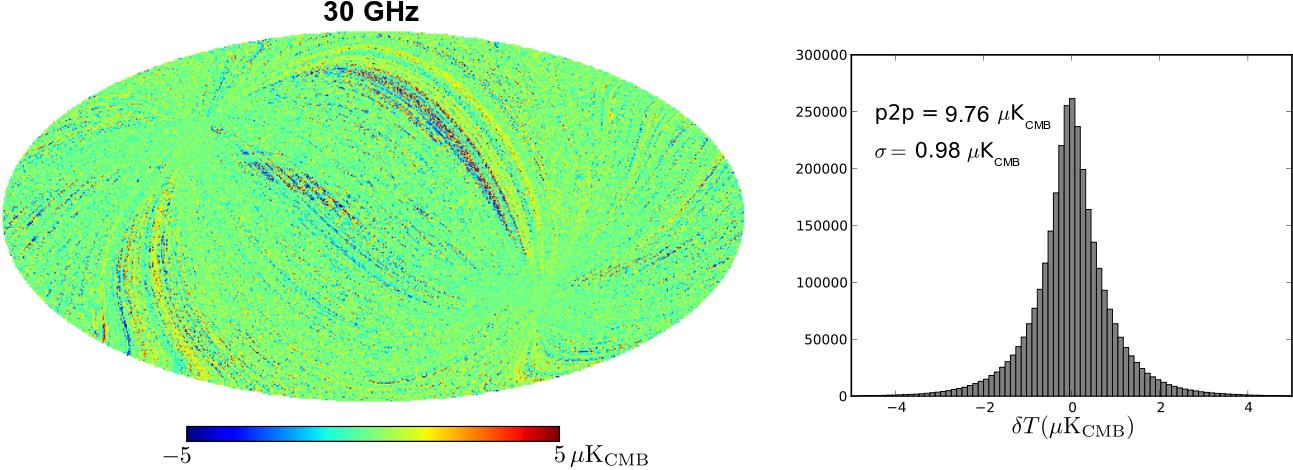}
            \includegraphics[width=18cm]{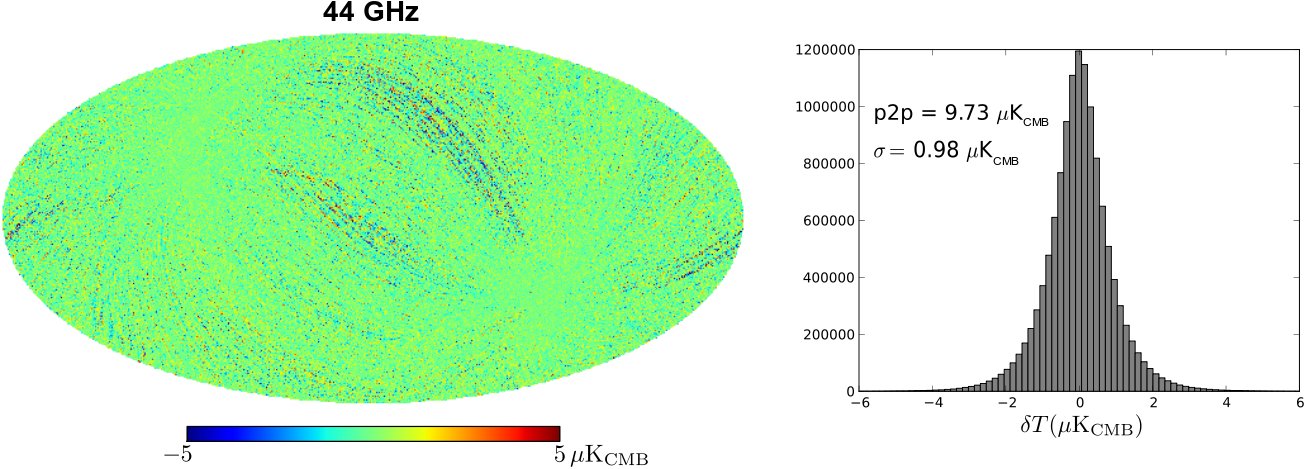}
            \includegraphics[width=18cm]{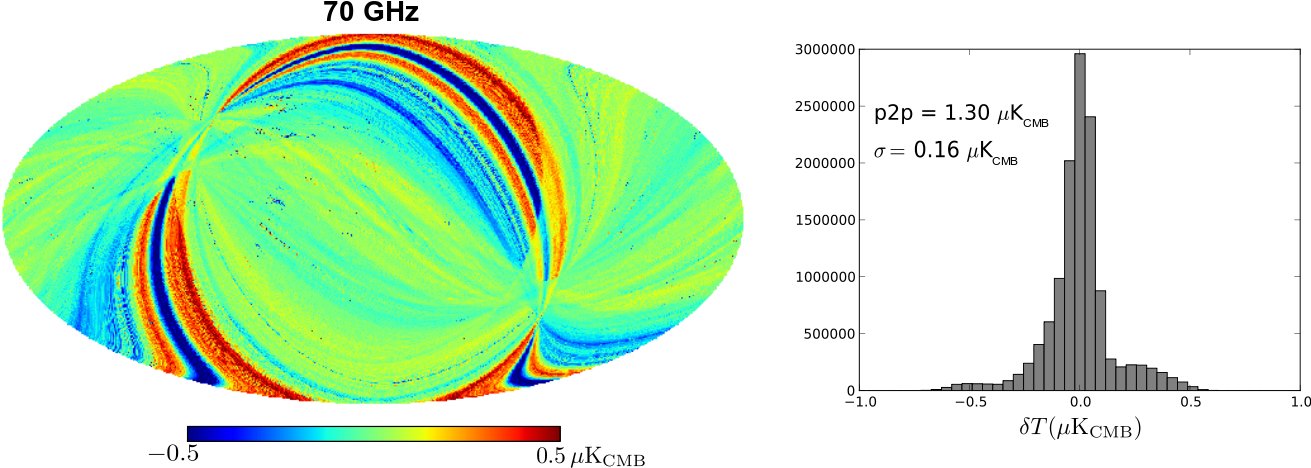}
        \end{center}
        \caption{Simulation of the systematic effects of temperature fluctuations of the 4\,K reference loads. 
            Maps are $N_{\rm side}=512$ at 30\,GHz, and $N_{\rm side}=1024$ at 44 and 70\,GHz. Note that the scale of 30 and 
            44\,GHz maps is 10 times larger than then scale of the map at 70\,GHz}
        \label{fig_4k_maps}
    \end{figure*}

    This difference is reflected in the maps shown in Fig.~\ref{fig_4k_maps}, where 
    the residual systematic effect in the 30 and 44\,GHz channels is $\sim 10$ times larger 
    than in the 70\,GHz channel.  Nevertheless, given the reduction provided by destriping, the residual systematic is 
    several times below the instrument noise, as shown by the angular 
    power spectra in Fig.~\ref{fig_condensed_ps}.  Further work will be aimed at correlating 
    radiometric and housekeeping data to further reduce the residual effect caused by these fluctuations. 
        
\section{Summary of main performance parameters}
\label{sec_performance_summary}

    Table~\ref{tab_summary_performance} gives a top-level summary of instrument performance parameters measured in 
flight. Optical properties have been successfully reconstructed using Jupiter transits and the main parameters are in agreement 
with pre-launch estimates. Also the white noise sensitivity agrees with ground measurements; for 
some channels sensitivity improved after in-flight bias tuning. Parameters describing the $1/f$ noise component 
are in line with ground measurements and the 50\,mHz requirement except at 30\,GHz.  That channel that has 
a knee frequency over 100\,mHz, which is, however, well-handled by destriping.  Absolute photometric calibration based on the CMB 
dipole yields an overall statistical uncertainty of $\lesssim 1\%$.  Variations due to slow instrumental 
variations are traced by the calibration pipeline, yielding an overall uncertainty ranging between 0.05\% 
and 1\%. The residual systematic uncertainty is of the order of 1\muK\ rms per pixel.  Average colour corrections are 
provided in Table~\ref{tab_colour_corrections}.

\begin{table*}                    
\begingroup
\newdimen\tblskip \tblskip=5pt
\caption{Summary of main LFI performance parameters.}
\label{tab_summary_performance}
\nointerlineskip
\vskip -3mm
\footnotesize
\setbox\tablebox=\vbox{
   \newdimen\digitwidth 
   \setbox0=\hbox{\rm 0} 
   \digitwidth=\wd0 
   \catcode`*=\active 
   \def*{\kern\digitwidth}
   \newdimen\signwidth 
   \setbox0=\hbox{+} 
   \signwidth=\wd0 
   \catcode`!=\active 
   \def!{\kern\signwidth}
\halign{\hbox to 2.7in{#\leaderfil}\tabskip=3em&
        \hfil#\hfil&
        \hfil#\hfil&
        \hfil#\hfil\tabskip=0pt\cr 
\noalign{\doubleline}
\omit\hfil Parameter\hfil&30\,GHz&44\,GHz&70\,GHz\cr
\noalign{\vskip 3pt\hrule\vskip 5pt}
Centre frequency [GHz]&\getsymbol{LFI:center:frequency:30GHz}&\getsymbol{LFI:center:frequency:44GHz}&\getsymbol{LFI:center:frequency:70GHz}\cr
\noalign{\vskip 3pt}
FWHM [\arcm]&\getsymbol{LFI:FWHM:30GHz}&\getsymbol{LFI:FWHM:44GHz}&\getsymbol{LFI:FWHM:70GHz}\cr
\noalign{\vskip 3pt}
Ellipticity&1.38&1.28&1.26\cr
\noalign{\vskip 3pt}
White noise sensitivity [$\,\mu\mathrm{K}_{\rm CMB}\, \mathrm{s}^{1/2}$]&146.8&173.1&152.6\cr
\noalign{\vskip 3pt}
$f_{\rm knee}$ [mHz]&113.7&56.2&29.5\cr
\noalign{\vskip 3pt}
$1/f$ slope&$-0.87$&$-0.89$&$-1.03$\cr
\noalign{\vskip 3pt}
Absolute calibration uncertainty[\%]&1&1&1\cr
\noalign{\vskip 3pt}
Relative calibration uncertainty [\%]&0.05&0.07&0.12\cr
\noalign{\vskip 3pt}
Systematic effects uncertainty [$\,\mu\mathrm{K}_{\rm CMB}$]&1.10&0.98&0.45\cr
\noalign{\vskip 5pt\hrule\vskip 3pt}}}
\endPlancktable
\endgroup
\end{table*}


\section{Conclusions}
\label{sec_conclusions}

    We have discussed the scientific performance of the \Planck\ Low Frequency Instrument after one year of operations. 

Since the start of \Planck\ nominal operations, the LFI has shown excellent stability in all measured 
parameters. The instrument uninterruptedly observed the microwave sky with negligible data loss and less 
than 1\% discarded data because of anomalies. Typical variations in the 
instrumental output were less than 1\% on time scales of several days and were mainly driven by slow thermal fluctuations.

The main beams have been reconstructed down to $-20$\,dB using Jupiter as a source, with results 
closely matching those expected from simulations.  In-flight measurements of white noise sensitivity 
are in very good agreement with ground results, with significant improvements in some channels thanks to 
in-flight bias tuning.  The impact of low frequency noise is very small, especially at 70\,GHz 
where the measured knee frequency is about 30\,mHz, almost twice better 
than the requirement.  At 30\,GHz the $1/f$ noise is higher than measured on the ground, 
with a knee frequency of about 100\,mHz.  Residual fluctuations in the timestreams, however, 
are effectively removed during map-making, as expected from several pre-launch simulation studies 
\citep{kurki-suonio2009,poutanen2004,keihanen2004}.  We have shown that the noise increase due to 
the $1/f$ component is a few percent at 30 and 44\,GHz, and only 0.2\% at 70\,GHz.

Photometric calibration is based on the CMB dipole via an iterative procedure explained in 
\citet{planck2011-1.6}. Excellent absolute calibration accuracy of $\lesssim 1\%$ will likely be 
improved still further in future analyses by using the orbital dipole as absolute calibrator.  Our current 
relative calibration traces gain variations on timescales larger than 5--10 days, 
yielding an overall statistical accuracy in the range 0.05--0.1\%.  Thermally-driven 
gain fluctuations on smaller timescales are currently not implemented in our gain model and contribute as a systematic 
uncertainty in the final maps.  A new version of the gain model, 
now being developed, will take into account the effect of such fluctuations to further reduce this residual uncertainty.

We have presented a preliminary analysis of all the systematic effects that are relevant at 
this stage of the analysis. Their impact on LFI science has been evaluated by 
projecting each effect on full-sky maps and angular power spectra through a dedicated pipeline that 
exploits radiometric and housekeeping data in conjuction with external parameters measured in ground or flight 
tests, such as thermal susceptibilities or spike characteristics. The combined residual effect from 
frequency spikes and thermal instabilities (in the 4\,K, 20\,K, and 300\,K stages) is several 
order of magnitudes below the instrumental noise at all angular scales smaller than 10\deg, and less than 10\% at larger scales.

The overall performance of LFI as measured in-flight demonstrates an
instrument with unprecedented combination of sensitivity, angular
resolution and suppression of systematic errors for full-sky imaging at
these frequencies. In particular, the control of systematic effects and
non-white noise components represent key challenges for accurate
extraction of the cosmological information encoded in the temperature and
polarisation maps. Our preliminary assessment shows that, even without
dedicated de-correlation of thermal effects, the LFI is already largely
immune to instrumental effects, with prospects of further suppression
after implementing a more representative gain model and temperature
fluctuation removal algorithms.

One more year of continuous observations is currently planned for \Planck, with a further LFI 
extension to the maximum lifetime allowed by the sorption cooler.  Everything to date suggests that the instrument 
will maintain its performance throughout the remaining period and provide rich and 
high-quality scientific data that will be explored for many years to come.


\begin{acknowledgements}
    \Planck\ is too large a project to allow full acknowledgement of all contributions by individuals, institutions, industries, and funding agencies. The main entities involved in the mission operations are as follows. The European Space Agency operates the satellite via its Mission Operations Centre located at ESOC (Darmstadt, Germany) and coordinates scientific operations via the \Planck\ Science Office located at ESAC (Madrid, Spain). Two Consortia, comprising around 50 scientific institutes within Europe, the USA, and Canada, and funded by agencies from the participating countries, developed the scientific instruments LFI and HFI, and continue to operate them via Instrument Operations Teams located in Trieste (Italy) and Orsay (France). The Consortia are also responsible for scientific processing of the acquired data. The Consortia are led by the Principal Investigators: J.-L. Puget in France for HFI (funded principally by CNES and CNRS/INSU-IN2P3) and N. Mandolesi in Italy for LFI (funded principally via ASI). NASA's US \Planck\ Project, based at JPL and involving scientists at many US institutions, contributes significantly to the efforts of these two Consortia. In Finland, the Planck LFI 70 GHz work was supported by the Finnish Funding Agency for Technology and Innovation (Tekes). This work was also supported by the Academy of Finland, CSC, and DEISA (EU).
    
    Some of the results in this paper have been derived using the HEALPix package \citep{gorski2005}.
    
    A description of the Planck Collaboration and a list of its members can be found at \burl{http://www.rssd.esa.int/index.php?project=PLANCK&page=Planck\_Collaboration}.

\end{acknowledgements}


\bibliographystyle{aa}

\bibliography{Planck_bib,early_papers,custom}

\raggedright
\end{document}